\documentclass[article,prc,epsfig,floatfix,nofootinbib,aps]{revtex4}
\usepackage{epsfig}
\begin{document}

\title{Shadowing Effects on $J/\psi$ and $\Upsilon$ Production
at the LHC}

\author{R. Vogt}

\address{{
Lawrence Livermore National Laboratory, Livermore, CA 94551, USA \break
and \break
Physics Department, University of California at Davis, Davis, CA 95616, 
USA} \break}

\begin{abstract}%new
\noindent{\bf Background:} 
Proton-nucleus collisions have been used as a intermediate
baseline for the determination of cold medium effects.  They lie between 
proton-proton collisions in vacuum and nucleus-nucleus collisions which are
expected to be dominated by hot matter effects.  Modifications of the quark
densities in nuclei relative to those of the proton are well established 
although those of the gluons in the nucleus are not well understood.  We
focus on the effect of these on quarkonium production in
proton-lead collisions at the LHC at a center of mass energy of 5.02 TeV.\\
{\bf Purpose:} We determine whether it is possible for the LHC proton-lead data
to be described by nuclear modifications of the parton densities, referred to
as shadowing, alone.  We compare our results to the nuclear modification factor
and to the forward-backward ratio, both as a function of transverse momentum,
$p_T$, and rapidity, $y$.\\
{\bf Methods:} We employ the color evaporation model of quarkonium production
at next-to-leading order in the total cross section and leading order in the
transverse momentum dependence.  We use the EPS09 NLO modifications as a 
standard of comparison.  We study the effect of the proton parton density and
the choice of shadowing parameterization on the $p_T$ and rapidity dependence
of the result.  We check whether or not the calculations at leading and 
next-to-leading order give consistent shadowing results.\\
We also investigate the size of the mass and scale uncertainties
relative to the uncertainty on the shadowing parameterization.  Finally, we
check whether the expected cold matter effect in nucleus-nucleus collisions 
can be modeled as the product of proton-nucleus results at forward and backward
rapidity.\\
{\bf Results:} We find that the rapidity and $p_T$ dependence of the nuclear
modification factor are generally consistent with the next-to-leading order
calculations in the color evaporation model.  The forward-backward ratio is
more difficult to describe with shadowing alone.  The leading and 
next-to-leading order calculations are inconsistent for EPS09 while other
available parameterizations are consistent.  The mass and scale uncertainties
on quarkonium production are larger than those of the nuclear parton 
densities.\\
{\bf Conclusions:} While shadowing is consistent with the nuclear suppression
factors within the uncertainties, it is not consistent with the measured 
forward-backward asymmetry, especially as a function of transverse momentum.
Data from $p+p$ collisions at the same energy are needed.
\end{abstract}

\maketitle

\section{Introduction}
\label{Sec:Intro}

In this paper we concentrate on comparison to the 2013 LHC $p+$Pb data on 
quarkonium production at $\sqrt{s_{NN}} = 5.02$ TeV.  The inclusive $J/\psi$ 
and $\Upsilon$ production data, binned as
a function of rapidity and $p_T$ ($J/\psi$) or rapidity alone ($\Upsilon$), 
come from the ALICE 
\cite{ALICEpPbpsi,ALICEpPbpsi_pT,ALICEpPbUps} and 
LHCb \cite{LHCbpPbpsi,LHCbpPbups}
collaborations.  We briefly discuss the detector acceptances for each 
experiment in turn here but provide a more complete description of the data
in Sec.~\ref{Sec:Data}.

Runs with mass asymmetric beams are different at the LHC than at 
RHIC because the LHC beams are not symmetric in 
energy.  
Instead, in the 2013 LHC $p+$Pb run a 4 TeV proton beam interacted with a 
$4(Z_{\rm Pb}/A_{\rm Pb}) = 1.58$ TeV/nucleon Pb beam.  The nucleon-nucleon
center of mass frame does not coincide with the laboratory frame in an 
energy-asymmetric collision system.  Instead the center of mass frame is
shifted by $\Delta y = 0.465$, taken to be in the direction of the proton 
beam, as defined by experiments.  In addition,
the ALICE and LHCb detector systems are not symmetric around the interaction
point, with muon spectrometers on only one side of midrapidity.  Therefore
the beams have to be run in two modes, Pb$+p$ and $p+$Pb.  In the
first case the lead beam is defined to move toward forward rapidity while 
in the second, the proton beam does.  Because the second configuration is 
most similar to fixed-target operation and corresponds to small parton momentum 
fractions in the nucleus, both setups are analyzed according to the convention 
that the proton beam moves to positive rapidity.  Thus the case of Pb$+p$
collisions corresponds to larger parton momentum fractions in the nucleus.

The ALICE $J/\psi$ measurement has been presented 
in both the central and forward/backward
regions.  Their muon spectrometer covers the pseudorapidity range 
$-4 < y_{\rm lab} < -2.5$ in the laboratory frame.  Due to the rapidity
shift in the asymmetric energy beams,
the backward rapidity range for dimuon coverage is $-4.46 < y_{\rm cms} < -2.96$
while the forward rapidity range for dimuon coverage is 
$2.03 < y_{\rm cms} < 3.53$.  The ALICE central detector includes dielectron
coverage for $|y_{\rm lab} < 0.8|$.  
The quarkonium measurements in this region correspond to 
the rapidity range $-1.37 < y_{\rm cms} < 0.43$.  

The LHCb detector covers $2 < y_{\rm lab} < 5$ in the laboratory frame.  
Like ALICE,
they define the forward direction as the direction of the proton beam so
that their backward coverage is $-5 < y_{\rm cms} < -2.5$ and the forward
coverage is $1.5 < y_{\rm cms} < 4$.

Because there is not yet a
measured $p+p$ baseline, the nuclear modification factor, 
\begin{eqnarray}
R_{p{\rm Pb}}(y,p_T) = \frac{d\sigma_{p{\rm Pb}}(y,p_T)/dy dp_T}{T_{p{\rm Pb}}
d\sigma_{pp}(y,p_T)/dy dp_T} \, \, ,
\label{eqn:rpa}
\end{eqnarray}
relies on an interpolation of the $p+p$ cross section.   
The rapidity shift was taken into account to obtain the $p+p$
cross section in the rapidity ranges of the $p+$Pb measurement.  The factor
$T_{p{\rm Pb}}$ in the denominator of Eq.~(\ref{eqn:rpa}) takes the centrality of
the collision into account and is calculated in a Glauber model 
\cite{ALICEpPbpsi}.  
In this paper, we concentrate only on the minimum
bias results.  See Ref.~\cite{FMV} for a discussion of the centrality dependence
of the $J/\psi$ measurement at RHIC.   Work is in progress on the centrality
dependence at the LHC \cite{TonyandMe}.

In addition to studying the nuclear modification factors in
these rapidity ranges, the forward-backward ratio,
\begin{eqnarray}
R_{FB}(y,p_T) = 
\frac{d\sigma_{p{\rm Pb}}(y>0,p_T)/dy dp_T}{d\sigma_{p{\rm Pb}}(y<0,p_T)/dy dp_T} 
\, \,  \, ,
\label{eqn:rfb}
\end{eqnarray}
has also been presented \cite{ALICEpPbpsi,LHCbpPbpsi}.
The unmeasured quantities $T_{p{\rm Pb}}$ and 
$\sigma_{pp}$ cancel for a rapidity range symmetric around $y_{\rm cms} = 0$.  
Therefore $R_{FB}$ is formed in the rapidity region where the
forward and backward acceptances completely overlap,
$2.96 < |y_{\rm cms}| < 3.53$ for ALICE and 
$2.5 < |y_{\rm cms}| < 4$ for LHCb.  
Some systematic uncertainties also cancel in the ratio. 
This ratio is perhaps a more faithful representation 
of cold matter effects on the 
$p+$Pb cross section.  However, for theoretical interpretation, 
$R_{p{\rm Pb}}$ is still desirable because even
a wrong model can produce the right ratio.

In a previous paper \cite{RV10}, predictions were made for $pA$ collisions at
$\sqrt{s_{_{NN}}} = 8.8$ and 5.5 TeV and ratios were formed both to $p+p$ 
collisions at the same energy and to $p+p$ collisions at the anticipated top
energy of $\sqrt{s} = 14$ TeV.  These calculations were made before the LHC
turned on and so did not employ the same energies at which data were ultimately
taken in LHC Run I: $\sqrt{s} = 2.76$, 7 and 8 TeV for $p+p$ collisions and
5.02 TeV for $p+$Pb.  These calculations assumed that the leading and 
next-to-leading order treatments of the modifications of the parton densities 
in nuclei, when employed consistently, would be identical, as discussed in more 
detail later.  

More recently, calculations were made for the nuclear modification 
factor as a function of $y$ and $p_T$ at the energy appropriate for the 
$p+$Pb run \cite{Albacete} but not taking the rapidity shift into account for
the $p_T$ acceptance.  In 
addition, the $p_T$ dependent ratio was presented for forward rapidity only.
These predictions, along with other, updated, CEM calculations with EPS09 NLO
nuclear parton densities (nPDFs)
were compared to the ALICE and LHCb $J/\psi$ and $\Upsilon$ data 
\cite{ALICEpPbpsi,ALICEpPbpsi_pT,ALICEpPbUps,LHCbpPbpsi,LHCbpPbups,RVHP,RVQM,RVGerry}.  In those calculations, the incorrect
factorization scale, $\mu_F$, was passed to the nuclear parton densities.
(The square of the scale was passed instead of the scale itself, as required
by most shadowing parameterizations, so that the overall 
cold matter effect was reduced
relative to the true value.  This was not the case for the LO predictions in
Ref.~\cite{RV10}.)  In this paper, this error is corrected and the 
$p_T$-dependent ratios calculated in the CEM are presented for the first time.

In the next section, Sec.~\ref{Sec:Data}, we provide a brief summary of the 
ALICE and LHCb $J/\psi$ and
$\Upsilon$ measurements to place our calculations in context.

We then summarize our calculation of quarkonium production in
$p+p$ collisions in Sec.~\ref{Sec:Vacuum}.  
We compare the $J/\psi$ and $\Upsilon$
distributions obtained with several different
sets of proton parton densities.  Comparison of these calculations to available
$p+p$ data can be found in Refs.~\cite{NVF,NVFinprep}.
A short summary of cold nuclear matter effects is given in Sec.~\ref{Sec:CNM}. 

The nuclear
parton density modifications used in this paper are described in 
Sec.~\ref{Sec:params}.
The calculations are compared to the
$p+$Pb data from ALICE and LHCb on $J/\psi$ and 
$\Upsilon$ production at $\sqrt{s_{_{NN}}} = 5$ TeV in Sec.~\ref{Sec:results}.
We calculate both $R_{p{\rm Pb}}$ and $R_{FB}$ as functions of rapidity and
transverse momentum.
The data are first compared to the EPS09 uncertainty bands.  We contrast the
leading order (LO) and next-to-leading order (NLO) 
results for order-by-order consistency.  Next, the data are
compared to all the nuclear parton densities discussed in Sec.~\ref{Sec:params} 
to see if
any parameterizations are particularly favored.  The mass and scale dependence
of the results is also shown.  Finally, we discuss how closely the $A+A$ 
calculations can be reproduced by a convolution of $p+A$ and $A+p$ collisions
and also compare to the RHIC data.

\section{Description of the $p+$Pb Quarkonium Data}
\label{Sec:Data}

In this section, we describe the quarkonium data for $J/\psi$ 
and $\Upsilon(1S)$ from ALICE and LHCb.  
Since there has been no $p+p$ run at $\sqrt{s} = 5$ TeV to date, 
the denominator of
$R_{p{\rm Pb}}$ has to be interpolated between available measurements at other
energies.  The $p+p$ cross sections also had to be adjusted to the rapidity 
ranges of the $p+$Pb measurement.

The interpolation
methods used by the collaborations depend on the quarkonium state,
the observable, and the previously available data.  However, in all cases,
the quarkonium states were assumed to be produced unpolarized.

To obtain $R_{p{\rm Pb}}(y)$ for the $J/\psi$, the ALICE Collaboration
used an energy interpolation between their $p+p$ measurements at 2.76 and 7 TeV
to obtain the $\sqrt{s}$ dependence of the $p+p$ cross section.  They presented
their $J/\psi$ $p+$Pb data in both a single rapidity
interval as well as in six rapidity bins 
measured for $p+p$ collisions (of course without the rapidity shift in $p+p$).
These first results were for $R_{p{\rm Pb}}$ at forward and backward rapidity
and $R_{FB}$ as a function of $p_T$ and $y$ \cite{ALICEpPbpsi}.  

Their energy interpolation was based on three assumed shapes: linear, power law
and exponential.  The central value of the result for each rapidity bin is an
average of the three shapes while the uncertainty is the quadrature sum of a
term related to the uncertainty on the data used for the interpolation and a
term related to the spread between results with different shapes.  

An additional small
systematic uncertainty was obtained by comparing the shapes with those of the
leading order CEM and the fixed-order next-to-leading logarithm (FONLL) approach
for inclusive open heavy flavor production\footnotetext{Note that the FONLL 
approach calculates the single 
inclusive heavy flavor distributions, not those of the $Q \overline Q$ pair
as in the CEM.}
\cite{ALICEpPbpsi}.  We note that the LO and NLO CEM energy dependence should be
similar if the same mass and scale parameters, as well as the same proton
parton densities are used. (The shapes will not be similar if {\it e.g.}
CTEQ6M is used at NLO and CTEQ61L is used at LO.)  Using the FONLL calculation
for the total $c \overline c$ cross section may produce a shape similar to the
CEM but the magnitude, of course, will be quite different.

In a later paper, the ALICE Collaboration presented results for the midrapidity
$R_{p{\rm Pb}}$, a bin around $-1.37 < y_{\rm cms} < -0.43$ to add to 
$R_{p{\rm Pb}}(y)$ as well as the ratios $R_{p{\rm Pb}}(p_T)$ at forward, backward
and midrapidity \cite{ALICEpPbpsi_pT}.  While the forward-backward ratio as 
a function of $p_T$ was published in Ref.~\cite{ALICEpPbpsi}, the separate
values of $R_{p{\rm Pb}}(p_T)$ were not yet available. 

The $p+p$ baseline for the $p_T$-dependent ratios was
obtained through interpolation.
At midrapidity, data from $\sqrt{s} = 0.2$, 1.96, 2.76 and 7 TeV were used.
The 1.96 TeV results from $p \overline p$ collisions from the Tevatron were
considered on the same basis as $p+p$ collisions because, at these high energies,
production is dominated by the $gg$ process.  Scaling in $x_T = m_T/\sqrt{s}$
was used to compare the disparate energies.  Only exponential, logarithmic and
power law dependencies were considered in this case because there is no $p_T$
dependence in the LO CEM and the FONLL approach is for single inclusive 
distributions, not pairs, so the $p_T$ slope is not available from these
calculations.

At forward rapidity, the only data available to include in the $\sqrt{s}$
interpolation are the 2.76 and 7 TeV data from ALICE.  (They did not employ
the LHCb results in their interpolation.)  In this region, the dependencies
were linear, power law and exponential.  The results for $R_{p{\rm Pb}}(p_T)$ 
are only shown up to 8 GeV because the $p+p$ data were limited to this $p_T$
range.  The non-prompt $J/\psi$ production from $b$ decays increases
with $p_T$, giving a $\sim 20$\% correction at $p_T \sim 8$ GeV 
\cite{ALICEpPbpsi_pT}. 

ALICE has also measured the $\psi'$ in $p+$Pb collisions,
finding significantly more suppression \cite{ALICEpPbpsip}.  Since this 
difference cannot be due to initial state effects alone, we do not address
that result in this work.

The ALICE Collaboration also measured the inclusive $\Upsilon$(1S) and
$\Upsilon$(2S) rates.  The rapidity dependence of $R_{p{\rm Pb}}$ was reported in
Ref.~\cite{ALICEpPbUps}.  
The $\Upsilon$ yields are not large so that
only one rapidity bin is reported.  To obtain the $p+p$ baseline for 
$R_{p{\rm Pb}}$, they used the LHCb results for $\Upsilon$ production at
$\sqrt{s} = 2.76$, 7 and 8 TeV, divided into rapidity bins.  They employed 21
different shapes, 15 from the LO CEM with different proton PDFs and 
factorization scale choices; 3 based on the FONLL $b$ quark distributions;
while linear, power-law, and exponential shapes rounded out the set.  
The agreement
of all the shapes with the data was generally poor so the fits with the worst
$\chi^2$/dof were discarded for the final fits.  In addition to the $J/\psi$
uncertainties described above, they also considered small rapidity shifts
between the ALICE and LHCb rapidity bins.  

They found that the $\Upsilon$ 
$R_{p{\rm Pb}}$ is quite similar to that of $J/\psi$ at forward rapidity while
at negative rapidity the $\Upsilon$ $R_{p{\rm Pb}}$ is compatible with unity but
lower than that of the $J/\psi$.  This is a fairly remarkable result since 
nuclear effects are generally expected to be smaller for the
$\Upsilon$ than for $J/\psi$
so that the $\Upsilon$ $R_{p{\rm Pb}}$ should be closer to unity at low $p_T$
where the difference in mass is the dominant effect.  At higher $p_T$, where
$p_T \gg m$, the results should be similar for the two quarkonium states.

They reported the $\Upsilon({\rm 2S})/\Upsilon({\rm 1S})$ ratios at forward
and backward rapidity, $0.26 \pm 0.09 \pm 0.04$ and $0.27 \pm 0.08 \pm 0.04$
respectively \cite{ALICEpPbUps}.  
The result is consistent with the $p+p$ ratio at 7 TeV.  This is
also consistent with shadowing being the dominant cold matter effect since
it affects the excited states the same way as the ground state.

The LHCb Collaboration has also measured $J/\psi$ \cite{LHCbpPbpsi}
and $\Upsilon$ \cite{LHCbpPbups} production in their muon spectrometer.
In addition to the inclusive $J/\psi$ result, they also separate 
$b \rightarrow J/\psi$ decays to present a non-prompt $J/\psi$ result.
Their primary functional dependence to interpolate between their $p+p$ results
at 2.76, 7 and 8 TeV is a power law, $\sigma(\sqrt{s}) = (\sqrt{s}/p_0)^{p_1}$.
They use linear and exponential dependencies to obtain a systematic uncertainty
on their interpolation.  They do not use any production models for the
energy interpolation \cite{LHCbpPbpsi}.  A similar, power-law-based method is 
used to obtain the $p+p$ baseline for $\Upsilon$ production \cite{LHCbpPbups}.

Recently, the ATLAS Collaboration has presented results on the forward-backward
$J/\psi$ ratio as a function of $y$ in the region $8 < p_T < 30$ GeV and
as a function of rapidity in the rapidity range $|y_{\rm cms}| < 1.94$
\cite{ATLASpPbpsi}.  Their results for $R_{FB}$ are consistent with unity 
within the uncertainties of the data.  CMS also recently presented
the prompt $J/\psi$ $R_{FB}$ as a function of $p_T$ and as a function of 
rapidity for high $p_T$ \cite{CMSpPbpsi}.  Their results are
consistent with those of ATLAS.

The LHCb $\Upsilon$ measurement includes low statistics for the $\Upsilon$(3S)
states, as well as the 1S and 2S states.  The ratios 
$\Upsilon({\rm 2S})/\Upsilon({\rm 1S})$ are $0.28 \pm 0.14 \pm 0.05$ in the
backward direction and $0.20 \pm 0.05 \pm 0.01$ at forward rapidity, both 
consistent with $p+p$ measurements \cite{LHCbpPbups}.  While the 
$\Upsilon({\rm 3S})/\Upsilon({\rm 1S})$ ratios are also consistent with
those in $p+p$ collisions, their low statistics reduces their significance.

Finally, we note that we do not discuss the intriguing 
CMS $\Upsilon(n{\rm S})/\Upsilon({\rm 1S})$
ratios measured as a function of both the number of tracks at midrapidity and
the transverse energy at forward rapidity \cite{CMSupsrats}.  
This effect, also seen in $p+p$
collisions, is not attributable to initial-state modifications of the parton
densities and, as such, is outside the scope of this work.

The quarkonium yields for all the ALICE and LHCb 
results discussed here are given in Table~\ref{ALICE-LHCb-yields}.

\begin{table}[t]
\begin{center}
\begin{tabular}{|c|c|c|c|c|c|}\hline
Experiment & $y$ acceptance & $N_{J/\psi}$ [Ref.] & $N_{\Upsilon ({\rm 1S})}$ [Ref.]  
& $N_{\Upsilon ({\rm 2S})}$ [Ref.] & $N_{\Upsilon ({\rm 3S})}$ [Ref] \\ \hline
ALICE & $2.03 < y_{\rm cms} < 3.53$ & $(6.69 \pm 0.05) \times 10^4$ 
\protect\cite{ALICEpPbpsi}  & $305 \pm 34$ \protect\cite{ALICEpPbUps} & 
$83 \pm 23$ \protect\cite{ALICEpPbUps} & - \\ 
      & $-4.46 < y_{\rm cms} < -2.96$ & $(5.67 \pm 0.05) \times 10^4$
\protect\cite{ALICEpPbpsi} & $161 \pm 21$ \protect\cite{ALICEpPbUps} & 
$42 \pm 14$ \protect\cite{ALICEpPbUps} & - \\
      & $-1.37 < y_{\rm cms} < -0.43$ & $465 \pm 37$ \protect\cite{ALICEpPbpsi_pT}
& - & - & - \\ \hline
LHCb  & $2.5 < y_{\rm cms} < 4.0$ & $25280 \pm 240$ \protect\cite{LHCbpPbpsi}
& $189 \pm 16$ \protect\cite{LHCbpPbups} & $41 \pm 9$ \protect\cite{LHCbpPbups}
& $13 \pm 7$ \protect\cite{LHCbpPbups} \\
      & $-4.0 < y_{\rm cms} < -2.5$ & $8830 \pm 160$ \protect\cite{LHCbpPbpsi}
& $72 \pm 14$ \protect\cite{LHCbpPbups} & $17 \pm 10$ \protect\cite{LHCbpPbups}
& $4 \pm 8$ \protect\cite{LHCbpPbups} \\ \hline
\end{tabular}
\end{center} 
\caption[]{The $J/\psi$ and $\Upsilon$ yields from the ALICE and LHCb
collaborations in their stated rapidity acceptance.  The results are integrated
over all $p_T$.}
\label{ALICE-LHCb-yields}
\end{table}

\section{Production in $p+p$ Collisions}
\label{Sec:Vacuum}

Following our previous work \cite{RV10,NVF}, we treat quarkonium production
within the color  
evaporation model (CEM).  In the CEM,
heavy flavor and quarkonium production are treated on an equal footing.  
The CEM has enjoyed considerable phenomenological success when 
applied at NLO in the total cross section and LO
in the quarkonium $p_T$ distribution \cite{Gavai:1994in,Amundson,SchulerV,NVF}. 
(See Ref.~\cite{NVF} for comparison of the $\sqrt{s} = 2.76$ and 7 TeV
ALICE data with the same CEM calculation employed here.)

\subsection{Color Evaporation Model Calculation}
\label{SubSec:CEM}

In the CEM, the quarkonium 
production cross section is some fraction, $F_C$, of 
all $Q \overline Q$ pairs below the $H \overline H$ threshold where $H$ is
the lowest mass heavy-flavor hadron.  Thus the CEM cross section is
simply the $Q \overline Q$ production cross section with a cut on the pair mass
but without any constraints on the 
color or spin of the final state. The color of the
octet $Q \overline Q$ state is
`evaporated' through an unspecified process which does not change the momentum.
The additional energy needed to produce
heavy-flavored hadrons when the partonic center-of-mass energy, 
$\sqrt{\hat s}$, is less than $2m_H$, the $H \overline H$
threshold energy, is nonperturbatively obtained from the
color field in the interaction region.
Thus the quarkonium yield may be only a small fraction of the total $Q\overline 
Q$ cross section below $2m_H$.
At leading order, the production cross section of quarkonium state $C$ in
a $p+p$ collision is
\begin{eqnarray}
\sigma_C^{\rm CEM}(s_{_{NN}})  =  F_C \sum_{i,j} 
\int_{4m^2}^{4m_H^2} d\hat{s}
\int dx_1 \, dx_2~ f_i^p(x_1,\mu_F^2)~ f_j^p(x_2,\mu_F^2)~ {\cal J}(\hat{s})
\hat\sigma_{ij}(\hat{s},\mu_F^2, \mu_R^2) \, 
\, , \label{sigtil}
\end{eqnarray} 
where $ij = q \overline q$ or $gg$ and $\hat\sigma_{ij}(\hat s)$ is the
$ij\rightarrow Q\overline Q$ subprocess cross section.  Here ${\cal J}(\hat{s})$
is a kinematics-dependent Jacobian.  At LO 
${\cal J}(\hat{s}) = \delta(\hat{s} - x_1 x_2 s)/s$, at NLO and for 
differential cross sections, the expressions are more complex.

The fraction $F_C$ must be universal so that, once it is fixed by data, the
quarkonium production ratios should be constant as a function of $\sqrt{s}$,
$y$ and $p_T$.  The actual value of $F_C$ depends on the heavy quark mass, 
$m$, the scale parameters, the parton densities and 
the order of the calculation.

We fit $F_{J/\psi}$ to both the full data set as well as to more limited sets.
Our final $J/\psi$
result is based on the total cross section data with only $p$, Be, Li,
C, and Si targets respectively.  In this way, we avoid uncertainties due to 
ignoring any cold nuclear matter effects which are on the order of a few percent
in light targets.  We also restricted ourselves to the forward cross sections
only, rather than include the $B_{ll} d\sigma/dy|_{y=0}$ data in the fits.  
The rapidity distributions calculated in the MNR code are subject to 
fluctuations about the mean, even with high statistics calculations. The
total cross sections, not subject to these fluctuations, are thus more accurate.
See Ref.~\cite{NVF} for more detail.

In the case of $\Upsilon$ production, however, most of the reported cross
section values are for $B_{ll} d\sigma/dy|_{y=0}$.  The branching ratio $B_{ll}$
here is a composite for the three $\Upsilon$ $S$ states which were not
separated at fixed-target energies.  For later experiments, with sufficient
resolution to separate the mass peaks, the individual $y=0$ cross sections
were multiplied by the PDG values of the branching ratios and summed 
\cite{NVFinprep}.  
The data in the $\Upsilon$ fits are from fixed-target energies,
$19.4 \leq \sqrt{s} \leq 44$ GeV, and collider data from the ISR, 
S$p \overline p$S and the Tevatron.  The $p + \overline p$ data from the 
S$p \overline p$S and the Tevatron are fit with the
same coefficient as the lower energy $p+p$ data.  At $\sqrt{s} = 630$ GeV, the
difference between the $p+p$ and $p + \overline p$ 
$b \overline b$ cross sections
is less than $0.5$\%, too small to affect the fit results.

We use the same values of the charm quark
mass and scale parameters as found in Ref.~\cite{NVF} to obtain the 
normalization $F_C$ for the $J/\psi$,
$(m,\mu_F/m, \mu_R/m) = (1.27 \pm 0.09 \, {\rm GeV}, 2.1^{+2.55}_{-0.85},
1.6^{+0.11}_{-0.12})$.  In the case of $\Upsilon$ production, we employ  
$(m,\mu_F/m, \mu_R/m) = (4.65 \pm 0.09 \, {\rm GeV}, 1.4^{+0.77}_{-0.49},
1.1^{+0.22}_{-0.20})$.  
We determine $F_C$ only for the central parameter set in each case and scale
all the other calculations by the same value of $F_C$ to
obtain the extent of the $J/\psi$ and $\Upsilon$ mass and scale uncertainty 
bands, as described in detail in Sec.~\ref{SubSec:M_muvar}.

We find $F_{J/\psi} = 0.020393$ for the central result
with $(m,\mu_F/m, \mu_R/m) = (1.27 \, {\rm GeV}, 2.1,1.6)$ employing the
CT10 parton densities \cite{NVF}.
We obtain $F_{\Sigma \Upsilon} = 0.0077$ for the central $y=0$ result 
for the combined $\Upsilon$ $S$ states with the CT10 parton densities and
$(m,\mu_F/m, \mu_R/m) = (4.65 \, {\rm GeV}, 1.4,1.1)$.
After separating the $S$ states, the inclusive $1S$ value is 
$F_{\Upsilon} = 0.022$ \cite{NVFinprep}.

Our CEM calculations use the NLO $Q \overline Q$ code of 
Mangano {\it et al.}~(MNR) \cite{MNRcode} with
the $H \overline H$ mass cut in Eq.~(\ref{sigtil}), as described in 
Refs.~\cite{Gavai:1994in,rhicii}.  Because the NLO $Q \overline Q$ 
code is an exclusive calculation, we take the mass cut on the
invariant average over kinematic variables of the $c$ and $\overline c$. Thus,
instead of defining $\mu_F$ and $\mu_R$ relative to the quark mass, they are
defined relative to the transverse mass, 
$\mu_{F,R} \propto m_T = \sqrt{m^2 + p_T^2}$ where 
$p_T$ is that of the $Q \overline Q$ pair, 
$p_T^2 = 0.5(p_{T_Q}^2 + p_{T_{\overline Q}}^2)$.

At LO in the total cross section,  the $Q \overline Q$ pair
$p_T$ is zero.  Thus, while our calculation is
NLO in the total cross section, it is LO in the quarkonium $p_T$ distributions. 
In the exclusive NLO calculation ~\cite{MNRcode}
both the $Q$ and $\overline Q$ variables
are integrated to obtain the pair distributions,
recall $\mu_{F,R} \propto m_T$.

Results on open heavy flavors indicate that some level of
transverse momentum broadening is needed to obtain agreement with the low $p_T$
data.  This is often done by including some intrinsic transverse momentum,
$k_T$, smearing to the initial-state parton densities.
The implementation of intrinsic $k_T$ in the MNR code is not handled in the 
same way as calculations of other hard processes due to the nature of the code.
In the MNR code, the cancellation of divergences is done numerically.  
Since adding additional numerical Monte-Carlo integrations would slow the
simulation of events, in addition to requiring multiple runs with the same
setup but different intrinsic $k_T$ kicks, the kick is added in the
final, rather than the initial, state. 
In Eq.~(\ref{sigtil}), the Gaussian function $g_p(k_T)$,
\begin{eqnarray}
g_p(k_T) = \frac{1}{\pi \langle k_T^2 \rangle_p} \exp(-k_T^2/\langle k_T^2
\rangle_p) \, \, ,
\label{intkt}
\end{eqnarray}
\cite{MLM1}, multiplies the parton
distribution functions for both hadrons, 
assuming the $x$ and $k_T$ dependencies in the initial partons completely
factorize.  If factorization applies, 
it does not matter whether the $k_T$ dependence
appears in the initial or final state if the kick is not too large.  
In Ref.~\cite{MLM1}, $\langle k_T^2 \rangle_p = 1$ GeV$^2$, along with the
Peterson fragmentation function with parameter $\epsilon = 0.06$,
was found to best 
describe fixed-target charm production.  We note that currently Peterson
fragmentation with $\epsilon = 0.06$ is considered too strong.  The FONLL
fragmentation scheme for open heavy flavor is softer \cite{CNV}.

In the code, the $Q \overline Q$ system
is boosted to rest from its longitudinal center-of-mass frame.  Intrinsic 
transverse
momenta of the incoming partons, $\vec k_{T 1}$ and $\vec k_{T 2}$, are chosen
at random with $k_{T 1}^2$ and $k_{T 2}^2$ distributed according to
Eq.~(\ref{intkt}).   A second transverse boost out of the pair rest frame
changes the initial transverse momentum of
the $Q \overline Q$ pair, $\vec p_T$, to $\vec p_T
+ \vec k_{T 1} + \vec k_{T 2}$.  The initial
$k_T$ of the partons could have alternatively been given to the entire
final-state system, as is essentially done if applied in the initial state,
instead of to the $Q \overline Q$ pair.  There is no difference if the
calculation is LO but at NLO an additional light parton can 
also appear in the final state so the correspondence is not exact.  
In Ref.~\cite{MLM1}, the difference between the two implementations is claimed 
to be small if $k_T^2 \leq 2$ GeV$^2$.  We note that the
rapidity distribution, integrated over all $p_T$, is unaffected by the
intrinsic $k_T$.

The effect of the intrinsic $k_T$ on the shape of the quarkonium $p_T$ 
distribution can be expected to decrease as 
$\sqrt{s}$ increases because the average $p_T$ also increases 
with energy.  However, the value of $\langle k_T^2 \rangle$ may increase with
$\sqrt{s}$ so that effect remains important at higher energies.  We assume
the form  $\langle k_T^2 \rangle = 1 + (1/n)\ln(\sqrt{s}/20)$~GeV$^2$. 
Using the RHIC $J/\psi$ data, we found that $n = 12$ gave the best 
description of the $p_T$ distribution both at central and forward rapidity
\cite{NVF}.
A larger value of $n$ and thus of $\langle k_T^2 \rangle$ is required for
the $\Upsilon$ $p_T$ distribution.  We set $n=3$ for $\Upsilon$
from comparison to the
Tevatron results at $\sqrt{s} = 1.8$ TeV \cite{NVFinprep}.
For this study, $n$ is not modified by the nuclear medium.

We note that most approaches to quarkonium production: the CEM; the color 
singlet model (CSM); and the Nonrelativistic QCD approach (NRQCD), assume the 
validity of collinear factorization which separates the initial, 
nonperturbative parton densities from the perturbatively-calculable
hard scattering that produces the final state.  We assume collinear 
factorization to hold for quarkonium production in the CEM.  
Factorization has been proved for quarkonium production 
in NRQCD at high $p_T$ \cite{Jianwei} but not at low $p_T$.  
The open charm flavor results at low $p_T$
at the LHC \cite{NVF} agree with calculations employing
collinear factorization \cite{NVF} better than calculations employing
$k_T$ factorization \cite{Alberico}.  Collinear factorization should work 
better for bottom since the factorization scale and the $x$ region probed
are both larger.

Since the start up of the LHC, several groups have performed global analyses
of the nonperturbative matrix elements in the NRQCD approach up to NLO,  
see Ref.~\cite{StrongDoc} and references therein 
for discussion and comparison of the results for
$J/\psi$ production data in $e^+ + e^-$, $e+p$, $p+p$ and $p+ \overline p$ 
collisions.  In Ref.~\cite{StrongDoc}, it is clear that the  
nonperturbative matrix elements are quite sensitive to the
minimum $p_T$ employed in the fits.  These matrix elements do not appear to
be universal since choosing different data sets to fit to result in quite
different values of the matrix elements.  Indeed, the agreement with
the $e^+ + e^-$ and $e+p$ data is poor unless the minimum $p_T$ is low, 
$p_T \sim 3$ GeV, but these fits cannot reproduce the measured high $p_T$
quarkonium polarization \cite{StrongDoc}.

In addition, if the fitted matrix elements are used to calculate the
$J/\psi$ cross section at $y =0$ in $p+p$ collisions as a function of 
$\sqrt{s}$, good agreement with neither the shape nor the
magnitude of the cross section can be obtained, see Ref.~\cite{JPLpriv}.  
The results
overshoot the measured cross sections significantly, sometimes by an order
of magnitude \cite{JPLpriv}.  This is not a surprising outcome because the
the integrated $y = 0$ cross section is dominated by low $p_T$ $J/\psi$
production.  In the same paper, the CSM cross sections are also compared to
these data.  It was found that only the LO CSM calculation produces a physical
$\sqrt{s}$ dependence that agrees relatively well with the data, at NLO some
values of $\mu_F$ give an unphysical energy dependence \cite{JPLpriv}.  They
also point out that the CEM produces the best agreement with the $\sqrt{s}$
dependence over the entire range.

We also note that, while the CSM and NRQCD make predictions
for the $p_T$ dependence at relatively high $p_T$, the region of interest for
nuclear effects on the parton densities, as addressed here, 
is at low $p_T$ and the CEM, through 
the application of $k_T$ smearing, is the only production model that 
addresses the entire $p_T$ range.  Since the average $k_T$ is an 
energy-dependent parameter, ideally this should be replaced by a low-$p_T$
resummation.

Some recent results may provide improvements for NRQCD, at least at collider
energies \cite{Jianwei,Mapp,MapA}.  References~\cite{Mapp,MapA} perform small
$x$ resummation in the color-glass condensate (CGC) in the NLO NRQCD approach 
with the nonperturbative matrix elements taken from the high $p_T$ fits of
Chao {\it et al.} \cite{Chao_etal}.  The CGC gluon distribution is employed in 
the CGC+NRQCD calculation at low $p_T$.  This matches well with the high $p_T$ 
NLO NRQCD calculation in the intermediate $p_T$ range.  Thus the entire $p_T$
range of the data can be described within the combined approaches, both in
$p+p$ and $p+A$ collisions for forward rapidities and sufficiently 
high $\sqrt{s}$ \cite{Mapp,MapA}.  The energy dependence
of $d\sigma/dy|_{y=0}$ is also reproduced quite well for $\sqrt{s} > 0.2$ TeV
athough agreement with the midrapidity RHIC data is rather poor
\cite{Mapp}.  This approach
is inapplicable at lower $\sqrt{s}$.
Reference~\cite{Jianwei} presents a factorized power expansion for quarkonium
production, including next-to-leading power (NLP) contributions to the 
perturbative
part.  This formalism requires fragmentation functions for heavy 
$Q \overline Q$ pairs as well as for light partons.  With the fragmentation
functions and the NLP contributions included in NRQCD, it was found
that the $^3S_1^{[1]}$ and $^1S_0^{[8]}$ components of the cross section are 
dominated by the NLP contributions over all $p_T$, independent of the 
nonperturbative matrix elements \cite{Jianwei}.  This formalism describes
$J/\psi$ production well at collider energies for $p_T > 10$ GeV.  The
possible dominance of the $^1S_0^{[8]}$ contribution in the total production
rate could explain the apparent unpolarized $J/\psi$ production.  This 
conclusion is consistent with the data-driven approach to polarization in
Ref.~\cite{Carlos_pol}.  

We note that the polarization has not yet
been calculated in the CEM.  While it should be straightforward at LO, to go
to NLO one would have to start from a polarized $Q \overline Q$ pair production
code.

\begin{figure}[t]
\begin{center}
%\vspace*{-0.05in}
\includegraphics[width=0.45\textwidth]{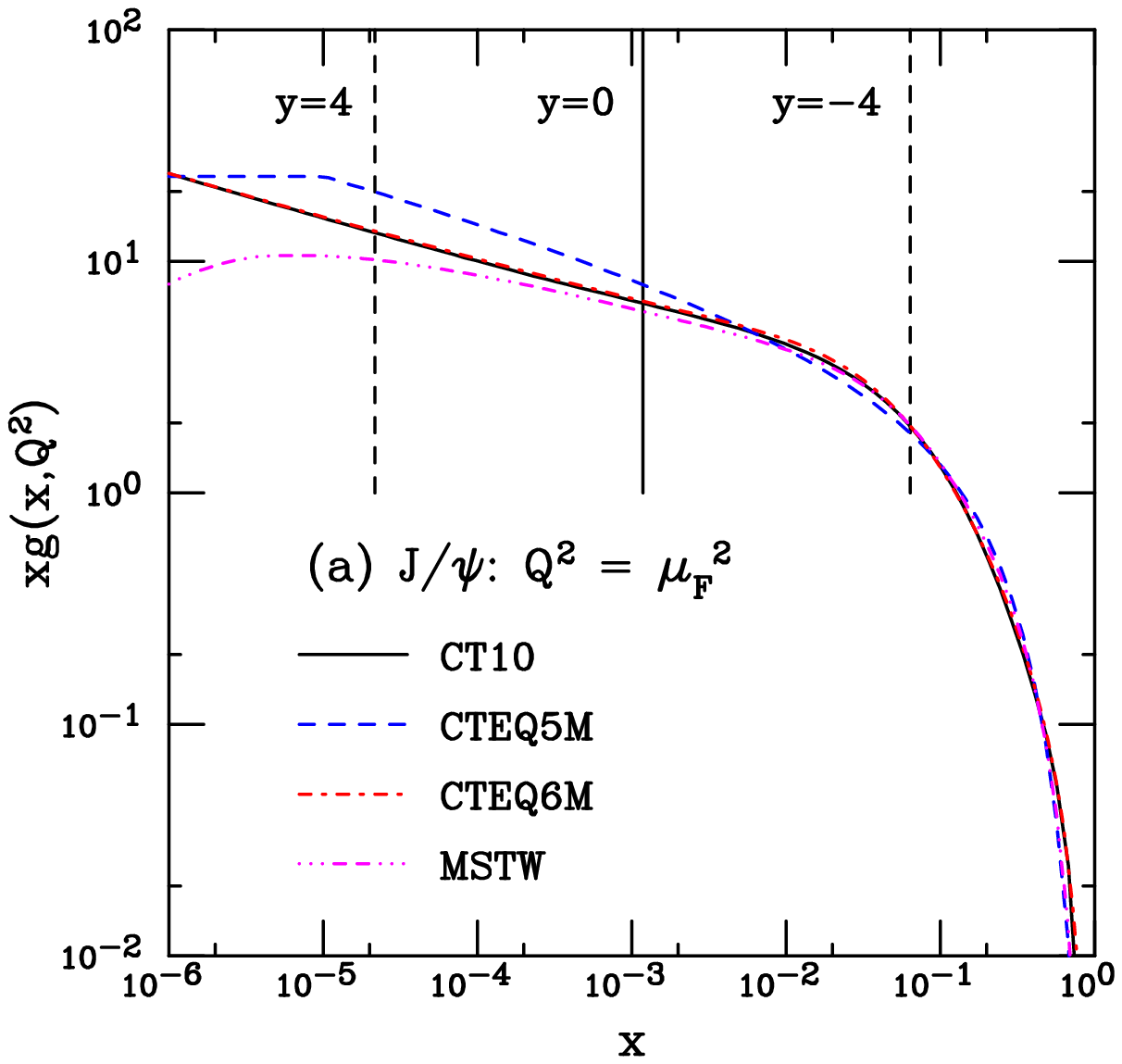}
\includegraphics[width=0.45\textwidth]{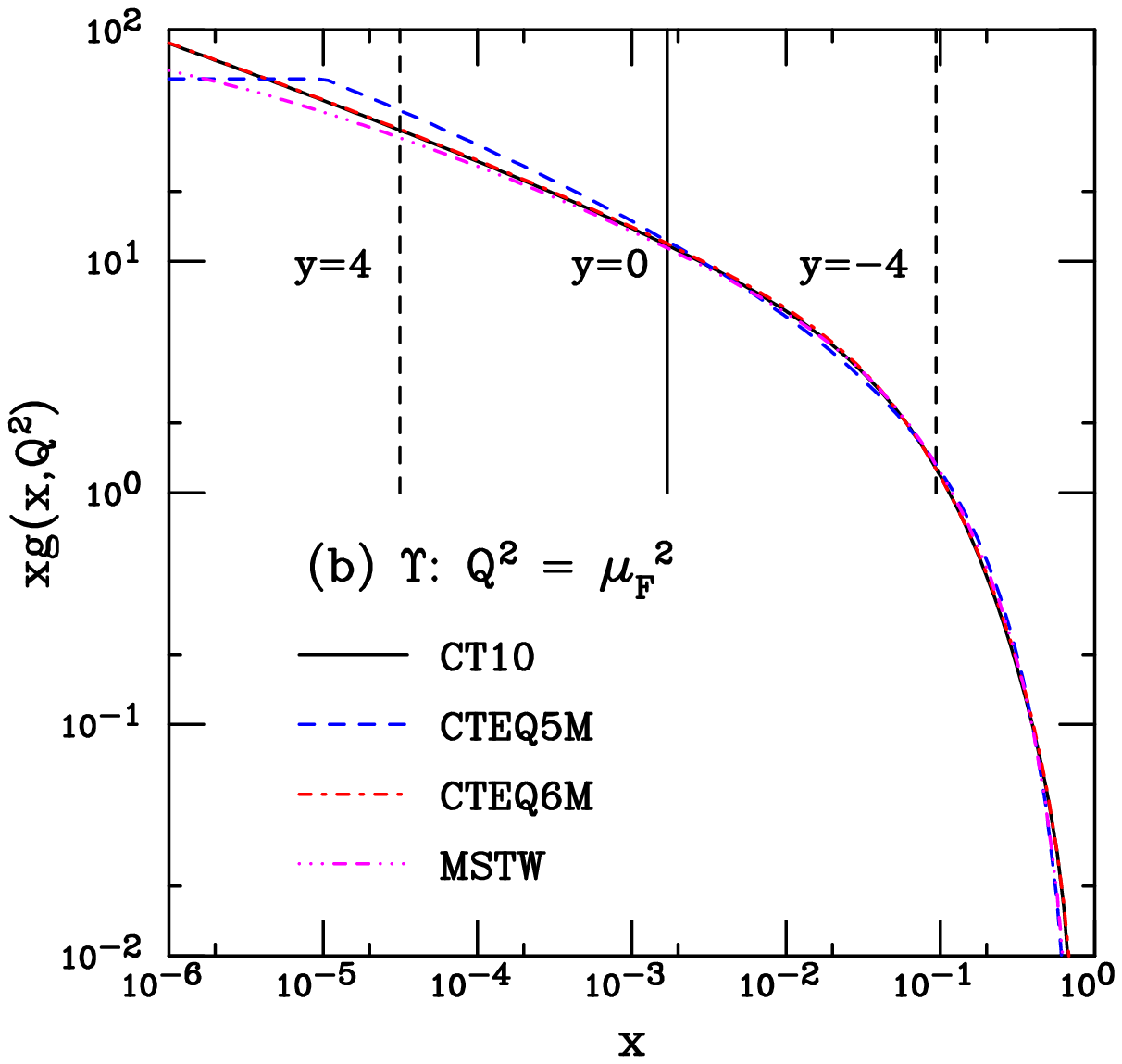}
%\vspace*{-0.4in}
\caption[]{(Color online)
The gluon distribution function at the factorization scale for $J/\psi$ (a)
and $\Upsilon$ (b) production.  The CT10 (black),
CTEQ5M (blue), CTEQ6M (red) and MSTW (magenta) are compared, all calculated
with the same input parameters.
}
\label{fig:gluonPDF}
\end{center}
\end{figure}

\subsection{Comparison of $p+p$ Results}
\label{SubSec:ppcomp}

We now turn to a comparison of the $J/\psi$ and $\Upsilon$ $p_T$ and rapidity
distribution in $p+p$ collisions for different proton parton densities.  Our
main results are obtained with the CT10 \cite{CT10} parton densities, also
used in our evaluations of the charm and bottom cross sections 
\cite{NVF,NVFinprep} where we find good agreement with the present
quarkonium data within the mass and scale uncertainties.  In
Fig.~\ref{fig:gluonPDF}, we compare the NLO CT10 gluon distributions
to those from CTEQ5M \cite{CTEQ5}, CTEQ6M \cite{CTEQ6} and MSTW \cite{MSTW}.
The MSTW central NLO set is a rather recent, frequently used parton density
shown as an alternative to CT10.  It is available at LO, NLO and NNLO and has
a low starting scale of $\mu_0^2 = 1$ GeV$^2$.  
The CTEQ sets and CT10, at LO and NLO, have a starting scale of $\mu_0^2 = X$
GeV$^2$.  The vertical lines at the top of the plots indicate the gluon
density at forward, mid- and backward rapidity.

The CTEQ6M and CTEQ5M distributions were chosen because they have been
used to extract the modifications of the parton densities in the nucleus.
The CTEQ6M distributions were used in a global
analysis of the nuclear modifications of the parton densities to obtain the
EPS09 NLO sets \cite{EPS09}. They were also used in previous estimates of the 
quarkonium cross sections at the LHC \cite{QQ_yellow}.  The corresponding 
EPS09 LO sets were obtained
based on the CTEQ61L densities, as we discuss in more detail later.
The CTEQ5M distributions were
used to extract the NLO nuclear modifications in the FGS sets \cite{FGS}.
Older distributions used to extract nuclear modifications are CTEQ4L 
\cite{CTEQ4} and GRV LO \cite{GRV} (EKS98 \cite{Eskola:1998iy,Eskola:1998df}) 
and GRV98 LO and NLO \cite{GRV98} (nDS \cite{deFlorian:xxx}).  
Since these proton PDFs are now 
outdated, we do not show them here.

Figure~\ref{fig:gluonPDF} shows the gluon densities in the proton at the
scales used to calculate $J/\psi$ (a) and $\Upsilon$ (b) production.  
The CT10 distributions, the most recent of all those
considered, follow the previous CTEQ6 global analysis.  The CT10 and CTEQ6M
gluon distributions are almost identical.  The earlier
CTEQ5M set is quite different: the lowest $x$ value included in the analysis
is $10^{-5}$ and, instead of extrapolating smoothly to lower values of $x$, 
the density is frozen such that
$xg(x<10^{-5},\mu^2) \equiv xg(x = 10^{-5},\mu^2)$.  It also tends to be higher 
than the other gluon PDFs for $10^{-5} < x < 10^{-2}$.  
The MSTW distribution is below the others for the $J/\psi$ factorization 
scale.  All the gluon distributions are more
similar at the $\Upsilon$ scale even though CTEQ5M is still somewhat higher
for $10^{-5} < x < 10^{-2}$.

The CTEQ6M and CT10 gluon
distributions are zero at the minimum scale, $x g(x,\mu_0^2) = 0$.  Thus they  
undergo rapid evolution at
low $x$ since they are based on a valence-like initial distribution, 
$x^\alpha(1-x)^\beta$.
On the other hand,
the MSTW sets allow a negative gluon density at low $x$.  Therefore the MSTW
gluon density exhibits a slower scale evolution.  It will thus produce
the smallest cross
sections for $J/\psi$ and $\Upsilon$ production while CTEQ5M will give the
largest.

\begin{figure}[t]
\begin{center}
%\vspace*{-0.05in}
\includegraphics[width=0.45\textwidth]{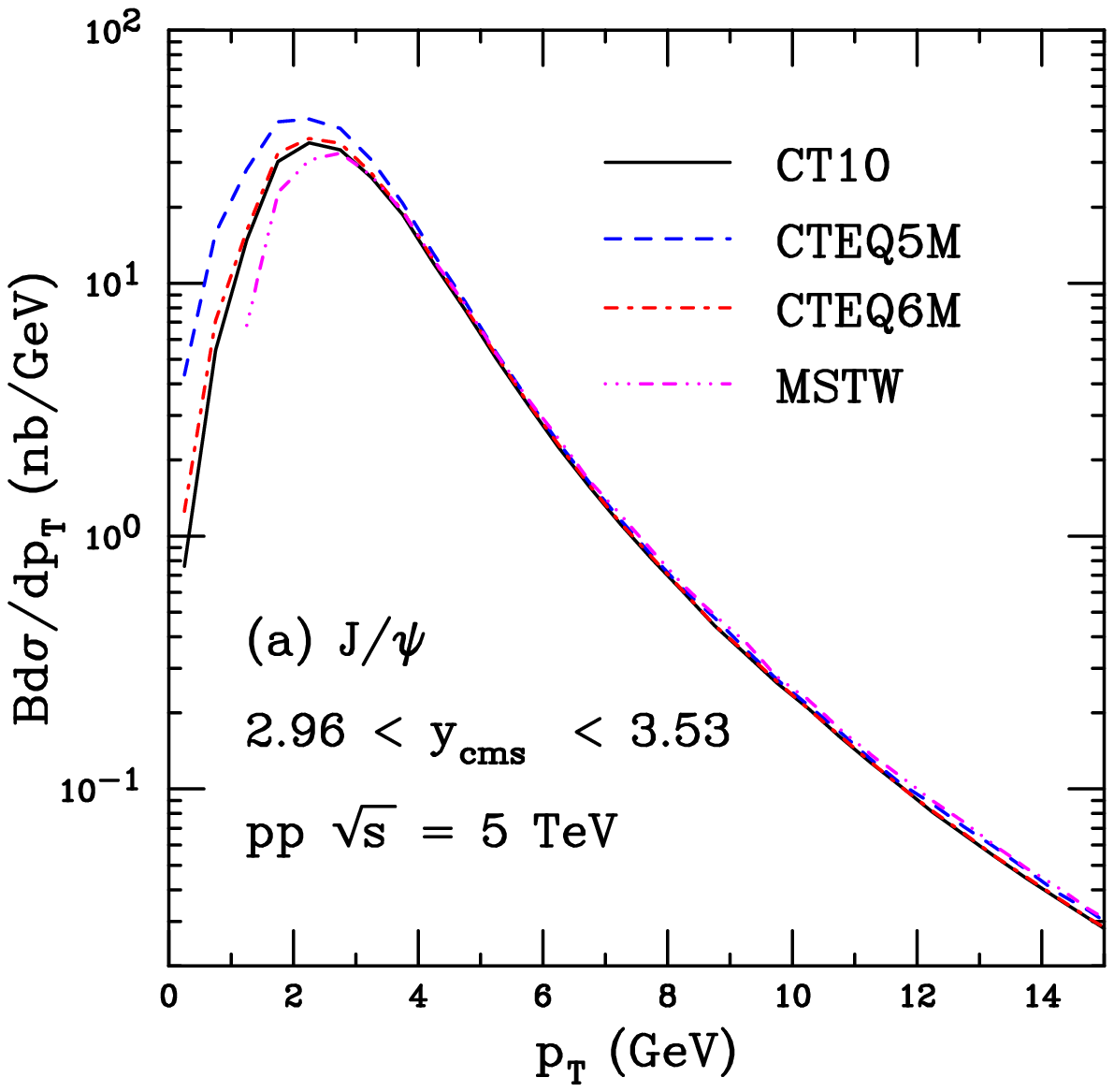}
\includegraphics[width=0.45\textwidth]{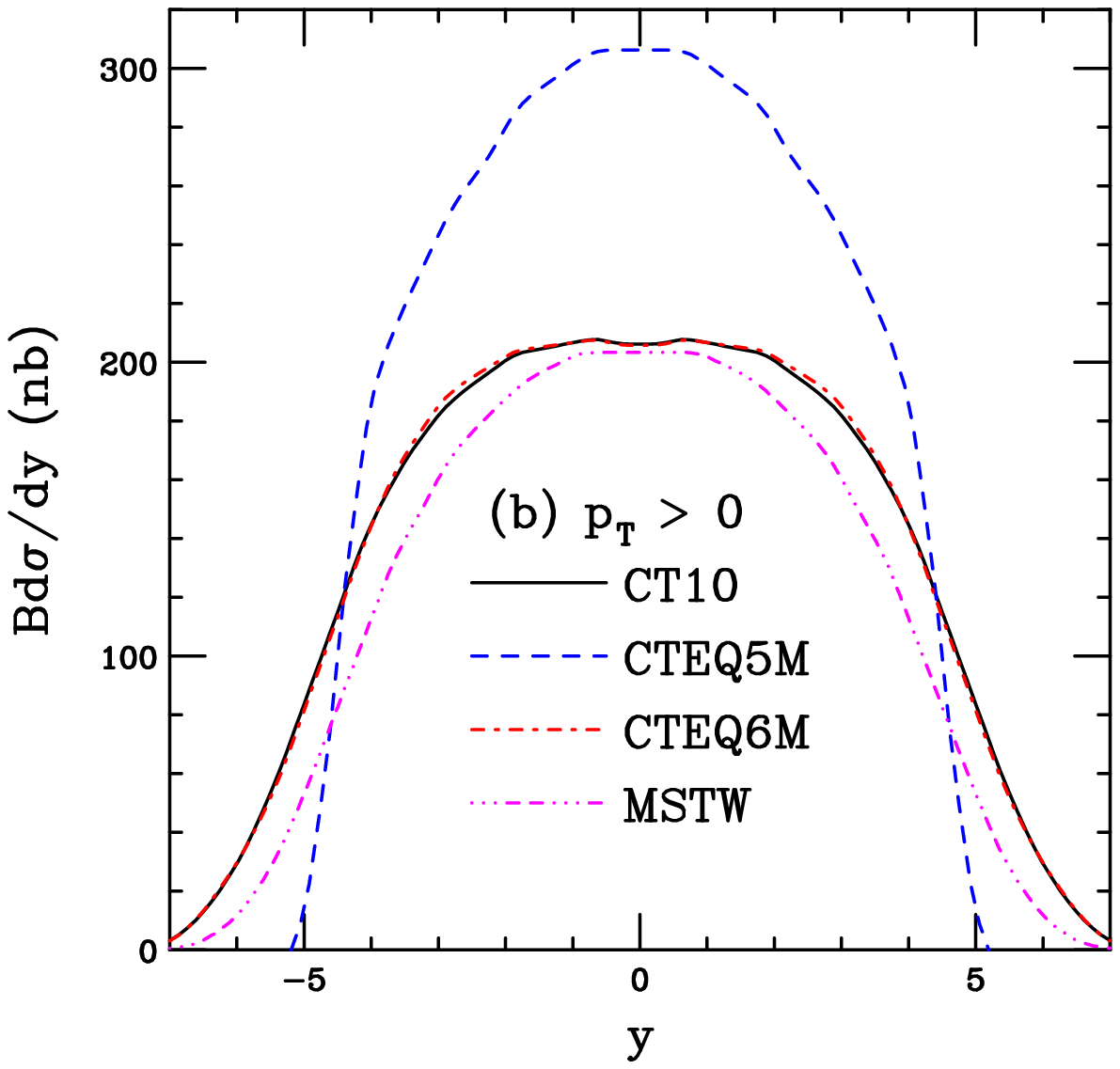}
%\vspace*{-0.4in}
\caption[]{(Color online)
The $J/\psi$ $p_T$ distribution at forward rapidity in $p+p$ collisions (a)
and the $p_T$-integrated $y$ distribution (b).  Results from CT10 (black),
CTEQ5M (blue), CTEQ6M (red) and MSTW (magenta) are compared, all calculated
with the same input parameters and using the same value of $F_C$ as for
CT10.
}
\label{fig:ppJpsi}
\end{center}
\end{figure}

In Fig.~\ref{fig:ppJpsi}, we show the shapes of the $J/\psi$ $p_T$ and $y$
distributions for the proton parton densities presented in 
Fig.~\ref{fig:gluonPDF}.  The $p_T$ distributions are shown in the region of
rapidity overlap with the shifted $p+$Pb range at forward rapidity,
$2.96 < y < 3.53$.  In $p+p$ collisions, the $p_T$ distributions at forward
and backward rapidity are identical because the $y$ distributions are symmetric.
 
The results are given with the same CEM normalization,
$F_C$, as found for the CT10 fits in Ref.~\cite{NVF}.  Because the low $x$
gluon distributions differ in shape and magnitude, the values of the cross
sections also differ by as much as $15-20$\%.  If fits of $F_C$ were made
with the three additional proton parton densities, the values of $F_C$ would
clearly differ.  By employing the same value of $F_c$ for all the sets, we
emphasize the differences in magnitude as well as shape.

\begin{figure}[t]
\begin{center}
%\vspace*{-0.05in}
\includegraphics[width=0.45\textwidth]{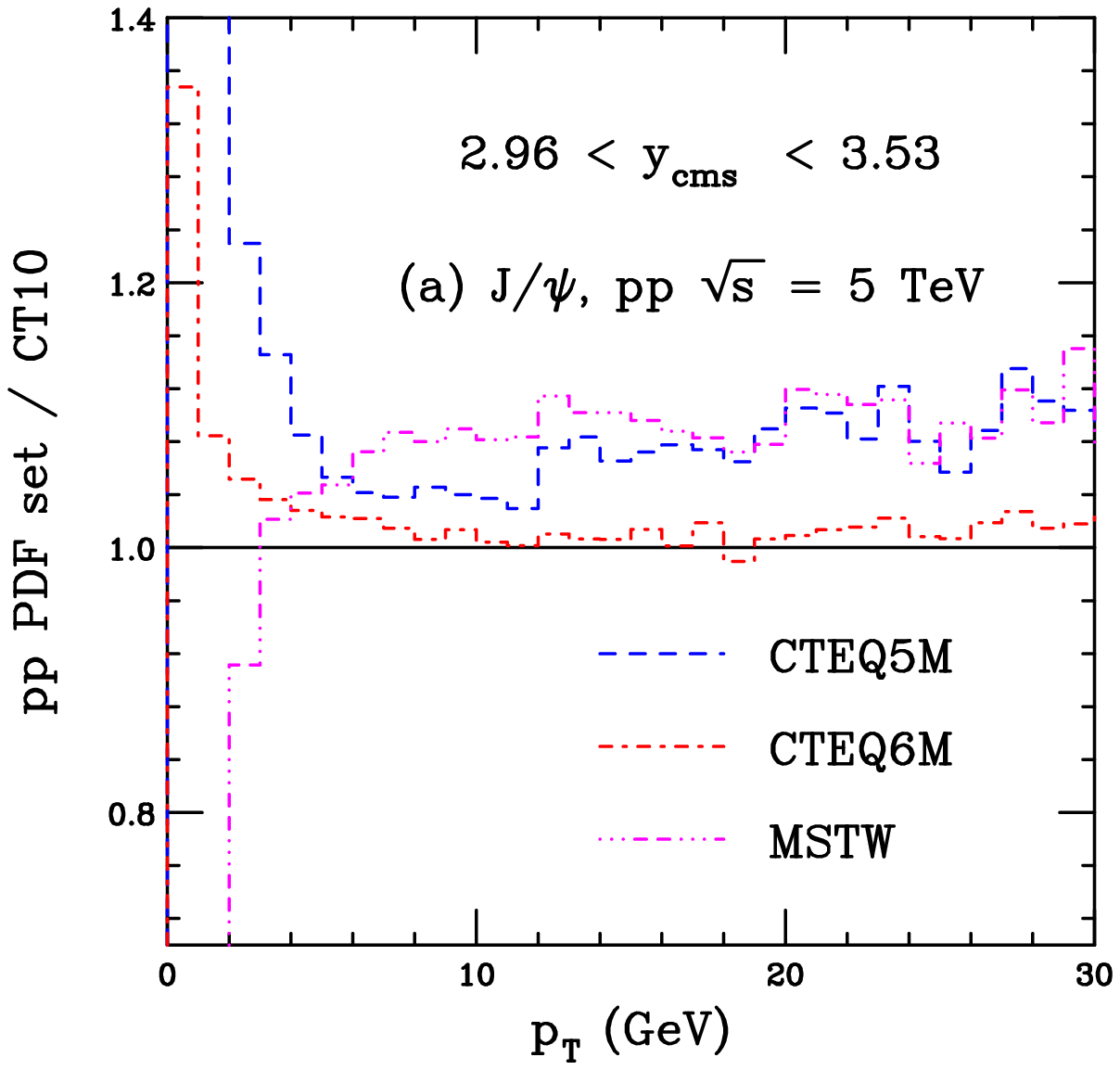}
\includegraphics[width=0.45\textwidth]{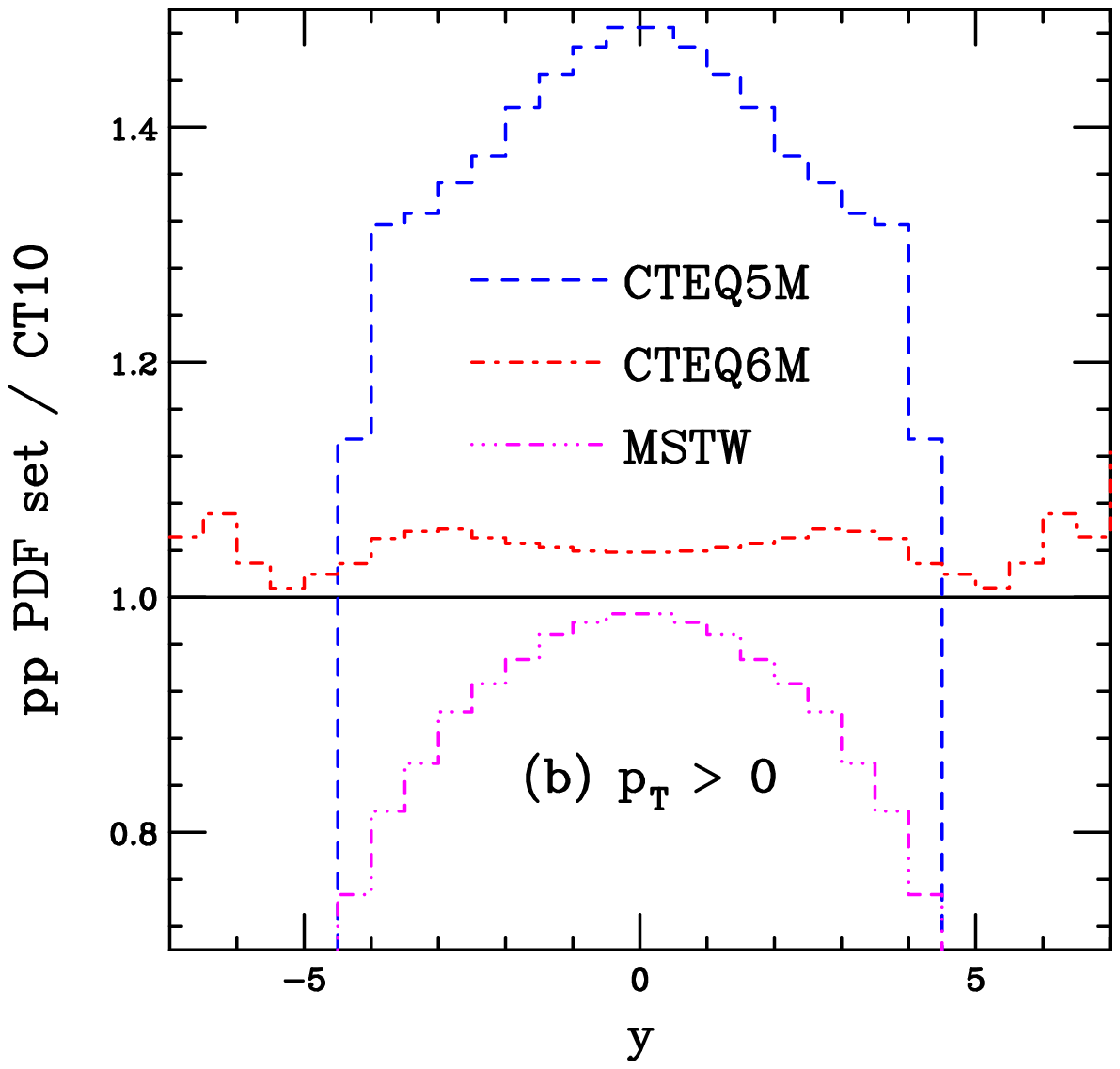}
%\vspace*{-0.4in}
\caption[]{(Color online)
The ratios of $J/\psi$ production as a function of $p_T$ at forward 
rapidity in $p+p$ collisions (a)
and $y$ integrated over all $p_T$ (b) relative to CT10 for
CTEQ5M (blue), CTEQ6M (red) and MSTW (magenta).
}
\label{fig:ppPsi_rat}
\end{center}
\end{figure}

The CTEQ5M distribution, higher than the other gluon densities at the $J/\psi$
scale for $x < 0.01$, results in a 20\% larger overall cross section.
This manifests itself at low $p_T$ where the integral difference is 32\%
and, most prominently, in the rapidity
distribution around $y \sim 0$ where it is $\sim 50$\% higher.  
The CTEQ5M gluon distribution, as already
noted, takes a constant value for $x < 10^{-5}$.  The rapidity distribution for
the PDF is consequently narrower than the other three shown.  The 
corresponding $p_T$ distributions do not reflect the $x < 10^{-5}$ behavior
because, in the rapidity range illustrated, the low $p_T$ $J/\psi$'s are at
higher $x$ values, see the $y=4$ line in Fig.~\ref{fig:gluonPDF}(a).

The similarity of the CT10 and CTEQ6M gluon distributions results in very 
similar $p_T$ and $y$ distributions.  The MSTW set gives a 15\% lower 
cross section for the $J/\psi$ with most of the difference at low $p_T$.  
The rapidity
distributions are also slightly narrower than CT10 but less than that of 
the CTEQ5M calculation.

Figure~\ref{fig:ppPsi_rat} presents the ratios of the CTEQ5M, CTEQ6M and
MSTW calculations to the CT10 results as a function of $p_T$ and $y$.  Aside
from the lowest $p_T$ bins, the ratios in Fig.~\ref{fig:ppPsi_rat}(a)
are practically independent of $p_T$
except for CTEQ5M which seems to show slow growth with $p_T$ after a strong
drop at $p_T < 5$ GeV.  For
$5 < p_T < 30$ GeV, the ratios to CT10 all agree within 10\%. The CTEQ6M
result differs from CT10 by $\sim 2$\% for $p_T > 5$ GeV.  
The $p_T$-integrated rapidity distributions in Fig.~\ref{fig:ppPsi_rat} are
all quite different.  While the CTEQ6M result is within 5\% of that of
CT10 over all rapidity, the other ratios are narrower.  Note that the $p_T>0$
ratio in Fig.~\ref{fig:ppPsi_rat}(b) is larger than the ratio at $p_T > 5$ GeV
in \ref{fig:ppPsi_rat}(a) because the $y$ distribution is dominated by low
$p_T$.  The sharp drop for
CTEQ5M is where $x \sim 10^{-5}$.  Since the MSTW gluon
distributions do not have the
same behavior, they are broader than CTEQ5M but the ratio is less than 0.7.

\begin{figure}[t]
\begin{center}
%\vspace*{-0.05in}
\includegraphics[width=0.45\textwidth]{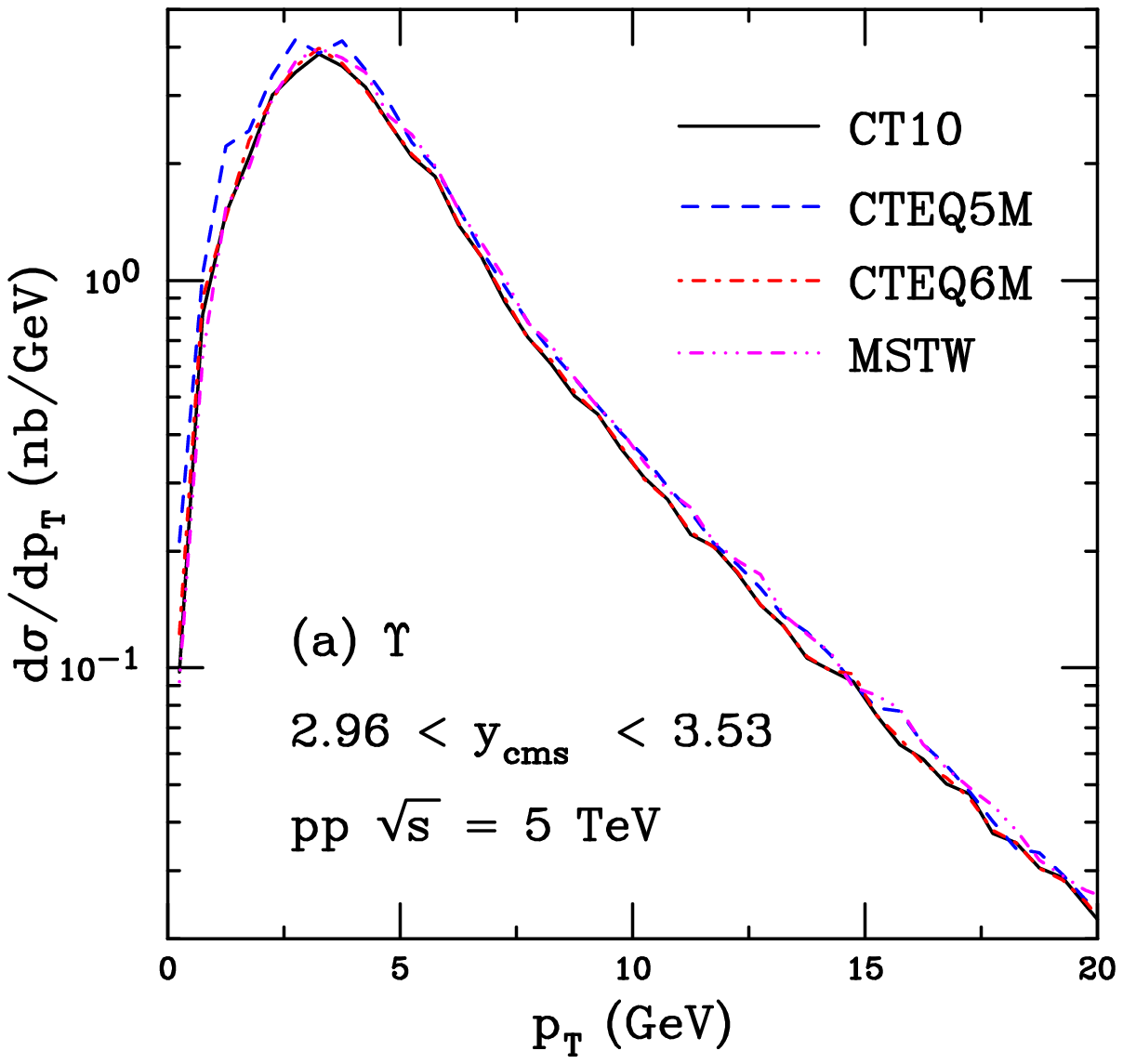}
\includegraphics[width=0.45\textwidth]{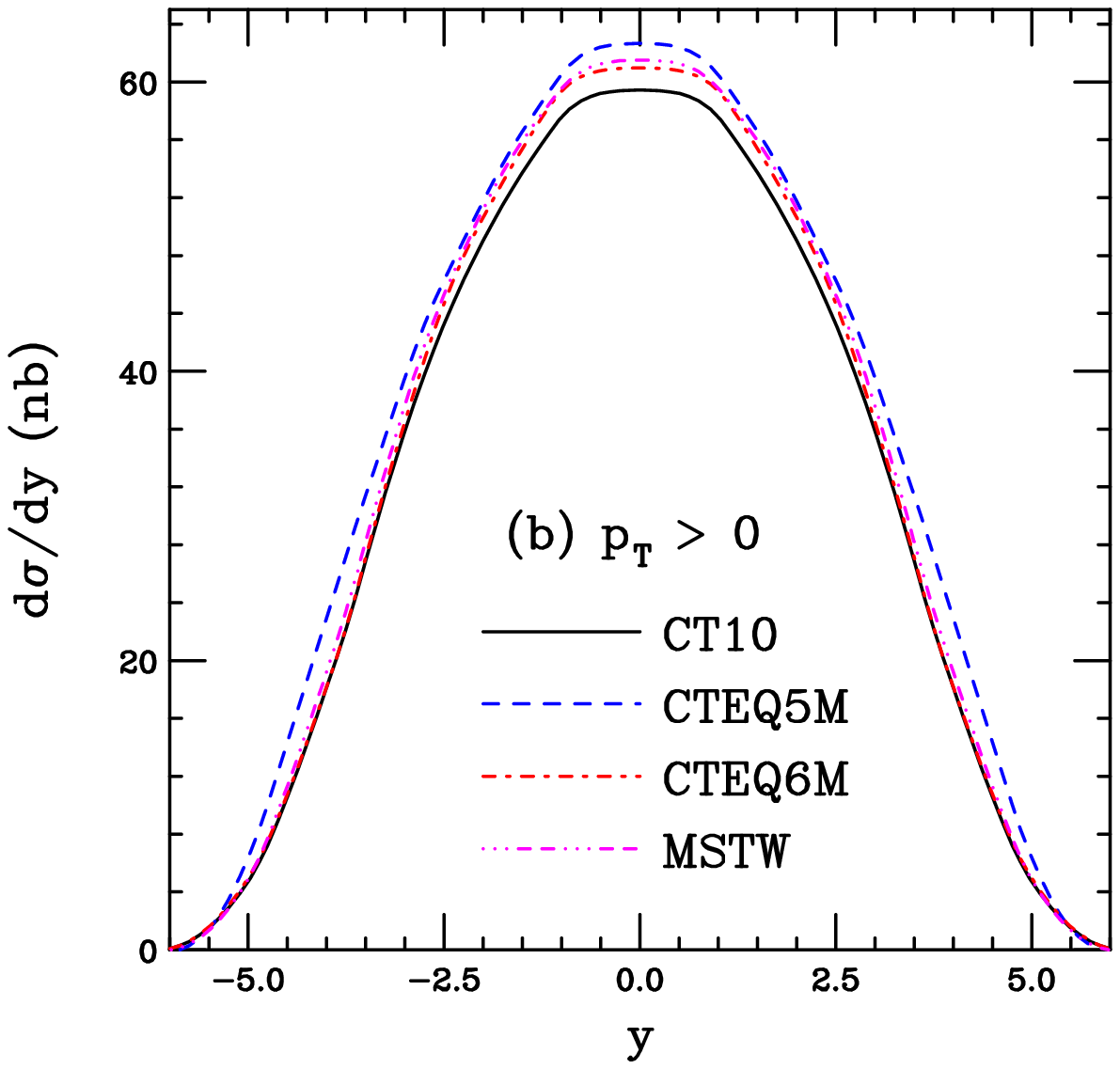}
%\vspace*{-0.4in}
\caption[]{(Color online)
The $\Upsilon$  distribution at forward rapidity in $p+p$ collisions (a)
and the $p_T$-integrated $y$ distribution (b).  Results from CT10 (black),
CTEQ5M (blue), CTEQ6M (red) and MSTW (magenta) are compared, all calculated
with the same input parameters and using the same value of $F_C$ as for
CT10.
}
\label{fig:ppUps}
\end{center}
\end{figure}

\begin{figure}[t]
\begin{center}
%\vspace*{-0.05in} 
\includegraphics[width=0.45\textwidth]{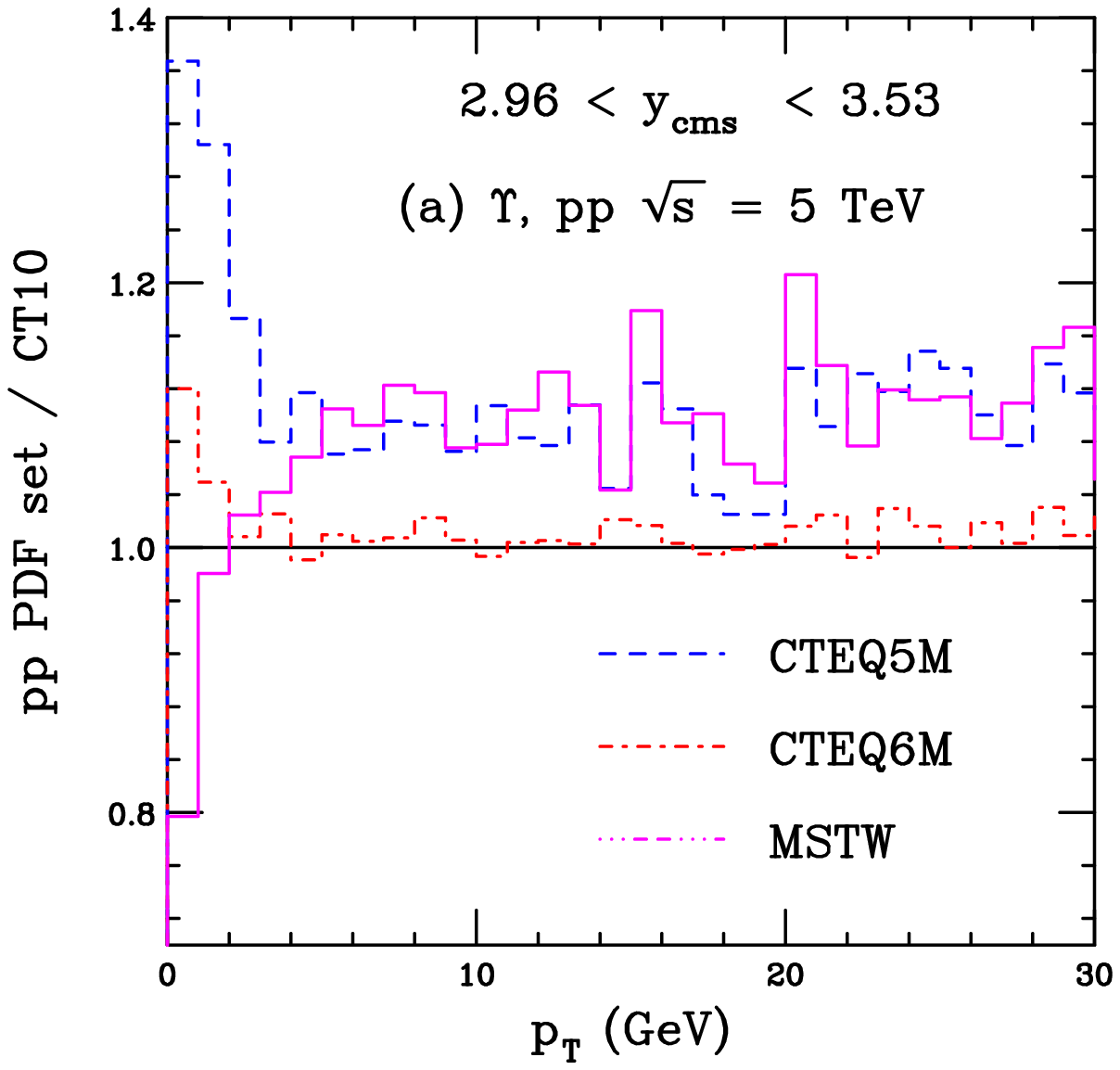}
\includegraphics[width=0.45\textwidth]{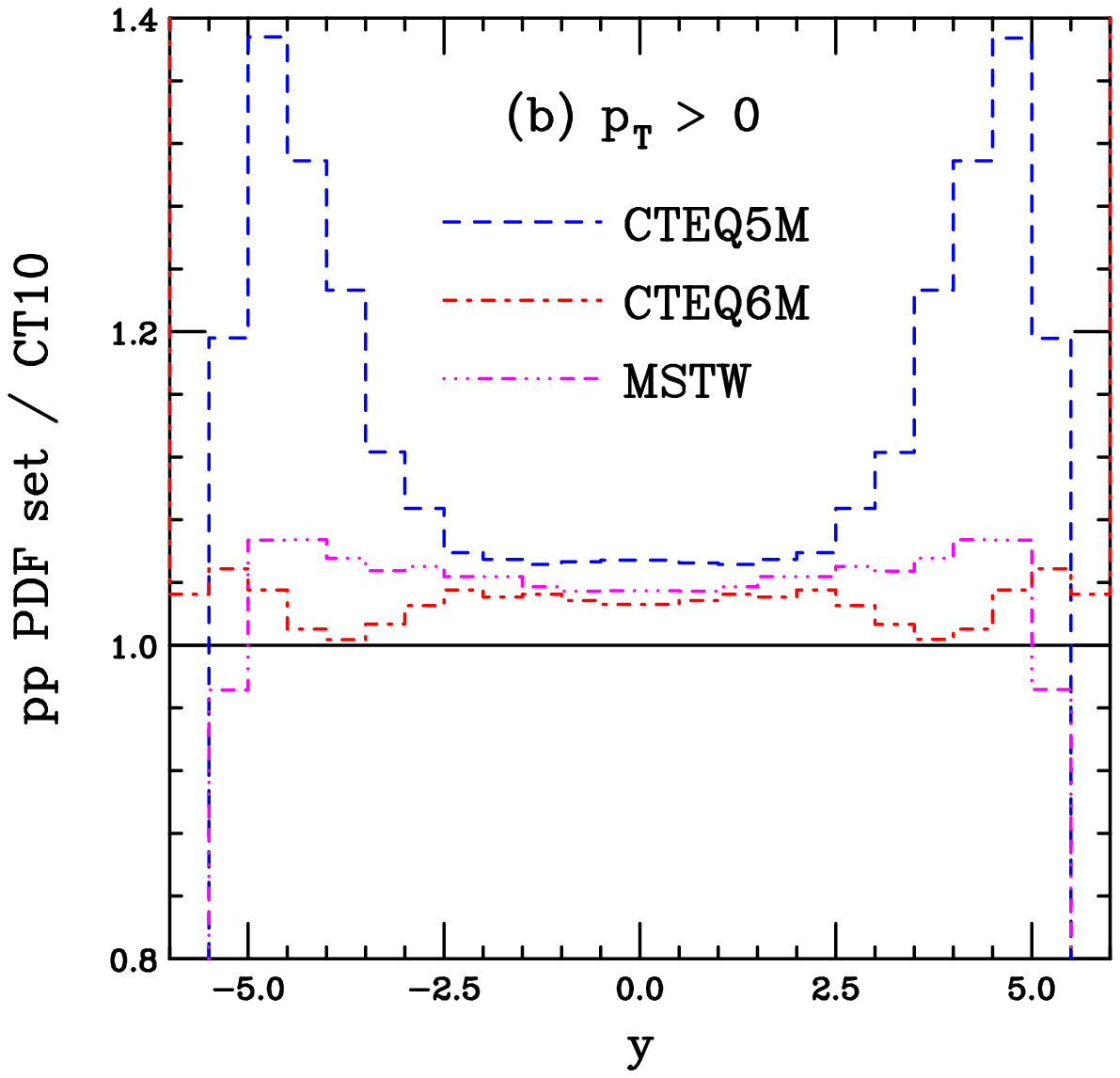}
%\vspace*{-0.4in}
\caption[]{(Color online)
The ratios of $\Upsilon$ production as a function of $p_T$ at forward 
rapidity in $p+p$ collisions (a)
and $y$ integrated over all $p_T$ (b) relative to CT10 for
CTEQ5M (blue), CTEQ6M (red) and MSTW (magenta).
}
\label{fig:ppUps_rat}
\end{center}
\end{figure}

The corresponding $\Upsilon$ results are displayed in Figs.~\ref{fig:ppUps}
and \ref{fig:ppUps_rat}.  The larger factorization scale makes the results
all very similar in magnitude for this case with the integrated
cross sections differing by
less than 10\%, even for CTEQ5M.  The ratios as a function of $p_T$ are all
relatively flat, especially between CTEQ6M and CT10.  The largest difference
among the results appears at large rapidities where there is a sudden 
increase in the ratio for $|y| \sim 5$, dropping off afterward where 
$x \sim 10^{-5}$ is reached for $\Upsilon$ production.

We have shown that the rapidity distributions are more dependent on the chosen
baseline proton parton densities than the $p_T$ distributions.  This arises
because of the differing behavior of the low $x$ gluon densities.  Due to the
larger scale, the dependence on the proton parton densities
is less important for the higher mass $\Upsilon$.

\section{Cold Nuclear Matter Effects}
\label{Sec:CNM}

To go beyond $p+p$ collisions, 
the proton parton densities must be replaced by those of
the nucleus.  
We assume that if $A$ is a nucleus, the nuclear parton densities, 
$f_i^A(x_2,\mu^2)$, factorize into
the nucleon parton density, $f_i^p(x_2,\mu^2)$, independent of $A$;
and a shadowing ratio, $S^i_{{\rm P},{\rm S}}(A,x_2,\mu^2)$ that
parameterizes the modifications of the nucleon parton densities in the nucleus.
The centrality dependence will be discussed elsewhere \cite{TonyandMe} so here
we deal with $S^i(A,x_2,\mu^2)$ alone.  (Note that we refer to $x_2$ here since
this is the 

We do not consider other cold matter effects in this paper because we want to
determine how far the data can be described by the assumption of shadowing
alone.  However, since lower energy $J/\psi$ production clearly requires
some effects beyond shadowing, we briefly mention these here.

At fixed-target energies, the $x_F$ dependence clearly shows that 
shadowing is not the only contribution to the $J/\psi$ nuclear 
dependence as a function of Feynman $x_F = 2p_z/\sqrt{s}$ for the final-state
particle \cite{e866,HERA-Bnew}.  
Indeed, the characteristic decrease 
of $\alpha(x_F)$ (in $d\sigma_{pA}/dx_F = (d\sigma_{pp}/dx_F) A^{\alpha(x_F)}$)
for $x_F \geq 0.25$ cannot be explained
by shadowing alone \cite{rv866}.  In fact, the data so far suggest
approximate scaling with $x_F$, not the target momentum fraction $x_2$ 
\cite{Mike}.
The PHENIX data are consistent with an increase of effective `absorption'
at forward rapidity, as discussed in Ref.~\cite{FMV}, 
similar to that seen in fixed-target experiments at
large $x_F$ \cite{QWG10}.  We will compare our results for shadowing alone with
the PHENIX data at the end of this paper.

Effects we do not consider here which affect the forward rapidity region in
particular are energy loss in cold matter
and intrinsic charm. 
The effect of energy loss, both with and without including shadowing, has been
discussed in detail in Ref.~\cite{ArleoPeigne1,ArleoPeigne2}.  
However, given the rather
simple power law dependence employed for the $p+p$ distribution, it would be
worth pursuing embedding this approach within a production model such as the
CEM.
Intrinsic charm is not considered here because
the interesting region, $x_F \geq 0.25$, is inaccessible for
quarkonium production in $|y| \leq 5$, see Ref.~\cite{RV10}.

We do not include absorption by nucleons or comovers, which could affect the
entire rapidity region, in this paper.
The effective $J/\psi$ absorption cross section 
has been seen to decrease with energy at midrapidity \cite{LVW}.  When
shadowing is included, the $\sqrt{s}$ dependence is somewhat stronger due to the
increased shadowing effect at low $x$ \cite{LVW}.  
If the nuclear crossing time
is shorter than the $J/\psi$ formation time, the effective absorption
decreases with $\sqrt{s_{_{NN}}}$ as the state remains small during the
entire time it spends in the target.  However, absorption may play a role at the
most backward rapidities when the quarkonium state is slow and can convert
from the pre-resonance state to the final-state $J/\psi$ in the target
\cite{FMV,Arleo}.  
The $\Upsilon$(1S) absorption cross section is expected to
be smaller than that of the $J/\psi$ because of its smaller radius.

While the $J/\psi$-comover cross section is smaller than
the nucleon absorption cross section, comovers may be more important for the
excited charmonium and bottomonium states.  (While these states have larger
radii than the ground state quarkonium, this does not affect their potential
absorption by nucleons since they still pass through the nuclear matter in
their pre-resonant state.)  Strong differences between the $J/\psi$ and $\psi'$
$R_{\rm dAu}$ at RHIC as a function of collision centrality \cite{psipabs}
would support the
importance of comovers since their density increases with centrality.
Recent comover-based $p+$Pb calculations show that this interpretation is
consistent with the data \cite{Elena}.

\section{Nuclear Modifications of the Parton Densities}
\label{Sec:params}

We use several parameterizations of the nuclear modifications in the parton
densities (nPDFs) to probe the possible range of
shadowing effects: EPS09 \cite{EPS09}, 
EKS98 \cite{Eskola:1998iy,Eskola:1998df}, nDS and nDSg \cite{deFlorian:xxx}, 
and the two FGS sets, FGS-H and FGS-L \cite{FGS}.  
Since $gg$ processes dominate quarkonium production over all measurable
rapidities at the LHC, we focus more on the behavior of the nuclear gluon
parton densities in this section.
Our main results in this paper are obtained with EPS09 NLO to make a fully
consistent NLO calculation with the NLO CEM cross sections.  However, in the
following section, we will
discuss the differences between the nPDF sets and between their LO and NLO 
manifestations on $J/\psi$
and $\Upsilon$ production in $p+$Pb collisions at $\sqrt{s_{_{NN}}} = 5$ TeV.

All these sets involve some data fitting, typically nuclear deep-inelastic
scattering (nDIS) data
with additional constraints from other observables such as Drell-Yan dimuon
production.  The fact that these sets also include Drell-Yan production and
do not include initial-state energy loss in matter in their global analyses,
they exclude the possibility of additional quark energy loss since 
incorporating both over counts
the effect on the sea quark densities.  (Of course, if energy loss is 
generated by gluon emission either from an initial-state gluon or from
the produced $Q \overline Q$ pair, there is no energy loss possible in the
Drell-Yan process, $q \overline q \rightarrow l^+ l^-$, 
since a virtual photon generates a lepton pair.)
None of these data provide any
direct constraint on the nuclear gluon density. It is thus obtained through
fits to the $\mu^2$ dependence of the nuclear structure function, $F_2^A$,
and the momentum sum rule.  The useful perturbative $\mu^2$ range of the
nDIS data is rather limited since these data are only available at fixed-target
energies.  Thus the reach in momentum fraction, $x$, is also limited and
there is little available data for $x < 10^{-2}$ at perturbative values of
$\mu^2$.  This situation is likely
not to improve until an $eA$ collider is constructed \cite{thomasqm08}.
We discuss the various nuclear modifications in order of their release.

The EKS98 parameterization, by Eskola and 
collaborators \cite{Eskola:1998iy,Eskola:1998df}, available for
$A > 2$, is a leading order fit using the GRV LO \cite{GRV} 
proton parton densities as a baseline.
The first EKS98 fits \cite{Eskola:1998iy} used the GRV LO proton densities
from 1992 \cite{GRV}.  This set employed a starting scale of   
$\mu_0^2 = 0.3$ GeV.  
The set eventually released as EKS98 was 
constructed with CTEQ4L \cite{CTEQ4} with initial scale of $\mu_0 = 1.6$ GeV.
To match the starting scale of
the EKS98 nPDF set, the CTEQ4L distributions were backward evolved from 
$\mu = 1.6$ GeV to 1.5 GeV \cite{Eskola:1998df}.  
They checked that the nPDF results from
these two proton PDFs were consistent even though the proton PDFs themselves
were quite different: the GRV LO set employed a valence-like gluon distribution
at the starting scale, $xg(x,\mu_0^2) \sim x^\alpha P(x)$ where $\alpha > 0$ and
$P(x)$ is a polynomial function, while CTEQ4L takes $\alpha < 0$, giving
$xg(x,\mu_0^2)$ a finite value as $x \rightarrow 0$.  The minimum scales of
the two sets are also very different.  
The kinematic range of EKS98, the only leading order set we consider since
no NLO set was obtained at the time, is $1.5 \leq \mu \leq 100$~GeV and 
$10^{-6} \leq x < 1$.  All the sets produced by Eskola and collaborators are
based on piecewise functions at the minimum scale of the set
that include small $x$ shadowing, antishadowing
at intermediate $x$, an EMC region (named for the effect first identified by
the European Muon Collaboration) with $R_g < 1$ at larger $x$ and Fermi motion
as $x \rightarrow 1$.

deFlorian and Sassot
produced the nDS and nDSg parameterizations \cite{deFlorian:xxx} at both
leading and next-to-leading order for $4<A<208$.  
Their results were based on the GRV98 proton PDFs at LO and NLO \cite{GRV98}, 
using the same starting scales as GRV98, $\mu_{0 \, {\rm LO}}^2 = 0.40$ GeV, 
and $\mu_{0 \, {\rm NLO}}^2 = 0.25$ GeV.  GRV98 assumed valence-like
input for the sea quarks and gluons at the minimum scales with 
$xc(x,\mu_0^2) =0$.  The GRV98 gluon densities are not significantly different
at low $x$ when evolved to higher scales.  The leading order set has a
somewhat higher gluon density at low $x$ \cite{GRV98}.  
Of the two deFlorian and Sassot sets, nDS and nDSg,
the first, nDS, is an unconstrained fit that gives the best fit, $\chi^2$.
To provide a set with stronger gluon shadowing, they constrained a second fit
to give $S_g^{\rm Au} \equiv 0.75$ at $x = 0.001$ and $\mu^2 = 5$ GeV$^2$.
This set, nDSg, was a much poorer fit to the New Muon Collaboration (NMC) 
and SLAC nDIS data and the
E772 Drell-Yan cross sections \cite{deFlorian:xxx}.  

They showed that their results for the
suppression factor for $\pi^0$ production at RHIC, $R_{\rm dAu}^{\pi^0}(p_T)$, at
both LO and NLO were in agreement with each other \cite{deFlorian:xxx}.  
This should be the case
for a consistent order-by-order extraction of the nPDFs: not that the 
shadowing ratios are similar but that they give similar observable results,
such as the nuclear suppression factors.
The kinematic reach in $x$ is the same as EKS98 while the
$\mu^2$ range is larger, $1 < \mu^2 <10^6$~GeV$^2$.  

The FGS sets by Frankfurt, Guzey and Strikman take the highest initial scale
of all the nPDF sets, $\mu = 2$ GeV \cite{FGS}.  
They only provide NLO sets.  Instead
of a general global analysis, their nPDFs are based on the leading-twist
approximation and rely on the diffractive PDFs.  
The Gribov-Glauber approach employed naturally introduces a
centrality dependence for the shadowing.  Their baseline proton parton 
densities are the
CTEQ5M distributions. Therefore, their
extraction is only good for $x > 10^{-5}$.  

They extracted two sets to represent an upper and a lower
bound on the shadowing ratios.  FGS-H is gives stronger shadowing based on
the two-nucleon scattering cross section.  FGS-L is based instead on the $\pi N$
scattering cross section.  Since this is larger than the two-nucleon scattering
cross section, the shadowing effect derived in this case is weaker.  Their
approach is only applicable for $x < 0.1$.  For 
$x > 0.1$, the sea quark ratio relative to the nucleon is fixed to unity and
gluon antishadowing is obtained through the momentum sum rule.  Hence, in this
high $x$ region, the two parameterizations are identical.  
Their approach does not allow 
predictions for shadowing on the valence quark distributions, these are taken
from the LO EKS98 set. 
The FGS sets are available
only for some specific nuclei where diffractive data exist: $^{12}$C, $^{40}$Ca,
$^{110}$Pd, $^{197}$Au and $^{208}$Pb.  The sets are valid in $10^{-5} < x < 0.95$
and $2 < \mu < 100$ GeV.

The EPS09 \cite{EPS09} LO and NLO parameterizations
included uncertainties on the global analyses, both at LO and NLO, by varying
one of the 15 fit parameters within its extremes while holding the 
others fixed.  The upper and lower bounds on EPS09 shadowing are obtained by
adding the resulting uncertainties in quadrature \cite{EPS09}.  The EPS09
central LO results are in quite good agreement with the older
EKS98 parameterization.  The EPS09 LO sets are based on the CTEQ61L densities
which are nearly flat at the initial scale as $x \rightarrow 0$.  The EPS09
NLO densities are based on the CTEQ6M densities.  Starting with CTEQ6M, a
valence-like shape for the gluon distribution was adopted
at the starting scale, in this case the charm mass
threshold of $\mu_0 = 1.3$ GeV \cite{CTEQ6}.
The EPS09 sets assume the same starting scale as the CTEQ6 sets and is valid
in the range $10^{-6} < x < 1$
\cite{EPS09}.  

\begin{figure}[t]
\begin{center}
%\vspace*{-0.05in}
\includegraphics[width=0.495\textwidth]{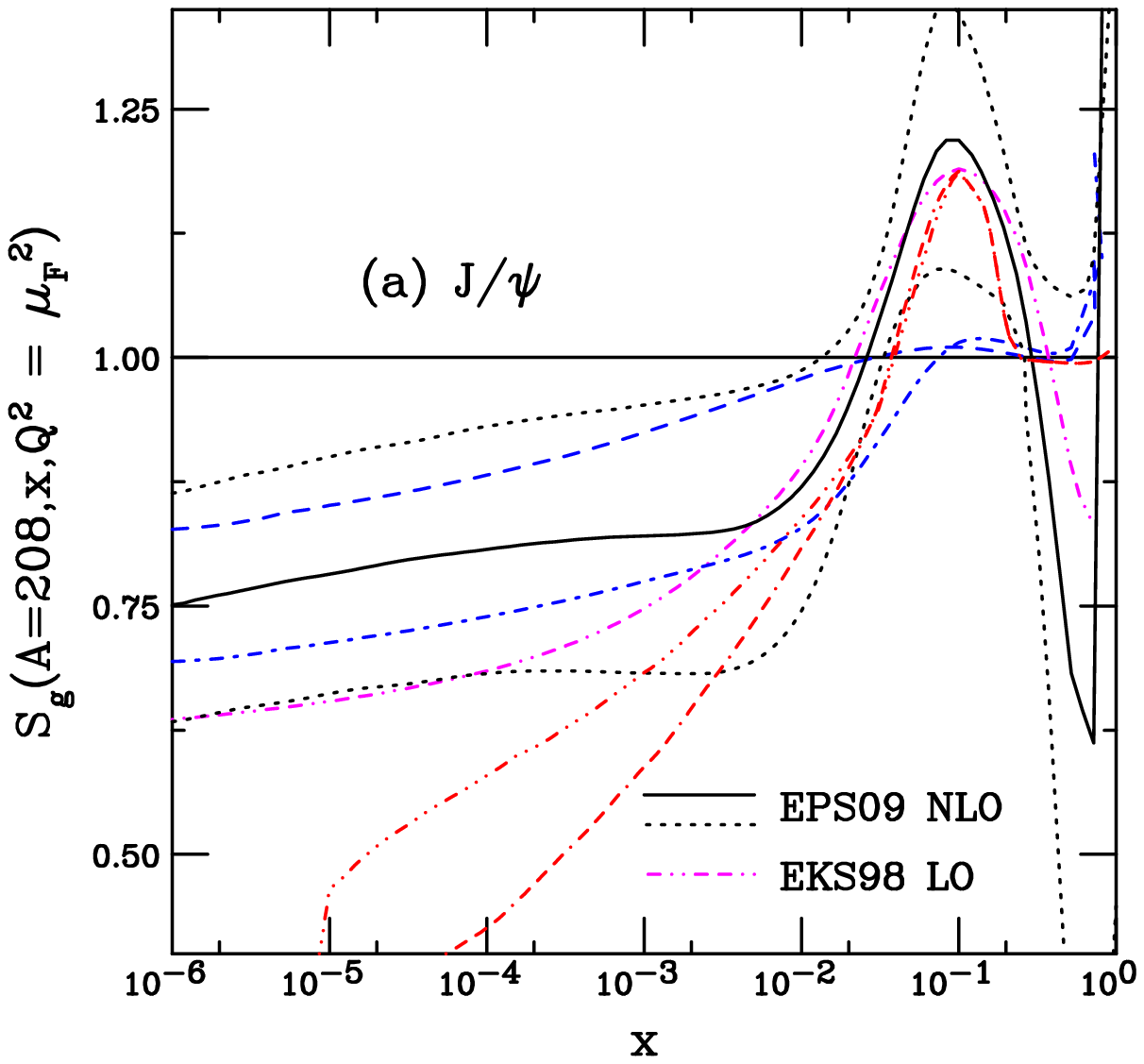}
\includegraphics[width=0.495\textwidth]{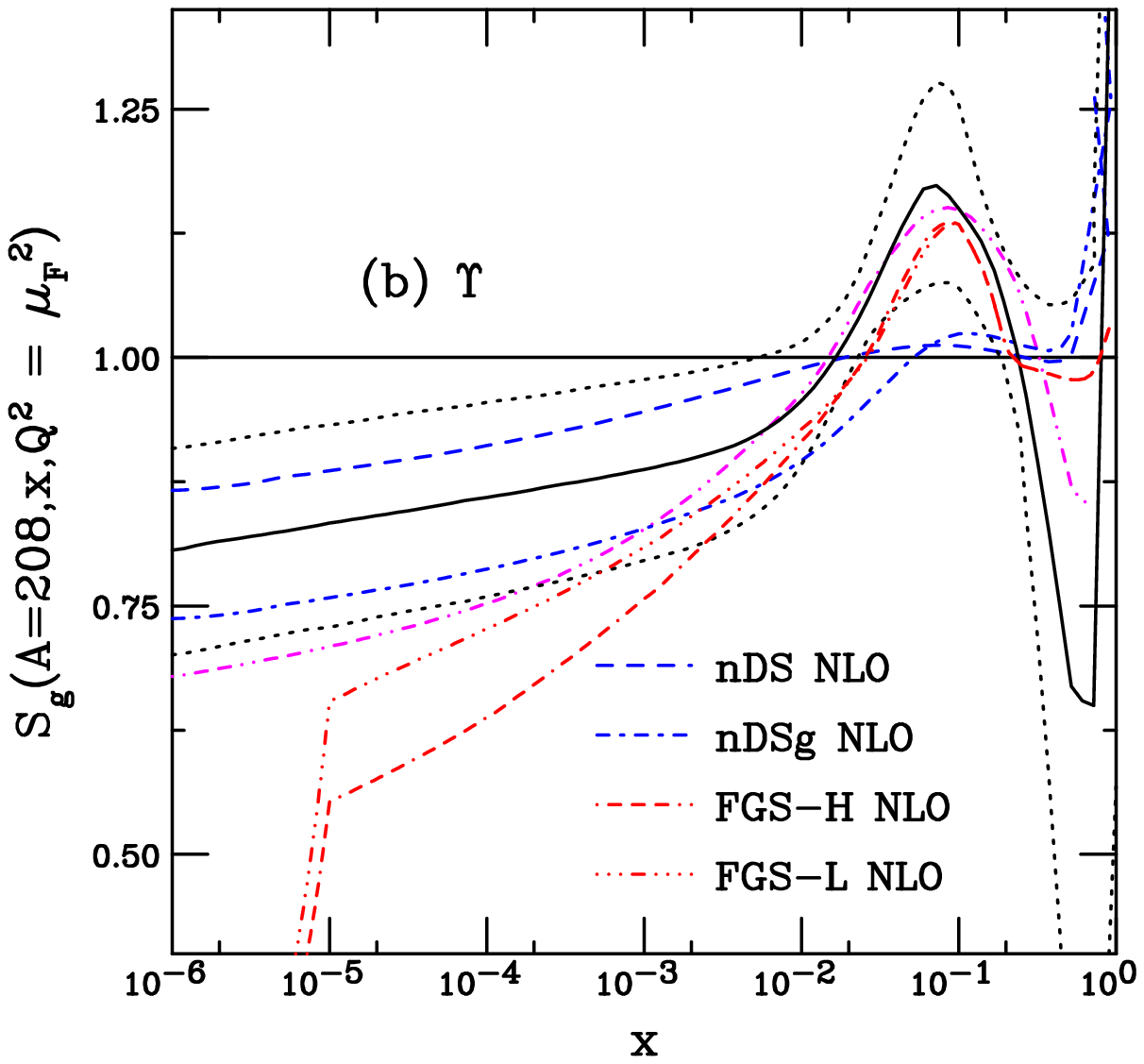}
\caption[]{(Color online)
Gluon shadowing ratios calculated for Pb nuclei ($A = 208$) calculated at the
central value of the fitted factorization scales for $J/psi$ (a) and
$\Upsilon$ (b).  The EPS09 NLO set is shown by the
black solid curve while the uncertainty band is outlined by the black dotted
curves.  The NLO nDS and nDSg parameterizations are given in the blue dashed
and dot-dashed curves respectively.  The LO EKS98 parameterization is
given in the magenta dot-dot-dash-dashed curve.  The NLO FGS-H and FGS-L
results are given by the red dot-dash-dash-dashed and dot-dot-dot-dashed
curves respectively.
}
\label{fig:nPDFs}
\end{center}
\end{figure}

Figure \ref{fig:nPDFs} shows the sets we have discussed in this section
at the scales of $J/\psi$ (a) and $\Upsilon$ (b) production.  The results are
shown for $x$ values as low as $10^{-6}$.
The EPS09 NLO
sets are shown in the black solid and dotted curves.  The solid curves show
the central results while the dotted curves outline the NLO bands.  The
EPS09 results for $J/\psi$ show a change in curvature, almost a dip, at
$x \sim 0.01$, with a small rise at lower $x$ followed by an eventual decrease.
There is a rather strong antishadowing peak from $0.02 < x < 0.2$
where the ratio drops below unity again, in the EMC region.

The EKS98 LO ratio is similar to that of the central EPS09 LO set but decreases
more smoothly, giving a stronger shadowing at lower $x$ than the NLO set,
a 40\% shadowing effect at $x = 10^{-6}$ rather than the 30\% effect at NLO.
At the lowest $x$ values, the central EPS09 LO is equivalent to the lower 
limit of the EPS09 NLO
band.

The FGS-H and FGS-L sets are similar to the others but are somewhat narrower 
in the antishadowing region.  However, they decrease faster with decreasing
$x$ than any of the other sets shown.  They drop sharply at
$x = 10^{-5}$.  Instead of either fixing the ratio at its value at 
$x = 10^{-5}$, as in the CTEQ5M proton PDF set, or trying to make a smooth
extrapolation to lower $x$, it simply falls off more like a step function.

The nDS set has somewhat stronger shadowing that the upper limit of the EPS09
NLO band while the nDSg set is between the EPS09 NLO central and lower limits.
These sets, of all the results shown, have no antishadowing at all.

The same features can be observed for the shadowing ratios at the $\Upsilon$
scale.  The EPS09 band now decreases more smoothly as $x$ decreases.  The FGS-H
ad FGS-L ratios show the steep drop at the limit where $x = 10^{-5}$ more 
clearly at the larger scale.  All the slopes of the other nPDF sets
become rather similar.

\section{Results}
\label{Sec:results}

We now show calculations of the $J/\psi$ and $\Upsilon$ production ratios,
$R_{p {\rm Pb}}$ and $R_{FB}$ as a function of rapidity and $p_T$, taking only
nuclear shadowing effects into account.  We begin with comparing results with
the same shadowing parameterization calculated with different proton PDFs
in Sec.~\ref{SubSec:pPDFs}.  These results are shown for the $J/\psi$ only and
compared to a subset of the available data.  In the remainder of this section,
we compare the calculations to the full $J/\psi$ and $\Upsilon$ data sets from
ALICE and LHCb.   We show the measurements compared to the EPS09 NLO uncertainty
bands in Sec.~\ref{SubSec:EPS09_NLO}.  Next, we compare and contrast the leading
and next-to-leading order shadowing results of EPS09 and nDS in 
Sec.~\ref{SubSec:LOvsNLO}.  Thereafter, the results from the nPDF sets shown in
Fig.~\ref{fig:nPDFs} are compared to the data and each other in 
Sec.~\ref{SubSec:nPDFs_comp}.  The mass and scale uncertainties on the central
EPS09 set are compared to the uncertainties in the shadowing parameterizations
themselves in Sec.~\ref{SubSec:M_muvar}.  Next we discuss how well the $A+A$
results factorize into a product of $p+A$ and $A+p$ collisions in 
Sec.~\ref{SubSec:Fact}.  Finally, we present the EPS09 NLO uncertainty bands 
for RHIC kinematics in Sec.~\ref{SubSec:RHIC}.

\subsection{Comparison of EPS09 NLO for Different Proton PDFs}
\label{SubSec:pPDFs}

To begin, we check whether or not the shadowing results depend on the chosen
set of proton parton distributions.  As demonstrated in 
Sec.~\ref{SubSec:ppcomp},
the choice of proton PDF has a strong effect on the shape of the individual
$p+p$ distributions.  We now want to find out of this also means a change in
the shape of the nuclear suppression factors since each shadowing 
parameterization is based on different proton PDFs.  We employ the central
EPS09 NLO set for this study.  

\begin{figure}[t]
\begin{center}
%\vspace*{-0.05in}
\includegraphics[width=0.45\textwidth]{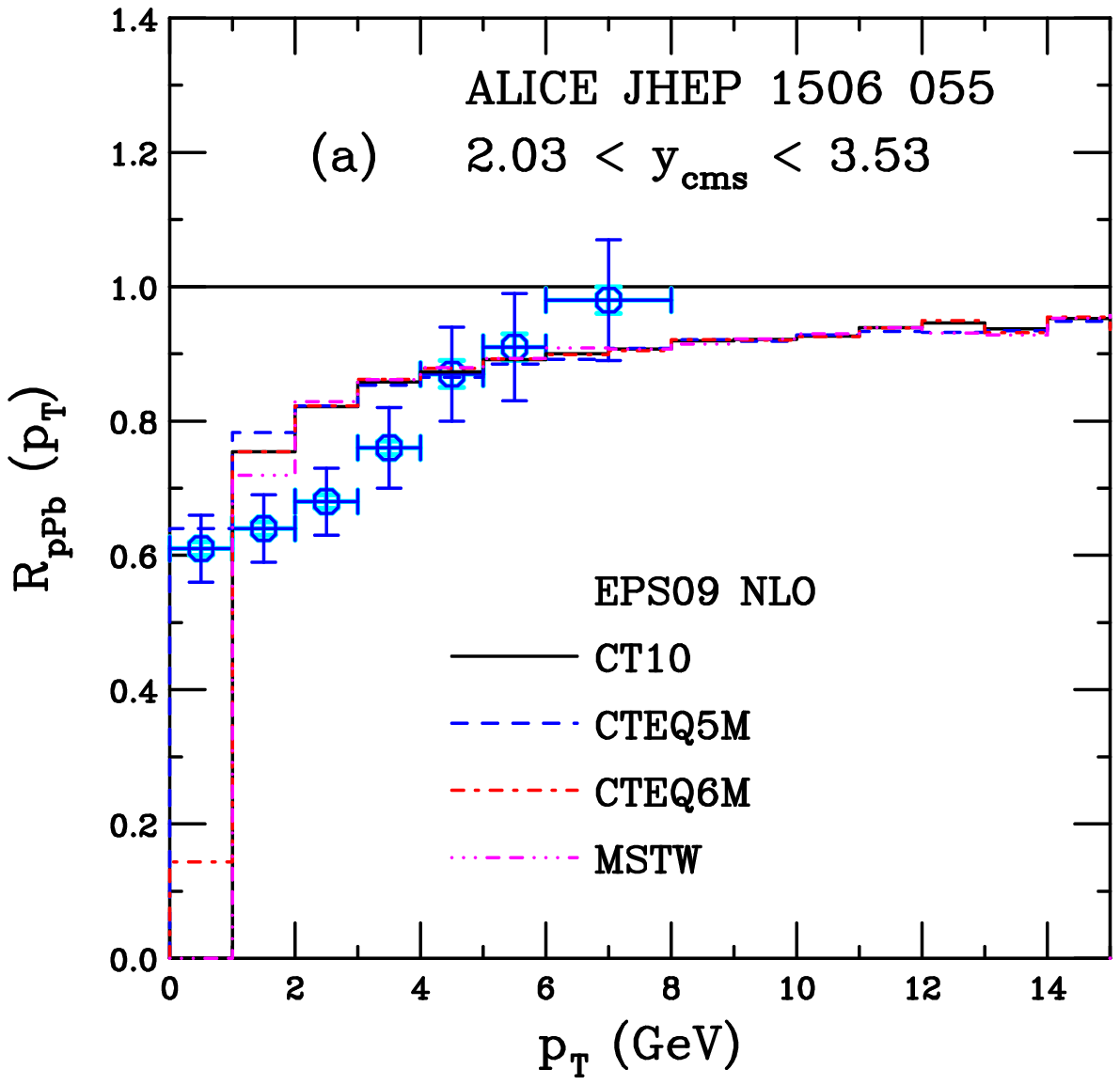}
\includegraphics[width=0.45\textwidth]{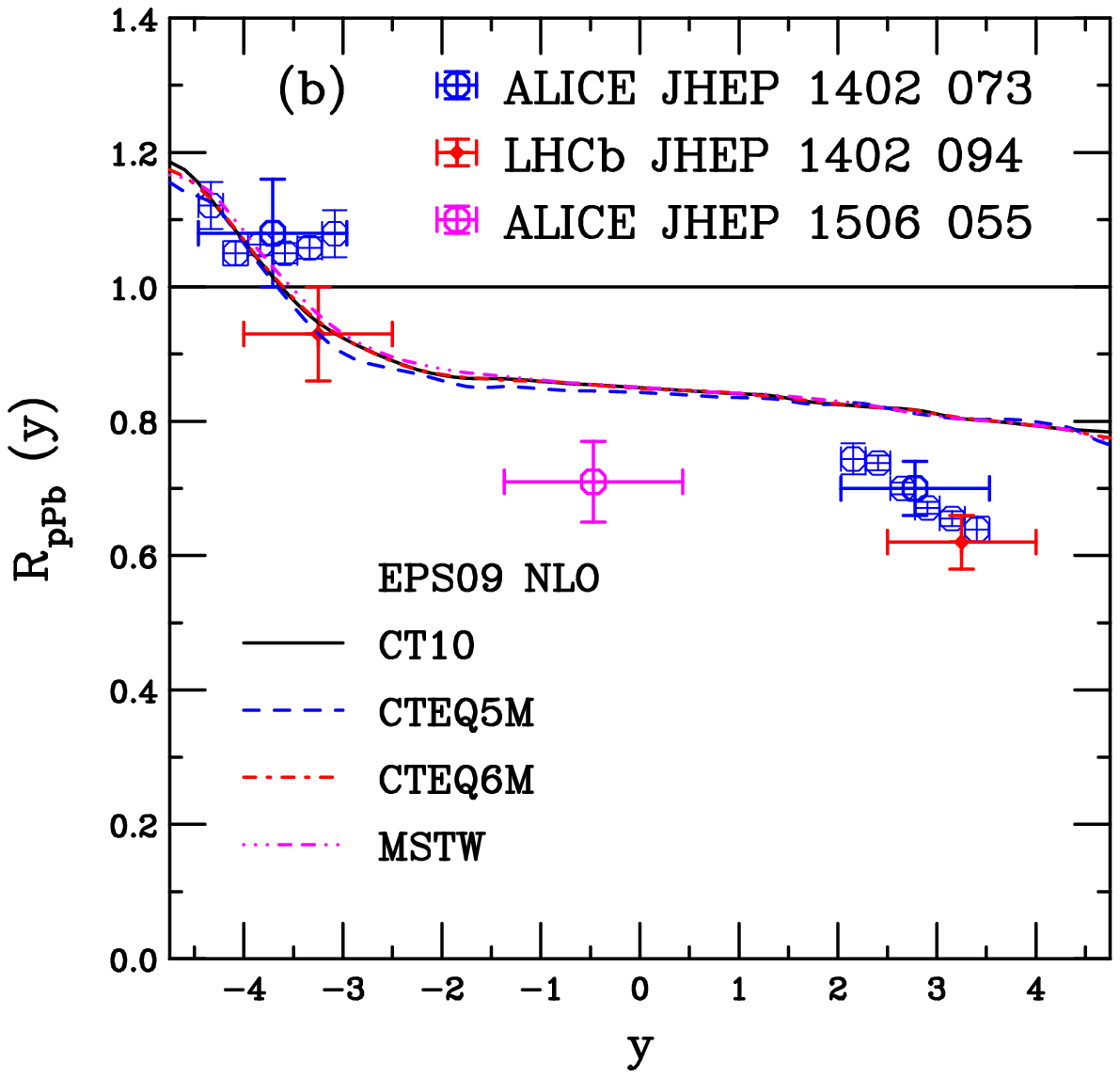}
\caption[]{(Color online)
The $J/\psi$ ratio $R_{p{\rm Pb}}(p_T)$ for ALICE at forward rapidity
\protect\cite{ALICEpPbpsi_pT} (a) and
$p_T$-integrated as a function of rapidity \protect\cite{ALICEpPbpsi} (b).  
The ratios are shown for CT10 (black),
CTEQ6M (blue), CTEQ5M (red) and MSTW (magenta).
}
\label{fig:pPDFs_comp}
\end{center}
\end{figure}

The results for $R_{p{\rm Pb}}(p_T)$ and 
$R_{p{\rm Pb}}(y)$ are shown in 
Fig.~\ref{fig:pPDFs_comp}.  The left hand side shows $R_{p{\rm Pb}}(p_T)$ 
for the ALICE forward rapidity bin, $2.03 < y_{\rm cms} < 3.53$ while the 
right-hand side displays the $p_T$-integrated $R_{p{\rm Pb}}(y)$.  The only 
visible difference between the results is for the lowest $p_T$ bin shown,
$1 < p_T < 2$ GeV, where the difference is $\sim 4$\% between the ratios
with CTEQ5M and MSTW relative to CT10.  There is no difference in the CT10
and CTEQ6M ratios.  At higher $p_T$, the difference is negligible.
Some variation on the order of the percent level is seen
as a function of rapidity, particularly in the backward region.  Since these
variations are significantly less than those between the nPDFs themselves, we
see that the choice of proton PDF has a negligible effect on the results.

\subsection{EPS09 NLO Uncertainties}
\label{SubSec:EPS09_NLO}

Here we present the EPS09 NLO uncertainties in $J/\psi$ and $\Upsilon$
production.  We will compare the calculated nuclear suppression factor, 
$R_{p{\rm Pb}}$, and the forward-backward ratio, $R_{FB}$, to the ALICE and LHCb
data.

\subsubsection{$J/\psi$}

\begin{figure}[t]
\begin{center}
\includegraphics[width=0.45\textwidth]{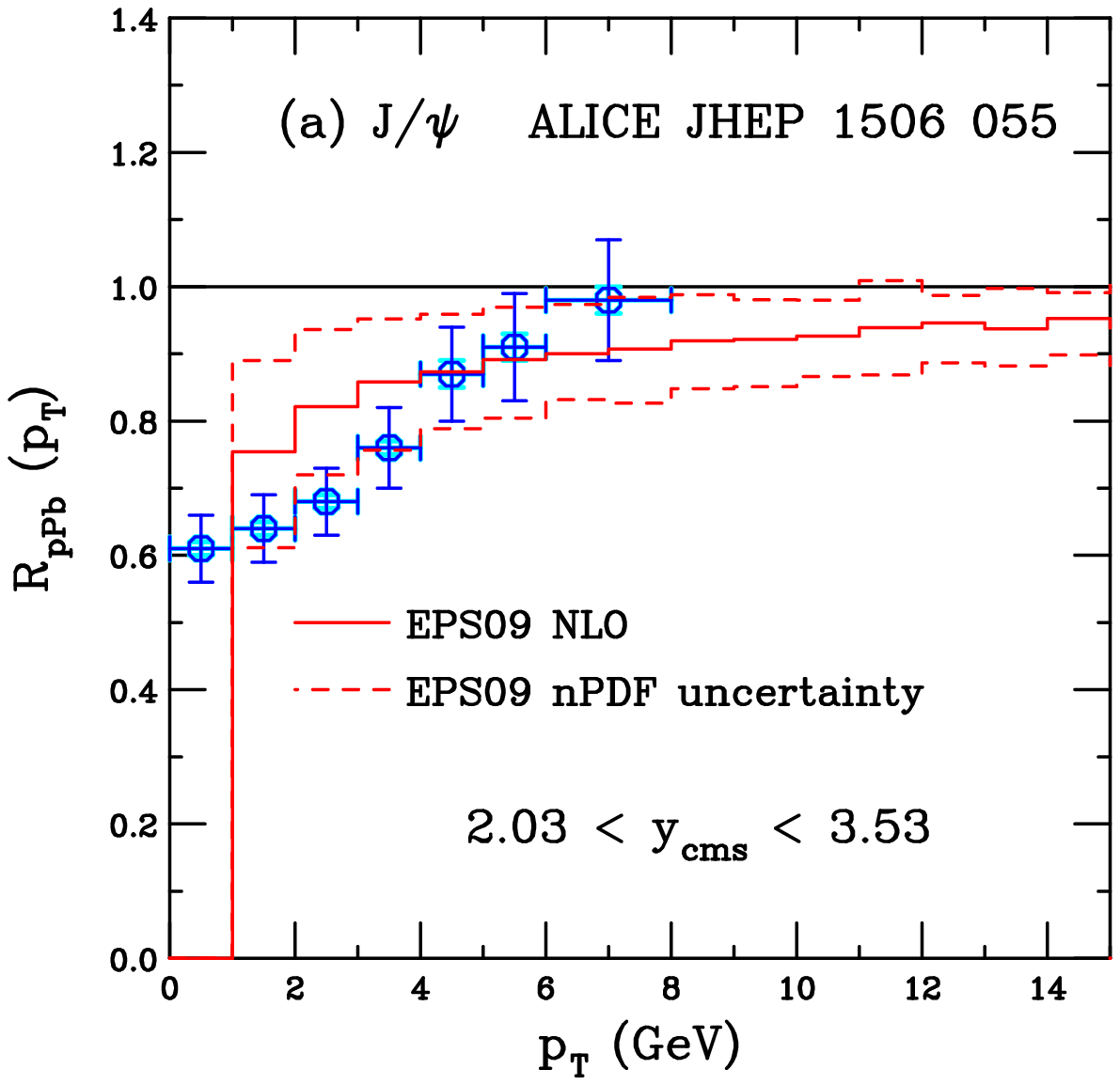}
\includegraphics[width=0.45\textwidth]{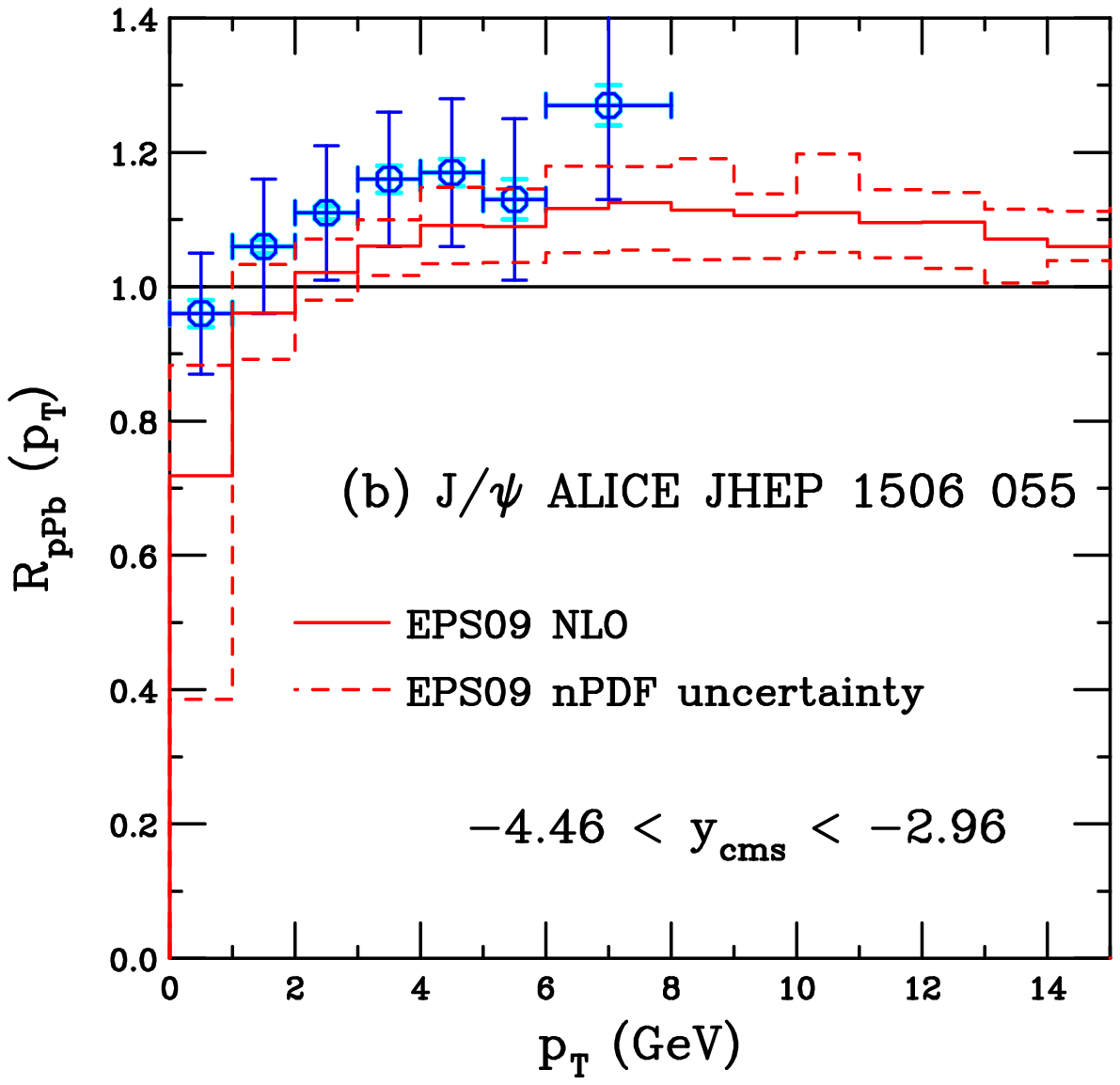} \\
\includegraphics[width=0.45\textwidth]{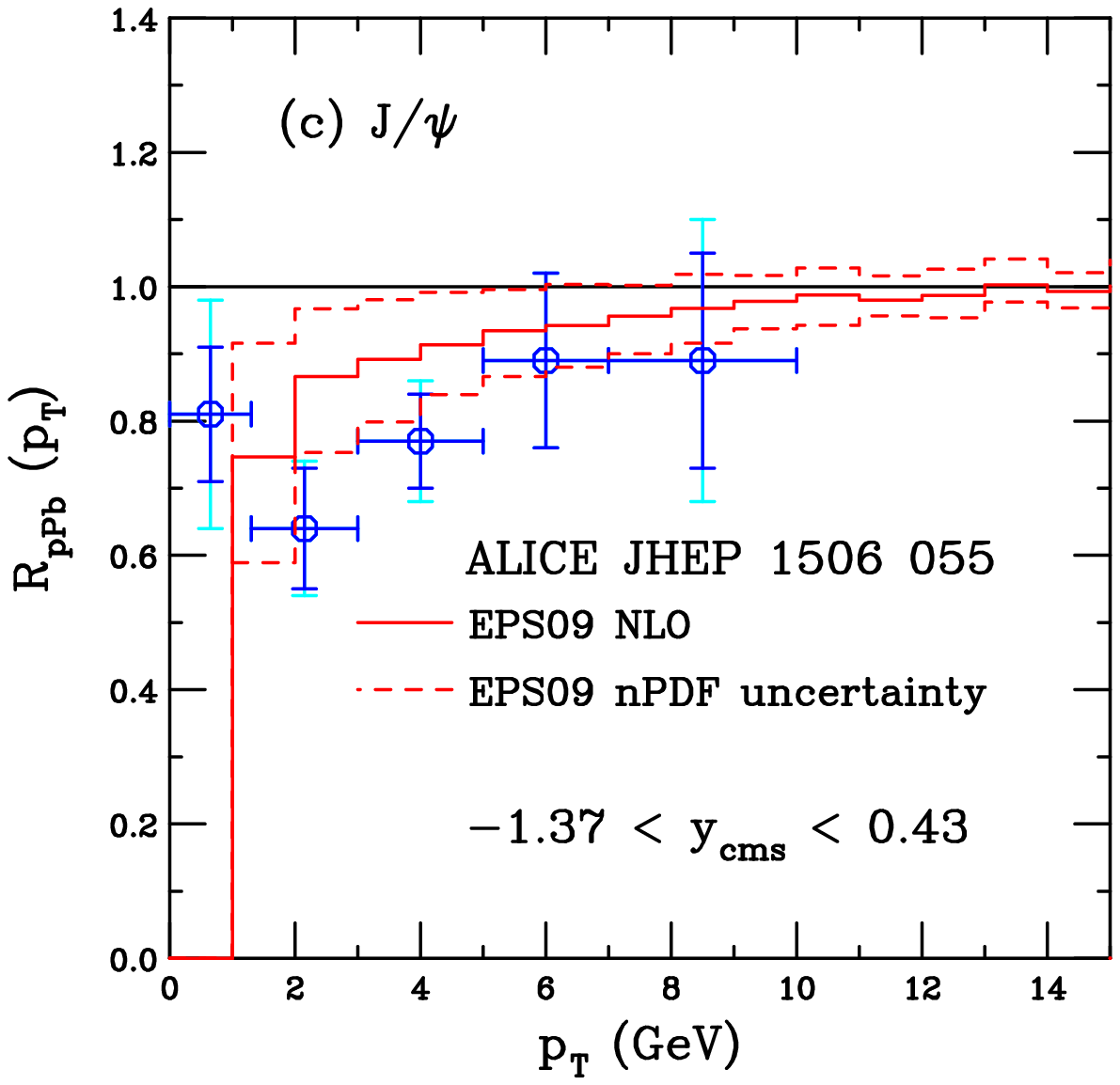}
\includegraphics[width=0.45\textwidth]{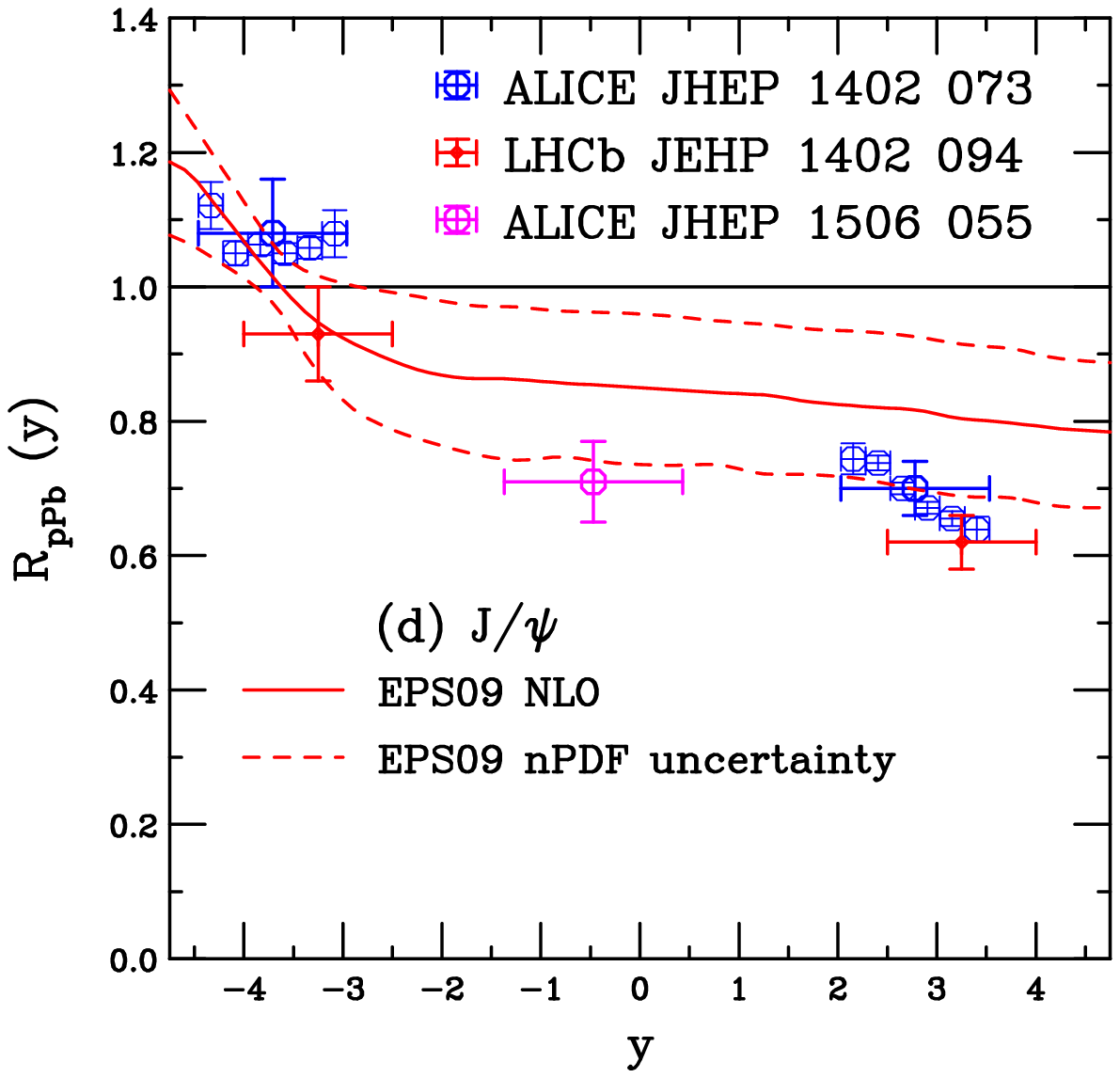}
\caption[]{(Color online)
The $J/\psi$ ratio $R_{p{\rm Pb}}(p_T)$ in the
ALICE acceptance at forward (a),
backward (b) and central (c) rapidity \protect\cite{ALICEpPbpsi_pT}.  
The ratio $R_{p{\rm Pb}}(y)$ \protect\cite{ALICEpPbpsi} is shown in (d).
The EPS09 NLO uncertainty band 
is shown.
}
\label{fig:EPS09RpPb_Psi}
\end{center}
\end{figure}

Figure~\ref{fig:EPS09RpPb_Psi} compares the suppression factors for
$J/\psi \rightarrow \mu^+ \mu^-$ as a function of $p_T$ at forward (a) and 
backward (b) rapidity as well as for $J/\psi \rightarrow e^+ e^-$ at 
midrapidity (c).  The rapidity dependent ratio is shown in (d).  The solid
curves show the EPS09 NLO central results while the dashed curves outline the
upper and lower limits of the EPS09 NLO uncertainty band.  In all cases, the
bands on the EPS09 NLO uncertainties are obtained by adding the differences
in the results obtained by varying each parameter separately by one standard
deviation of the fit in quadrature.

In general, the $p_T$-dependent data are in relatively good
agreement with the EPS09 NLO bands, if not with the central values themselves.
In the forward region, the measured $R_{p {\rm Pb}}$ is compatible with the lower
edge of the uncertainty band for $p_T<6$ GeV.  The data
at backward rapidity also suggest a somewhat stronger cold matter effect, 
compatible with the upper edge of the band.  While the uncertainties
on the data at midrapidity are larger than those on the calculation, the
data again indicate a somewhat stronger effect than suggested by EPS09 NLO.
However, within the uncertainties of both the data and the calculations, 
the results are, in general agreement.  

A similar conclusion can be drawn from the $p_T$-integrated rapidity dependence.
The ALICE data at forward and backward rapidity are shown in two ways: a single
$p_T$-integrated point over the entire rapidity acceptance and with each broad
$y$ bin broken up into six separate bins.  While the smaller bins are almost
independent of $y$ in the backward region, there is a decrease in 
$R_{p{\rm Pb}}(y)$ in the forward region.  The overall observed dependence of
$R_{p{\rm Pb}}(y)$ is in good agreement with the $p_T$-dependent results in
Fig.~\ref{fig:EPS09RpPb_Psi}(a)-(c).  While the LHCb are within one standard
deviation of the ALICE data, they are somewhat lower.

\begin{figure}[t]
\begin{center}
%\vspace*{-0.05in}
\includegraphics[width=0.45\textwidth]{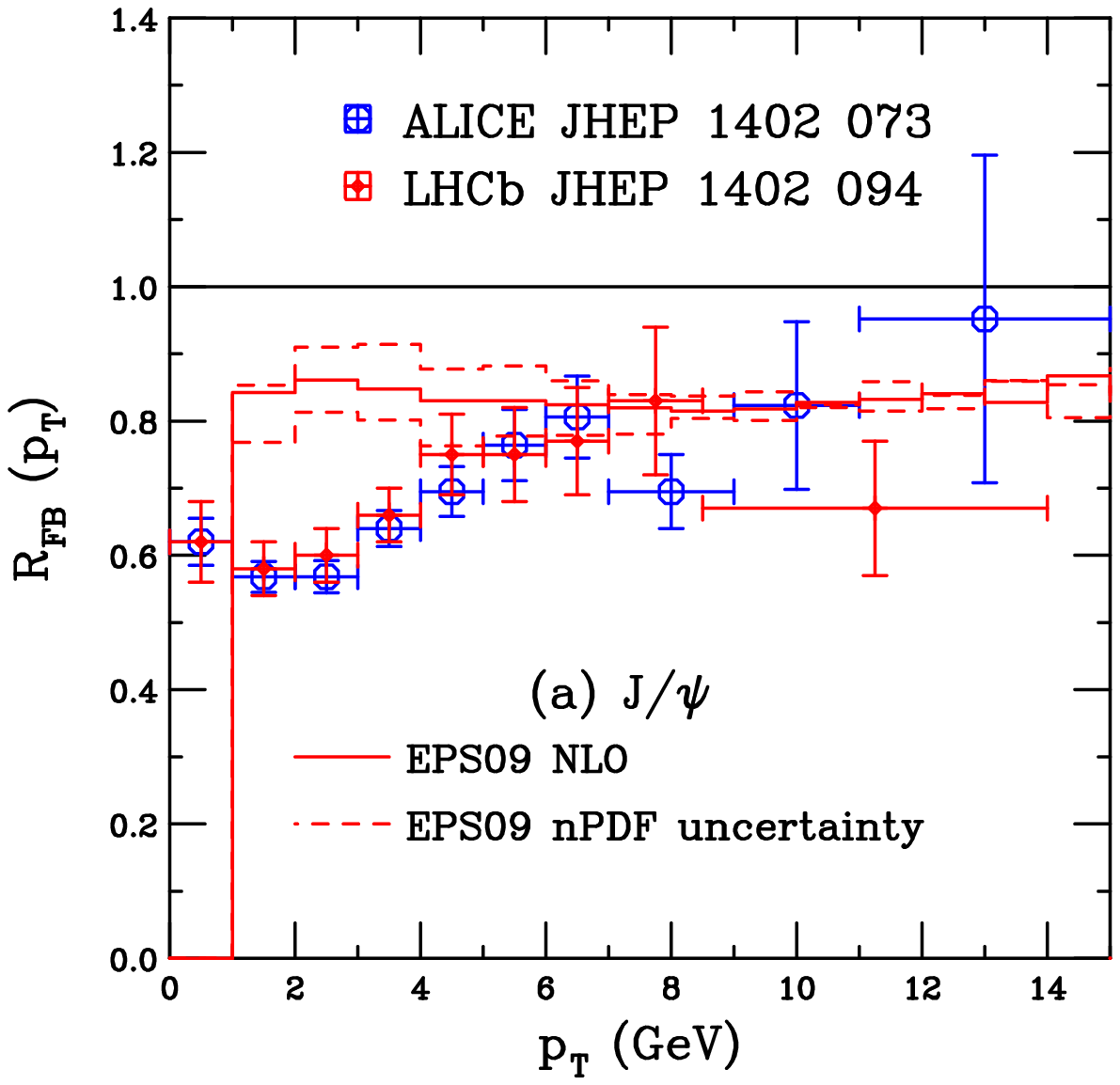}
\includegraphics[width=0.45\textwidth]{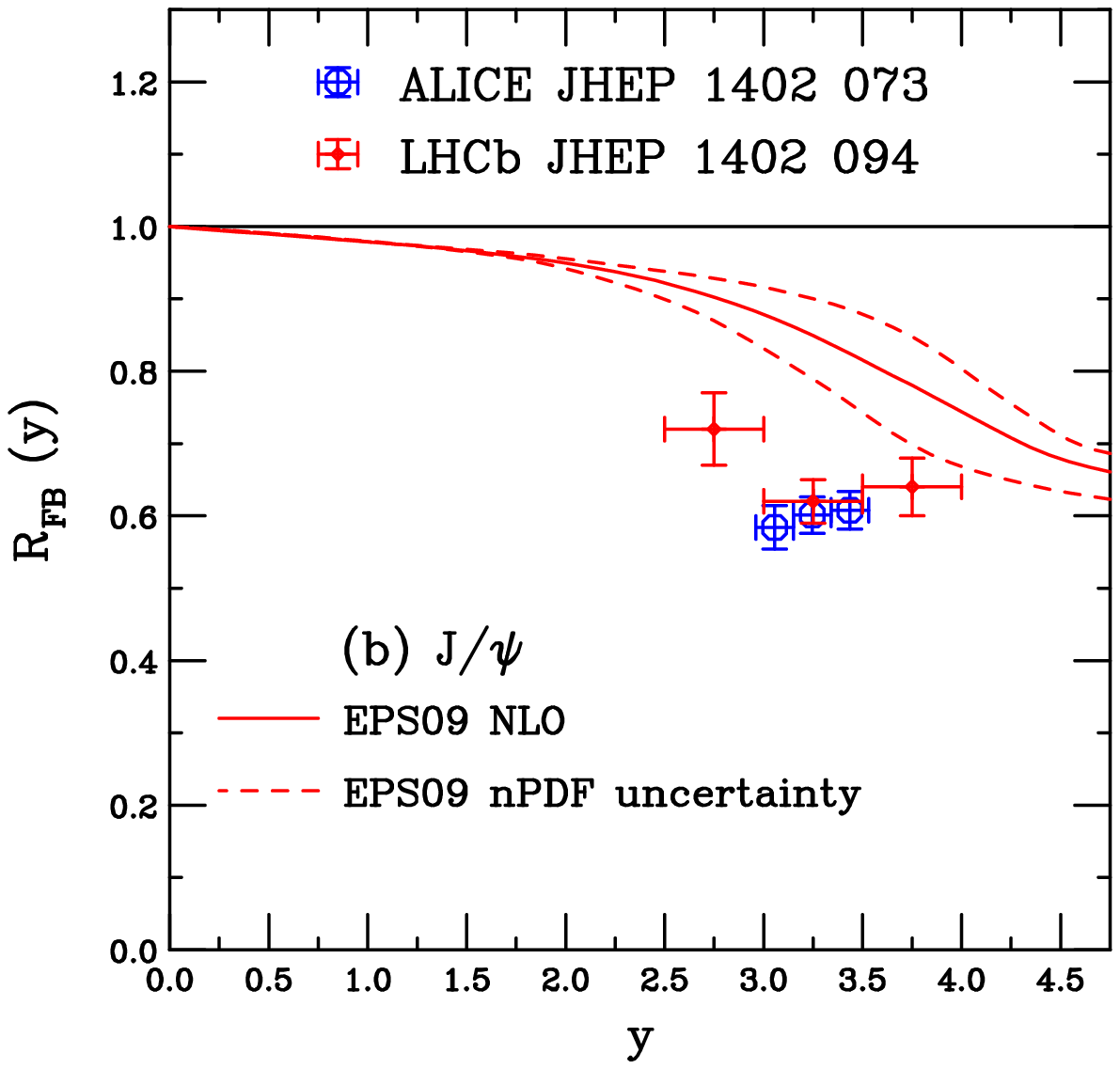}
\caption[]{(Color online)
The $J/\psi$ forward-backward ratio $R_{FB}(p_T)$ in the ALICE overlap region,
$2.96 < y_{\rm cms} < 3.53$ 
(a) and $R_{FB}(y)$ (b).  The EPS09 NLO uncertainty band
is shown with the ALICE \protect\cite{ALICEpPbpsi} and 
LHCb \protect\cite{LHCbpPbpsi} data.
}
\label{fig:EPS09RFB_Psi}
\end{center}
\end{figure}

As discussed previously, the suppression factor $R_{p{\rm Pb}}$ is 
somewhat artificial since there is 
currently no $p+p$ reference measurement at the same energy as the $p+$Pb
data.  Instead the reference is obtained from an interpolation between
the $p+p$ measurements at 2.76 and 7 TeV, as described in Sec.~\ref{Sec:Data}.  
The forward-backward ratio reduces the systematic uncertainty.  

The results are compared
to the $J/\psi$ calculations in Fig.~\ref{fig:EPS09RFB_Psi}.
We note that the uncertainties on the ratio are narrower than those of the
forward and backward regions separately, as shown in 
Fig.~\ref{fig:EPS09RpPb_Psi}.  The uncertainty bands are formed by
taking the forward-backward ratios for each of the 31 EPS09 sets individually
and then adding the uncertainties in quadrature.  The calculated
ratios are almost
independent of $p_T$.  This behavior is due to the fact that the forward
ratio is less than unity and increasing with $p_T$ while the backward ratio
is greater than unity over most of the $p_T$ range.
The data instead show a minimum at low $p_T$, increasing to
$R_{p{\rm Pb}} \sim 0.8$ at $p_T \sim 8$ GeV, albeit with large statistical
uncertainties.  

The calculated ratios as a function of rapidity are very narrow for $y < 2.5$
before broadening at larger $y$.  This behavior is obvious from looking
at $R_{p{\rm Pb}}(y)$ since, for $y > -2.5$ the uncertainty band is parallel to
the central value and almost linear so that when the forward-backward ratio
is formed, the band is compressed.  The ratio $R_{FB}(y)$ broadens and decreases
at $y > 3$, where, at backward $y$, $R_{p{\rm Pb}}(y)$ enters the antishadowing
region.  The forward-backward ratio narrows again at $y \sim 4$, near the
`pinch' in $R_{p {\rm Pb}}(y)$, where the data are best constrained, and finally
broadens again in the antishadowing regime where the constraints are poorer.
The data are below the calculated band because, while the backward region is 
rather well described, the central value is above the data at forward rapidity.
We note that while the ALICE $R_{FB}(y)$, in a smaller rapidity range, is almost
flat, the LHCb ratio varies more.  However, in the bin where the rapidity ranges
of the two experiments overlap, they are in good agreement.

Finally we remark on the recent ATLAS forward-backward measurement of 
$R_{FB}(p_T)$ in the central region, $|y_{\rm cms}| < 1.94$
and $R_{FB}(y)$ for $8 < p_T < 3  0$ GeV \cite{ATLASpPbpsi}.  Their results are 
consistent with unity in both $p_T$ and $y$ and agree well with our EPS09 NLO
calculations.  The results shown in Fig.~\ref{fig:EPS09RFB_Psi}(a) are in the 
forward region while the ATLAS measurement is at midrapidity.  Given the
higher scale implicit in the $p_T$ range covered by ATLAS, as well as the
higher $x$ in the rapidity range covered by ATLAS, it is unsurprising that
the EPS09 NLO result is consistent with unity here.  Similarly, given that
the EPS09 NLO result integrated over $p_T$ in Fig.~\ref{fig:EPS09RFB_Psi}(b)
is already similar to unity for $|y| < 1.5$, we can expect that the same
ratio for $p_T > 8$ GeV has a more shallow slope and deviates less from unity
than the results compared to ALICE and LHCb here.

\subsubsection{$\Upsilon$}

\begin{figure}[t]
\begin{center}
\includegraphics[width=0.45\textwidth]{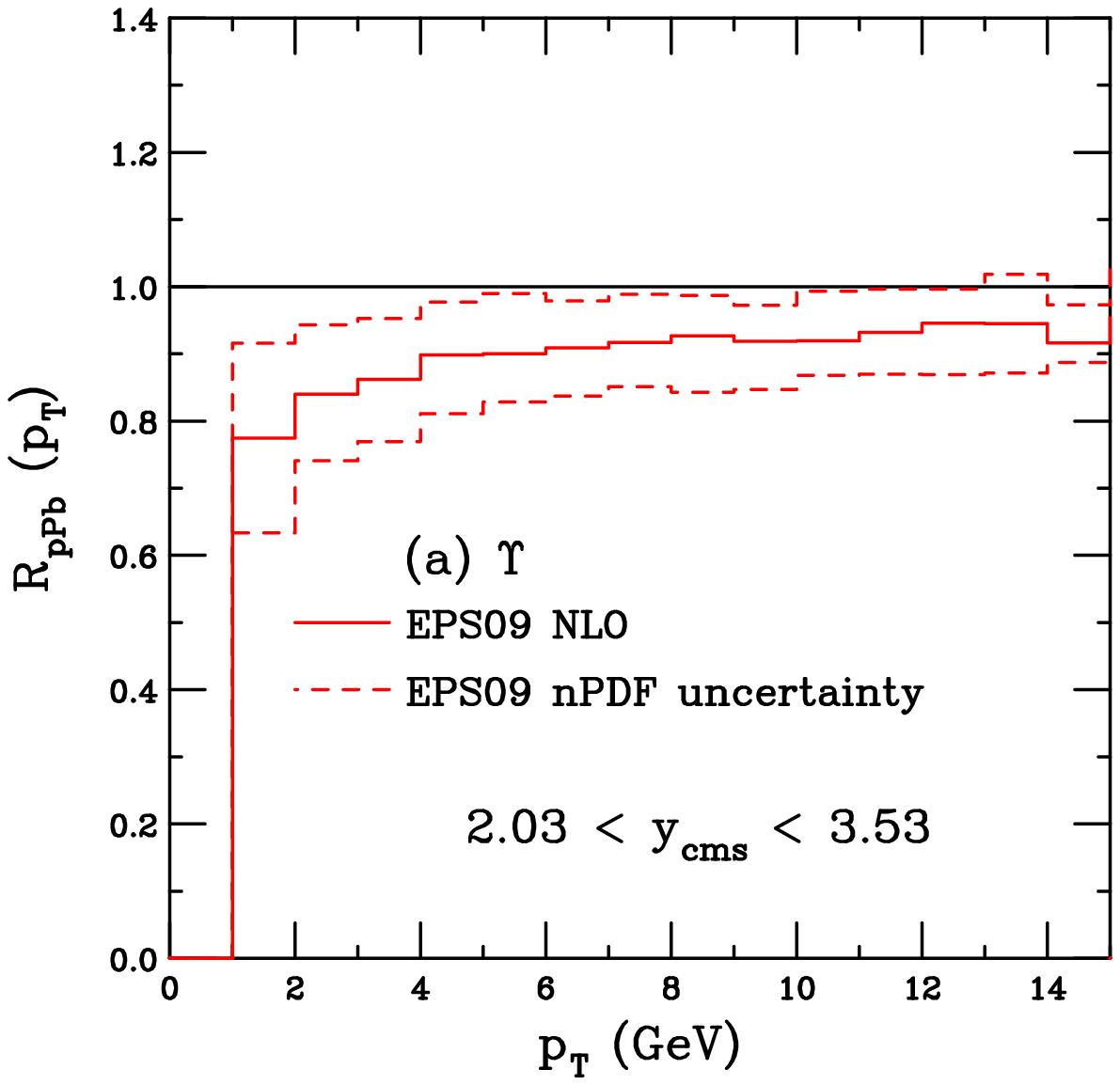}
\includegraphics[width=0.45\textwidth]{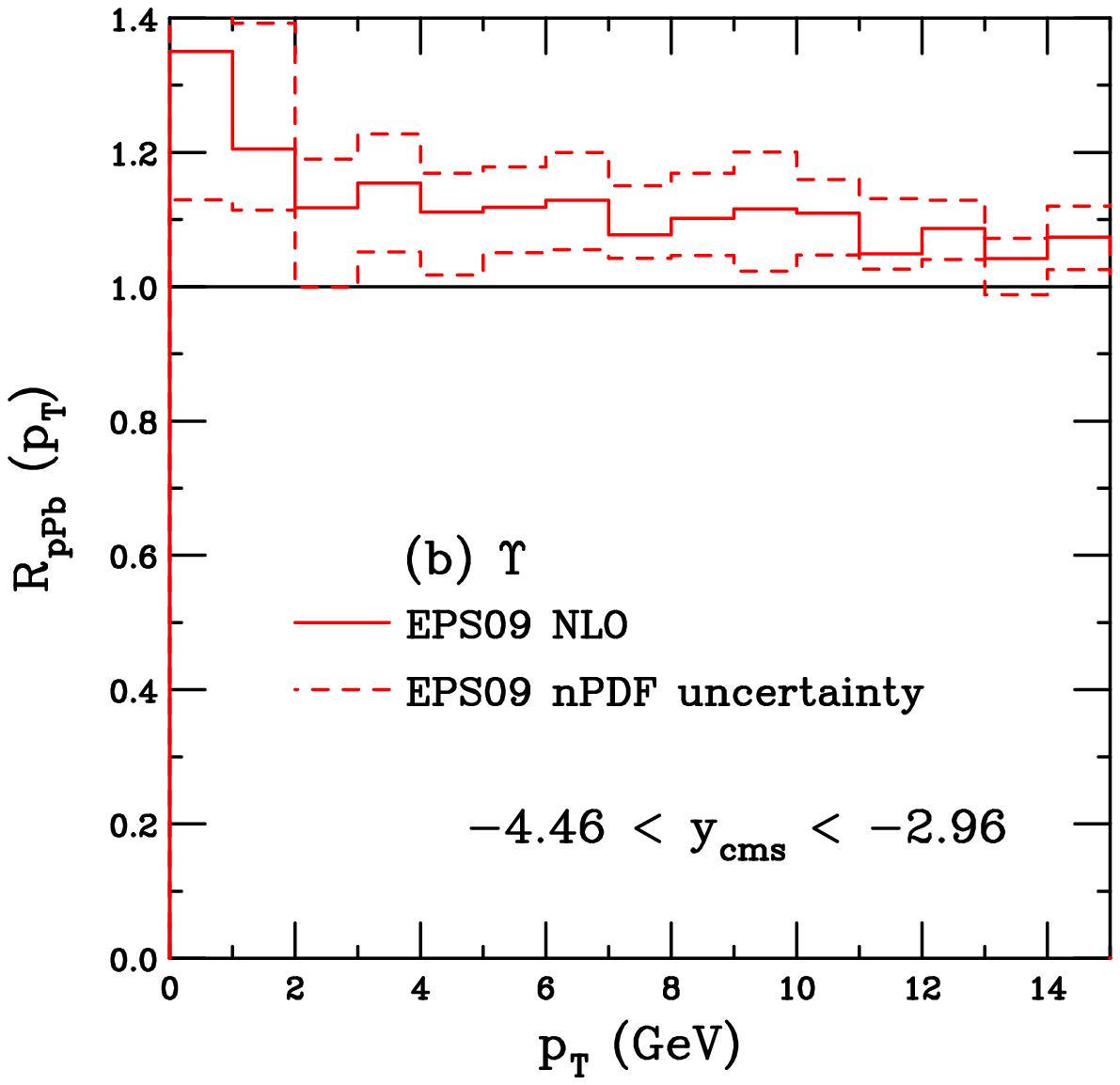} \\
\includegraphics[width=0.45\textwidth]{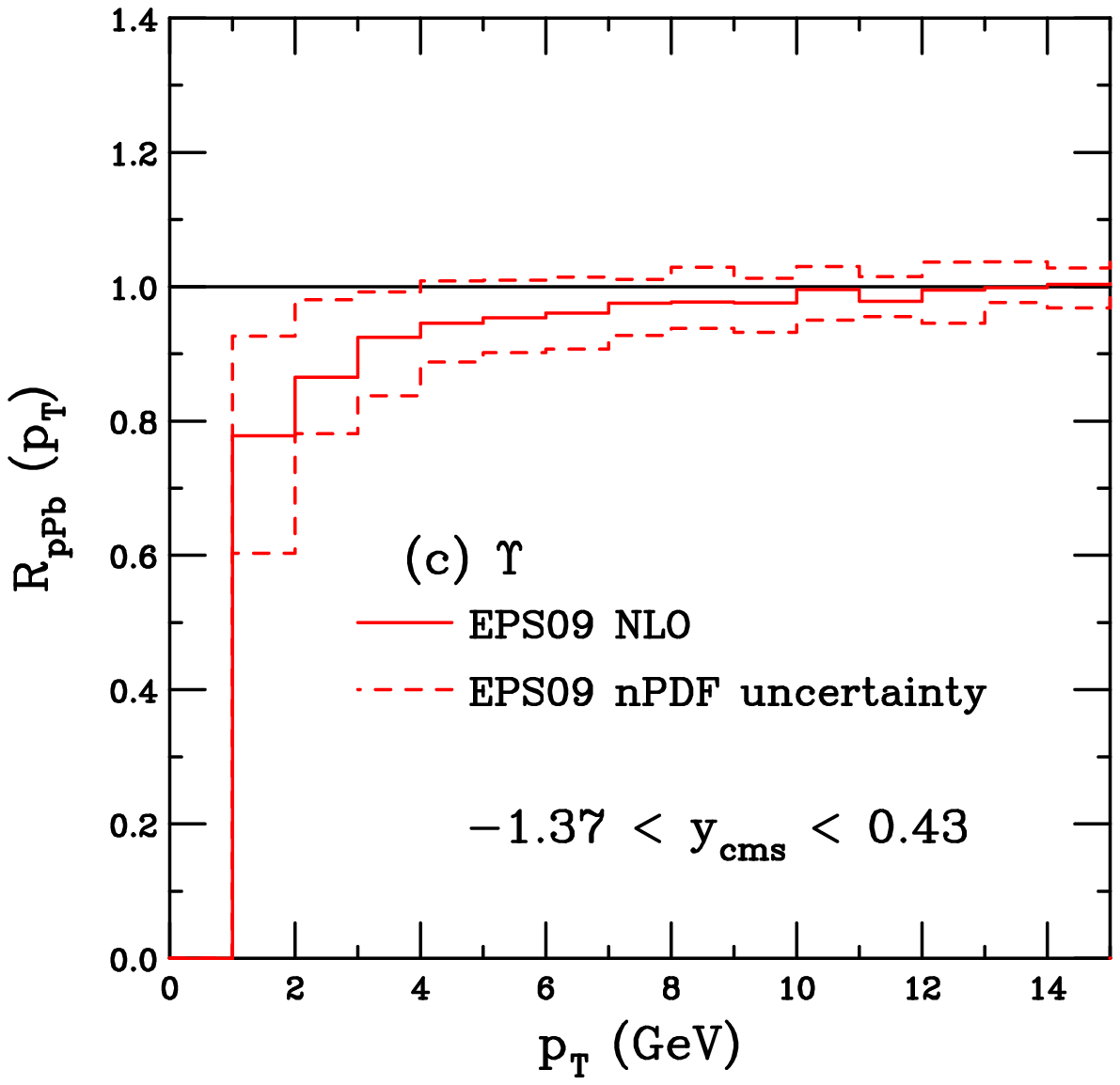}
\includegraphics[width=0.45\textwidth]{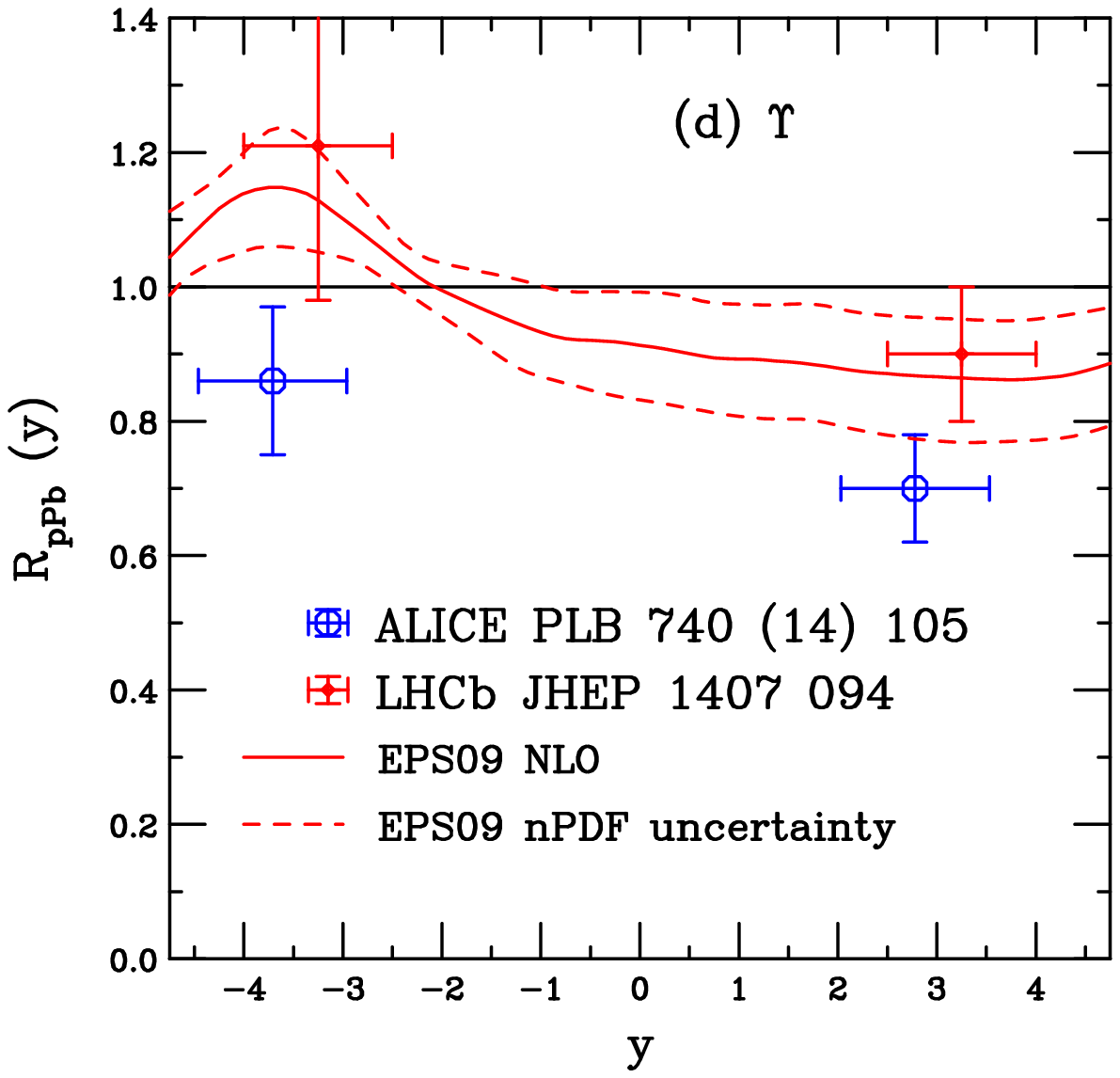}
\caption[]{(Color online)
The $\Upsilon$ ratio $R_{p{\rm Pb}}(p_T)$ in the
ALICE acceptance at forward (a),
backward (b) and central (c) rapidity.  The ratio 
$R_{p{\rm Pb}}(y)$ \protect\cite{ALICEpPbUps} is shown in (d).
The EPS09 NLO uncertainty band 
is shown.
}
\label{fig:EPS09RpPb_Ups}
\end{center}
\end{figure}

The results for the $\Upsilon$ suppression factor, $R_{p{\rm Pb}}$, are shown in 
Fig.~\ref{fig:EPS09RpPb_Ups}.  
In this case,
the data are insufficient to form $R_{p{\rm Pb}}(p_T)$ or $R_{FB}(p_T)$.  In
addition, no midrapidity $e^+ e^-$ $\Upsilon$ measurement has been reported.
It is also not possible to separate $R_{p{\rm Pb}}(y)$ into smaller bins due to
the low statistics.  Finally, the ALICE and LHCb $\Upsilon$ data, in the forward
and backward rapidity bins, while within
one standard deviation of each other, do not appear to be in very good agreement
with each other.  Therefore, it is more difficult to draw conclusions about
the $\Upsilon$ results.

The $p_T$-dependent ratios differ from those
for $J/\psi$ most at low $p_T$.  In particular, the backward rapidity region
shows an antishadowing effect already at low $p_T$.  At higher $p_T$ values, the
scales are similar since $p_T > m$ in both cases.  The EPS09 NLO 
uncertainty band
is somewhat narrower, especially as a function of rapidity, where the ALICE
and LHCb $\Upsilon$ data are both shown.  The agreement with the LHCb data
is quite good while the agreement with ALICE is rather poor.  The two data
sets are approximately within one standard deviation of each other but LHCb
is indicative of antishadowing at the approximate rapidity of the EPS09 NLO
antishadowing peak while the ALICE result is not.  
At forward rapidity, the ALICE
$J/\psi$ and $\Upsilon$ results for $R_{p {\rm Pb}}$ are similar.

\begin{figure}[t]
\begin{center}
%\vspace*{-0.05in}
\includegraphics[width=0.45\textwidth]{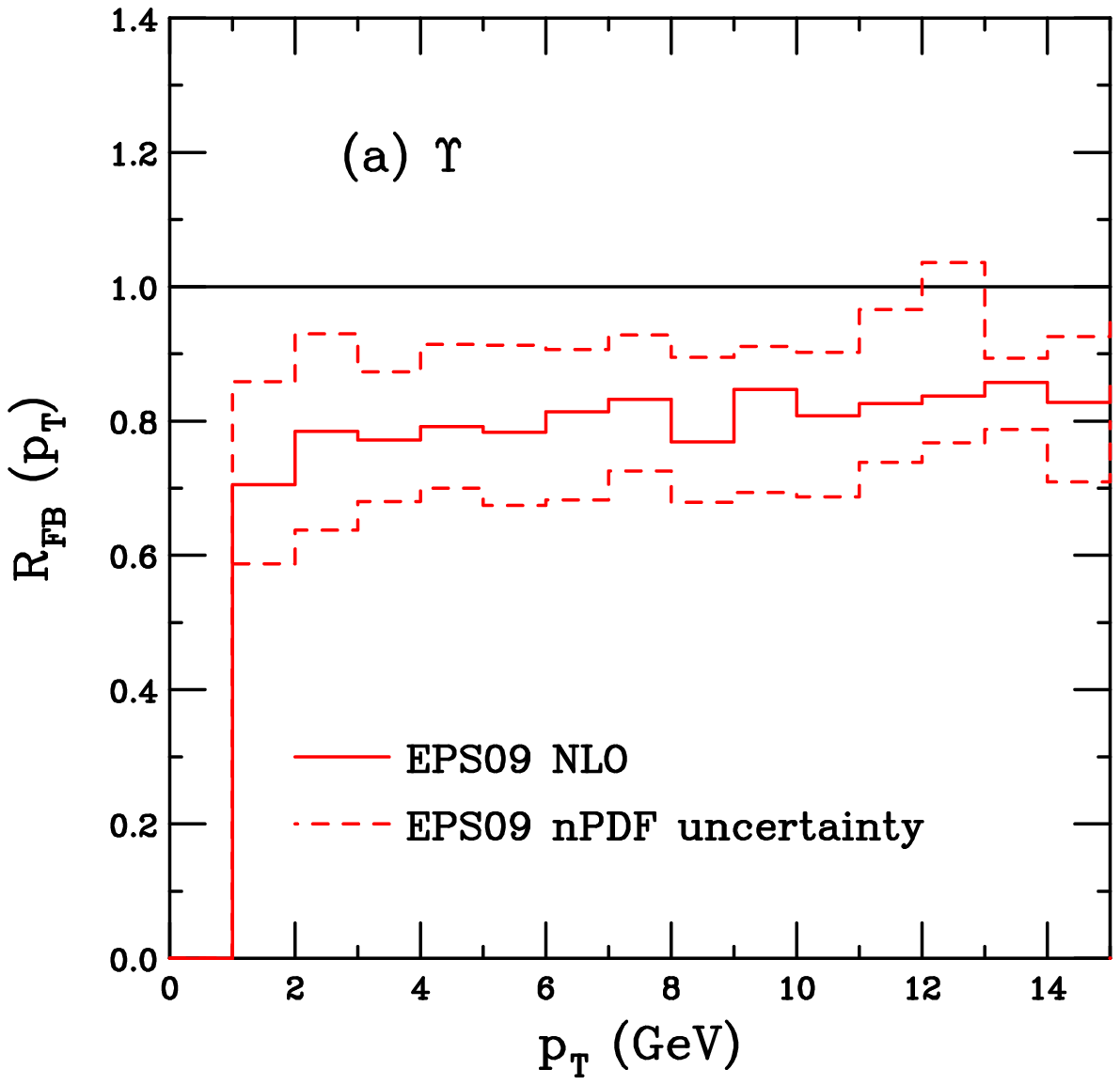}
\includegraphics[width=0.45\textwidth]{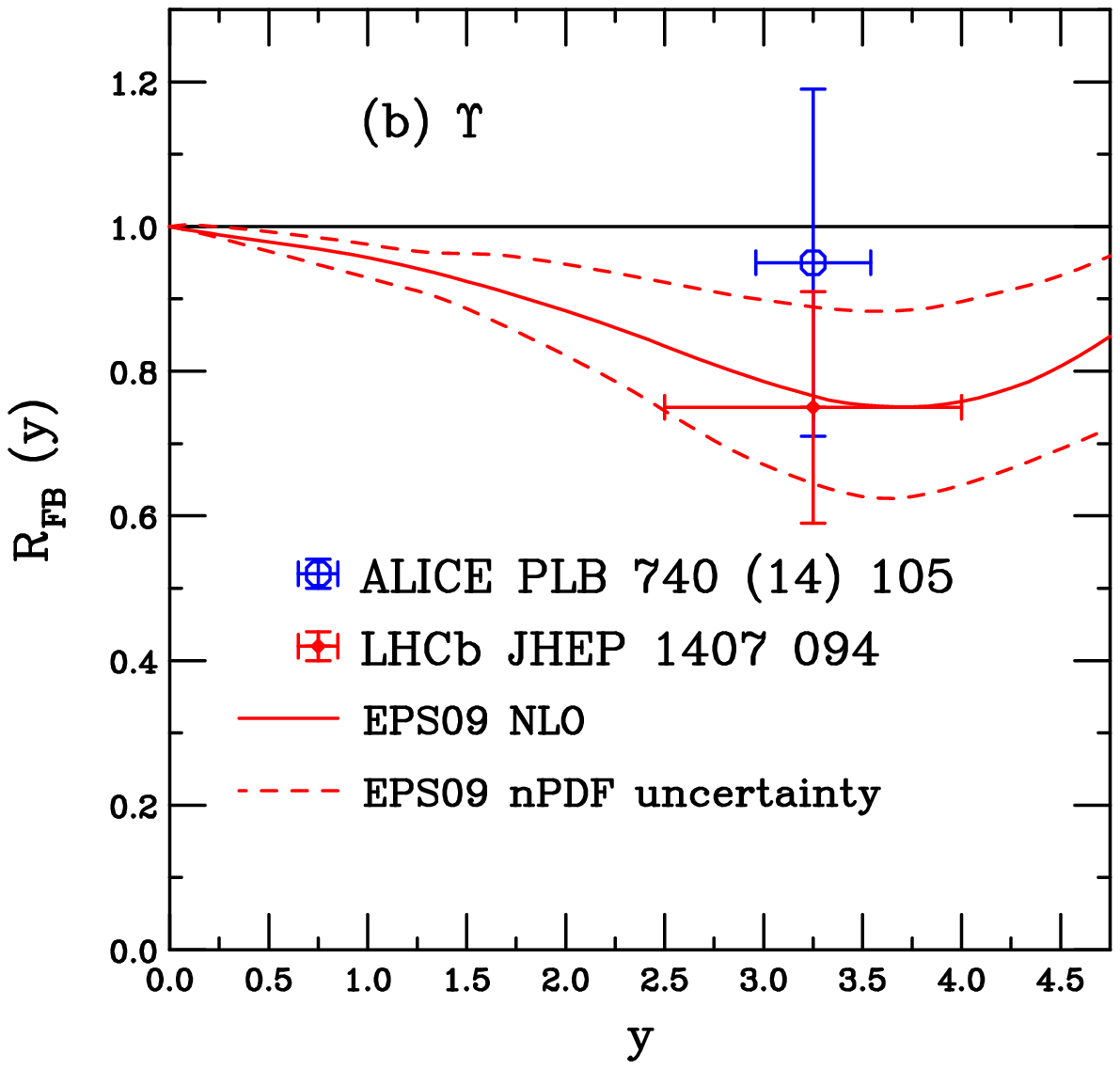}
\caption[]{(Color online)
The $\Upsilon$ forward-backward ratio $R_{FB}(p_T)$ in the ALICE overlap region 
(a) and $R_{FB}(y)$ (b).  The EPS09 NLO uncertainty band
is shown, along with the ALICE \protect\cite{ALICEpPbUps} and 
LHCb \protect\cite{LHCbpPbups} data.
}
\label{fig:EPS09RFB_Ups}
\end{center}
\end{figure}

The $\Upsilon$ forward-backward ratio is shown in Fig.~\ref{fig:EPS09RFB_Ups}.
The $p_T$-dependent ratio is lower than that of $J/\psi$ due to the greater
$\Upsilon$ antishadowing.  The uncertainty is larger because the 
uncertainty on the backward rapidity $J/\psi$ $R_{p{\rm Pb}}$ is narrower than
for $\Upsilon$ since the ALICE $J/\psi$ region is near the `pinch' between
the shadowing and antishadowing regions while the $\Upsilon$ rapidity is
directly in the region where the uncertainties are larger for gluons.  This
is also reflected by the rapidity dependence.

\subsection{Comparison of Leading and Next-to-Leading Order Results}
\label{SubSec:LOvsNLO}

Previously we reported that the shadowing parameterizations gave the same
results for the LO and NLO cross sections \cite{SQM04}.  That result was based,
however, on employing the LO EKS98 set in the LO and NLO CEM calculations.
In addition, other authors have suggested that the order of the calculation
does not matter and NLO nPDF sets can be used with LO quarkonium calculations.
It is worth checking these assumptions in detail.  

\begin{figure}[t]
\begin{center}
\includegraphics[width=0.425\textwidth]{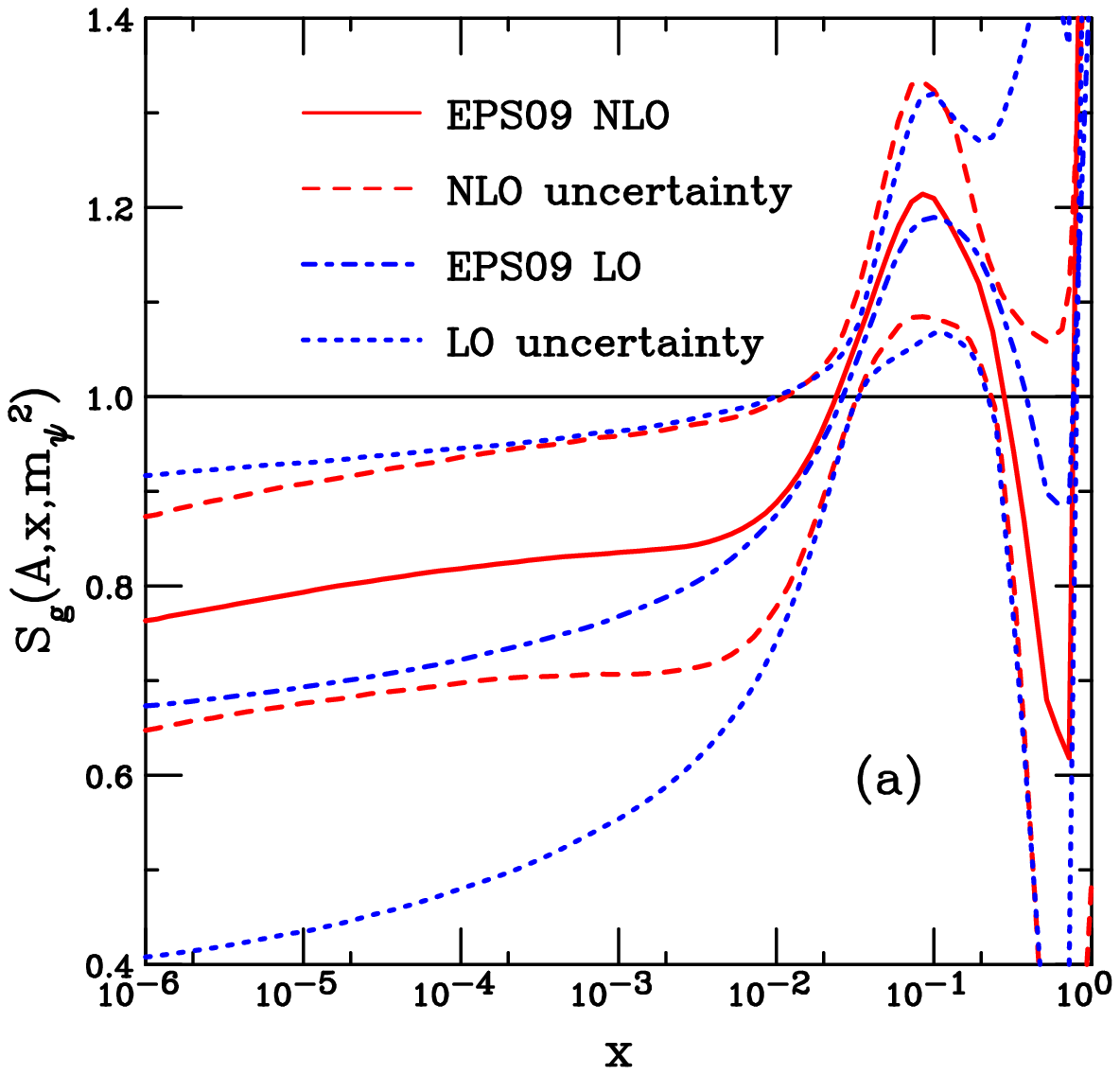}
\includegraphics[width=0.425\textwidth]{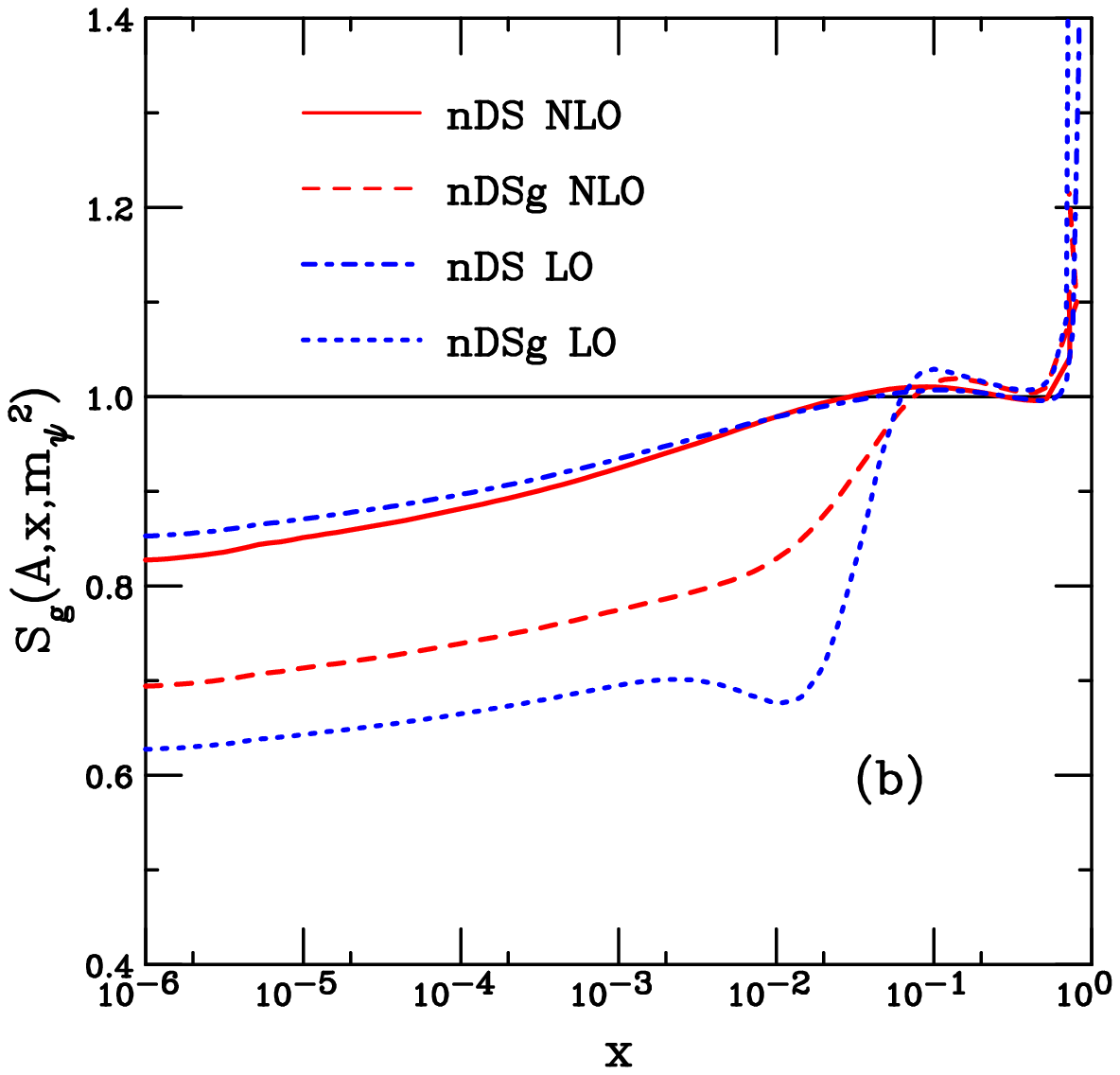} 
\\
\includegraphics[width=0.425\textwidth]{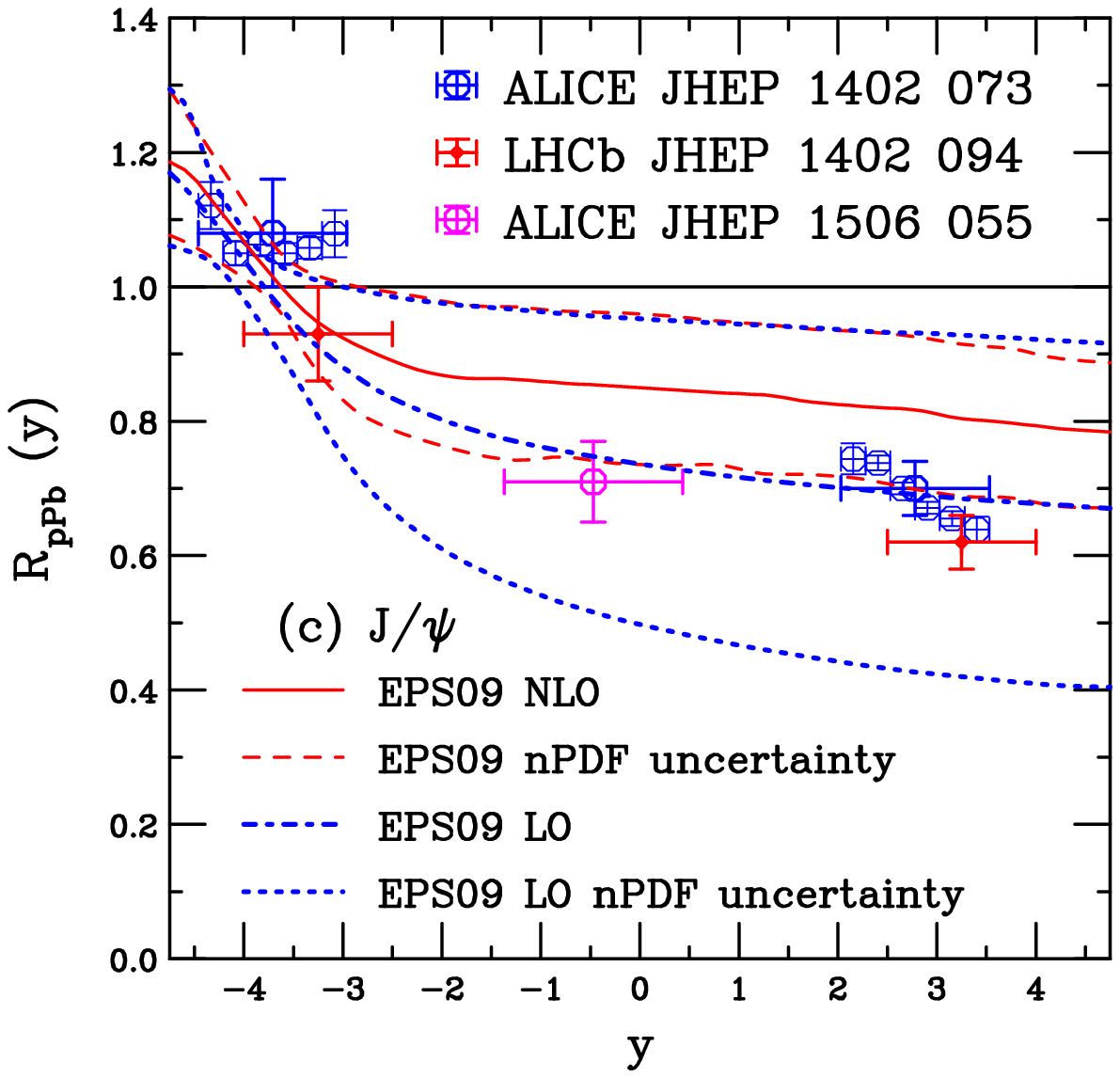} 
\includegraphics[width=0.425\textwidth]{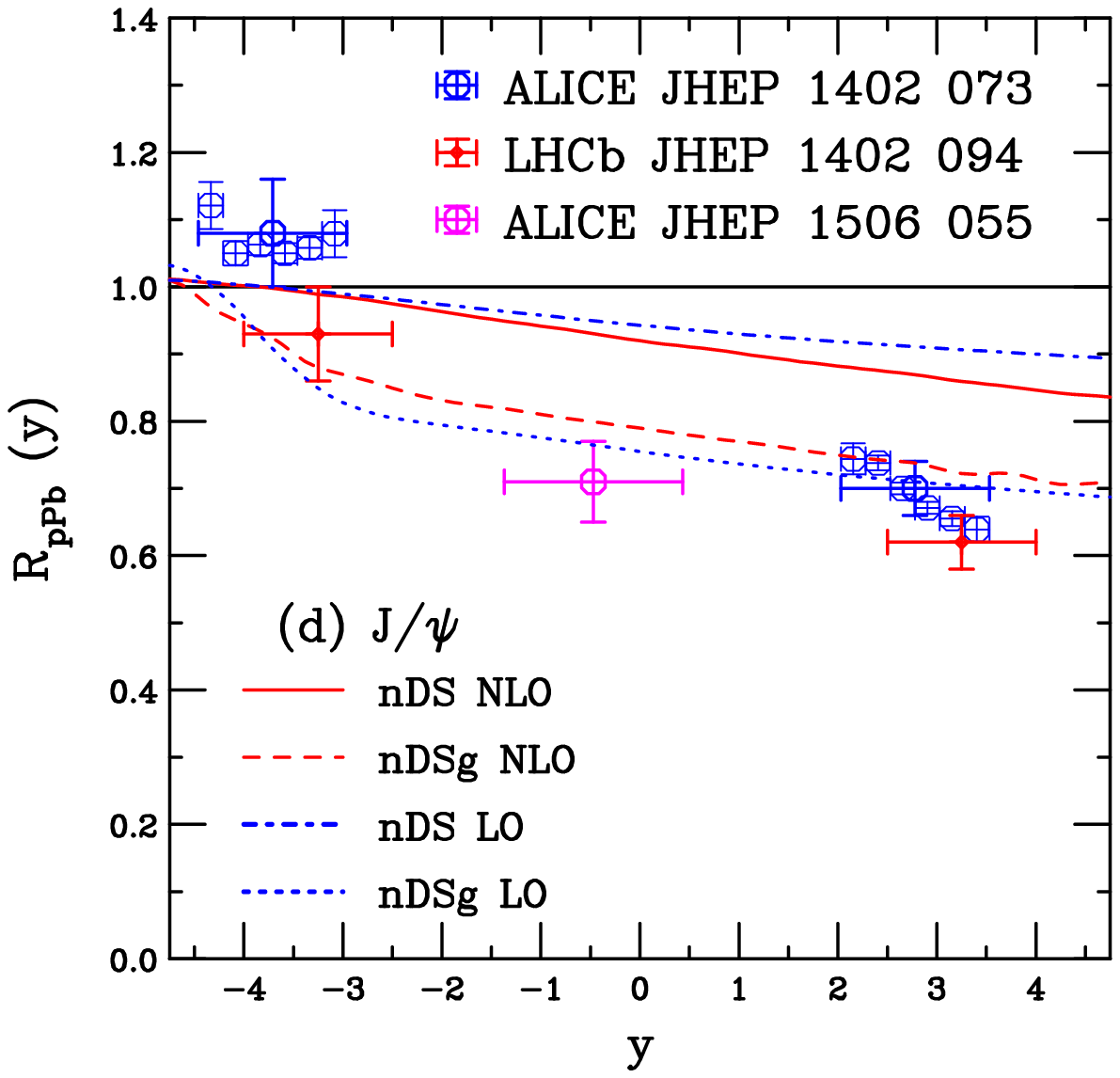}
\\
\includegraphics[width=0.425\textwidth]{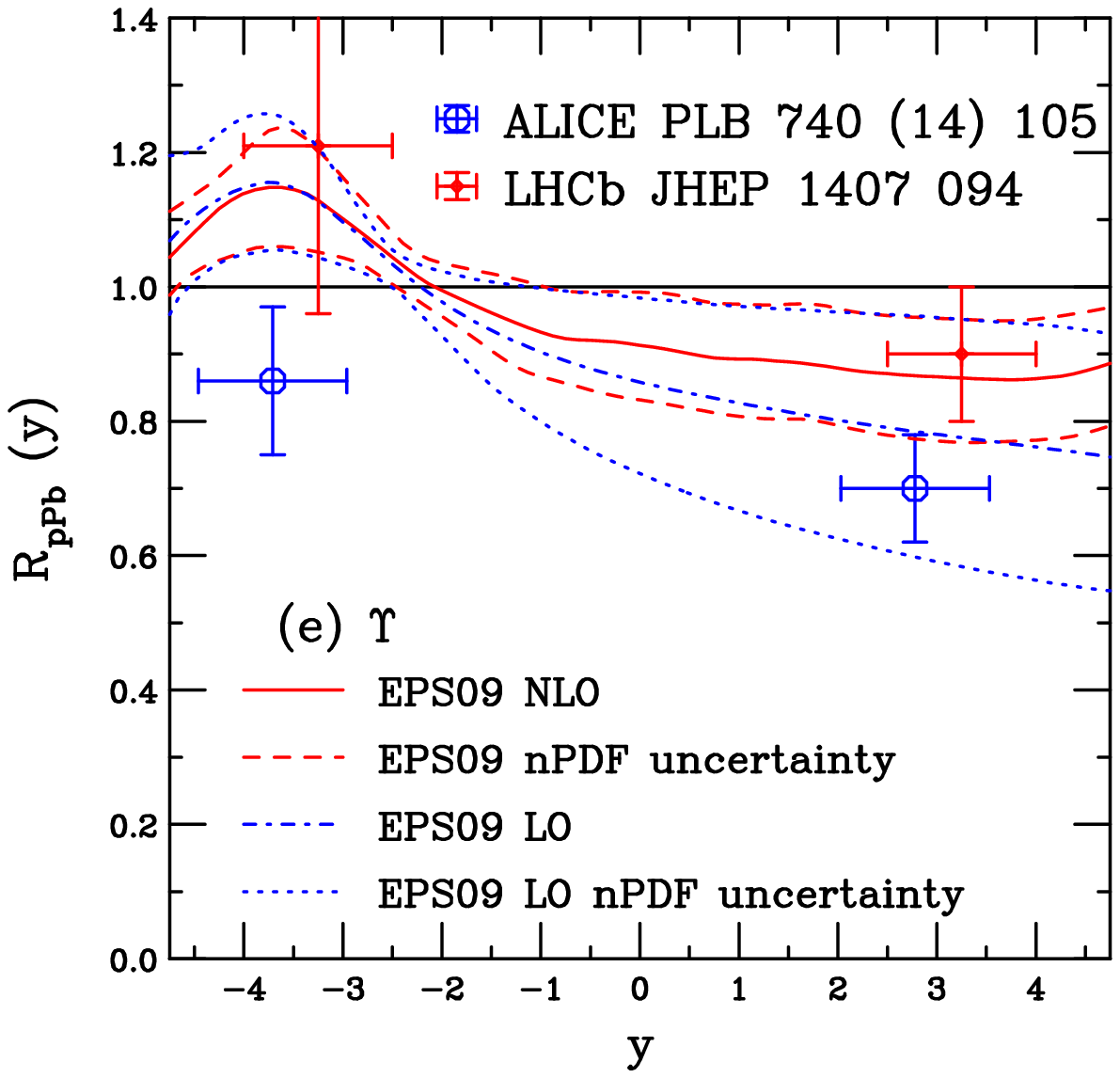} 
\includegraphics[width=0.425\textwidth]{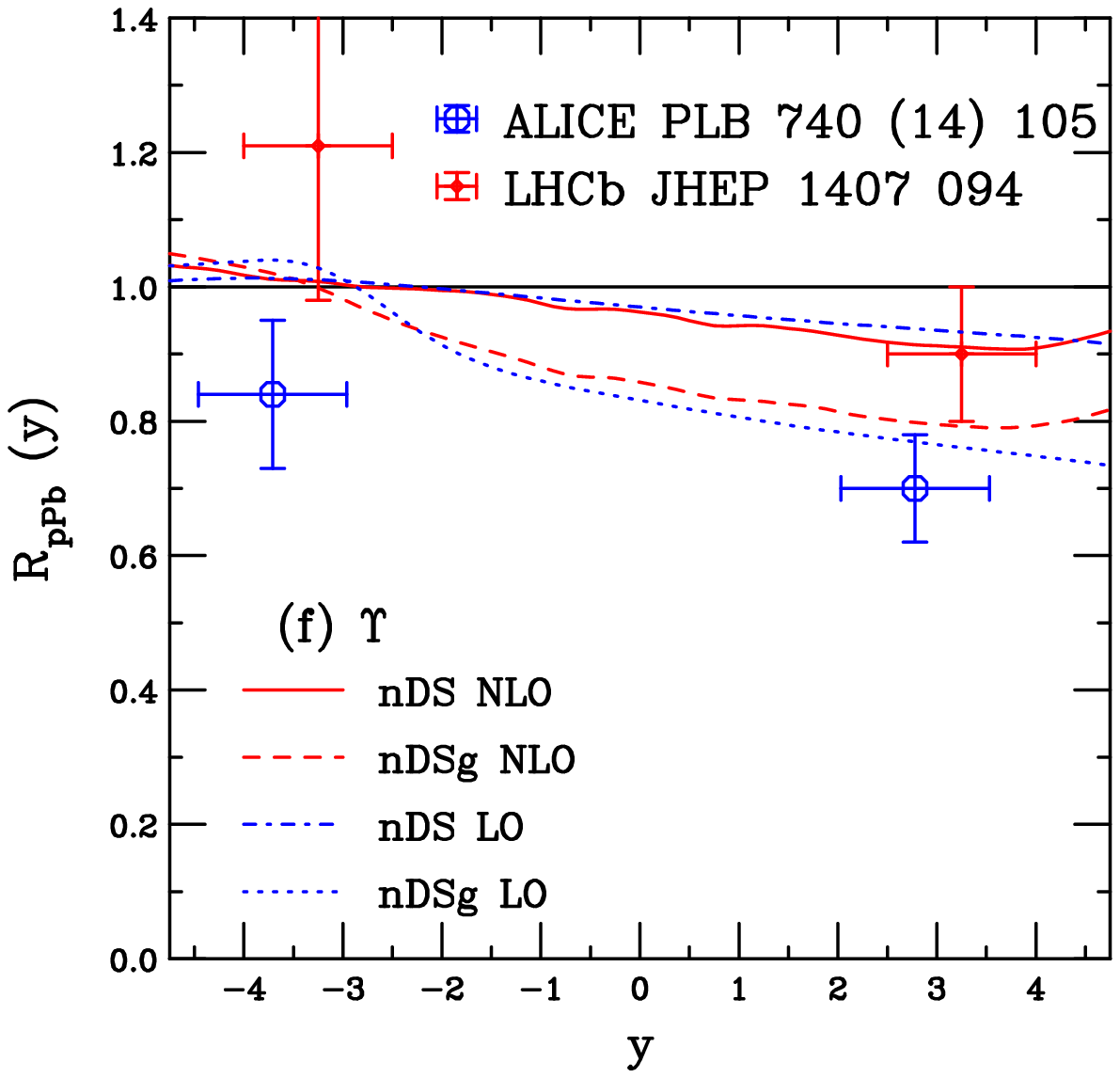} 
\end{center}
%\vspace*{-0.2in}
\caption[]{(Color online)
The EPS09 central results, as well as the uncertainty bands, are shown in (a),
(c) and (e).  In all cases, the solid
red curve shows the central NLO result while the dashed red curves delineate the
NLO uncertainty band.  The dot-dashed blue curve is the LO central result
while the dotted curves outline the LO uncertainty band. 
The nDS and nDSg results are presented in (b), (d) and (f).
The red solid curves show the NLO nDS result while the dashed red curves are the
nDSg results.  The blue dot-dashed and dotted curves show the nDS and nDSg LO
results.
(Top) The LO (blue) and NLO (red) gluon shadowing parameterizations. 
(Middle) The calculated $J/\psi$ $R_{p {\rm Pb}}(y)$ at 
$\sqrt{s_{_{NN}}} = 5$ TeV compared to the ALICE \protect\cite{ALICEpPbpsi} and 
LHCb \protect\cite{LHCbpPbpsi} data. 
(Bottom) The calculated $\Upsilon$ $R_{p {\rm Pb}}(y)$ at 
$\sqrt{s_{_{NN}}} = 5$ TeV compared to the ALICE \protect\cite{ALICEpPbUps} and 
LHCb \protect\cite{LHCbpPbups}.
}
\label{fig:LOvsNLO}
\end{figure}

As discussed in the previous section, the nDS and nDSg LO and NLO
sets were checked for
consistency for the PHENIX $\pi^0$ data \cite{deFlorian:xxx}.  We now do the
same for $J/\psi$ and $\Upsilon$ production as a function of rapidity for
EPS09 LO and NLO and for nDS and nDSg LO and NLO in Fig.~\ref{fig:LOvsNLO}.

The shadowing parameterizations are compared in  Fig.~\ref{fig:LOvsNLO} (a)
and (b).  There are considerable differences between the EPS09 bands at
$x < 0.01$.  The NLO result  has a different curvature than the LO one, changing
slope and becoming flatter, decreasing slowly as $x$ is lowered.  On the other
hand the LO bands decrease smoothly.  There is, however, very little difference
in the upper limit of the band, with the weakest shadowing effect.  
The difference is
more pronounced for the central sets and striking for the lower limit of the
band, with the strongest shadowing.  In this case, the lower NLO limit is
close to the central EPS09 LO set.  There is also some difference between the
nDS and nDSg parameterizations.  While the difference is essentially 
negligible for the nDS set, there is a stronger deviation between the nDSg
curves, particularly around $x \sim 0.01$.  For $x < 0.002$, the ratios are
parallel.

We now turn to whether these differences affect $R_{p {\rm Pb}}(y)$ for
quarkonium production.  The EPS09 LO calculation is made with a fully LO
CEM calculation in $2 \rightarrow 1$ kinematics.  The NLO calculation is with
the NLO matrix elements in the CEM.

Results are shown for 
in Fig.~\ref{fig:LOvsNLO} (c) and (d).  There is clearly a strong effect for
EPS09 with the central LO result passing through the ALICE and LHCb data while
only the lower limit of the NLO band reaches the data, an ap+proximate 15\%
difference between the two results.  However, the nDS and nDSg results, while
clearly not on top of each other, as they are for the $\pi^0$ calculation
\cite{deFlorian:xxx}, agree within $\sim 3$\%.  

Why is one derivation nearly 
consistent order by order and the other is not?  We suspect that the primary 
cause is the low $x$ behavior of the baseline proton parton densities, as 
already mentioned in Sec.~\ref{Sec:params}.  
The CTEQ6M, GRV98 LO and GRV98 NLO gluon distributions are
all based on valence-like initial distributions at $\mu_0$, $x^\alpha P(x)$.
However, the CTEQ61L gluon density takes an almost constant
finite value at $\mu_0^2$, {\it i.e.} $\alpha \sim 0$.
This initial difference carries forward over all $\mu^2$ 
because the NLO set must
have stronger $\mu^2$ evolution to fit the same sets of DIS data as at LO.
Thus the CTEQ6M distribution evolves much more rapidly in $\mu^2$ than does
CTEQ61L.  At the factorization scale of $\Upsilon$ production, the difference
between CTEQ6M and CTEQ61L is still substantial while the GRV98 LO and NLO
gluon distributions, with similar starting behavior, are closer together.
The EPS09 LO shadowing thus has to be stronger at low $x$ to produce the same
behavior at higher scales as the EPS09 NLO shadowing.  The compensation does
not have to be as large for the nDS sets since the GRV98 LO and NLO sets are
more similar at low $x$ and the scale relevant for quarkonium production.

The agreement between the EPS09 LO and NLO results is not improved for the
higher scale of $\Upsilon$ production, as shown in 
Fig.~\ref{fig:LOvsNLO} (e) and (f).  The higher scale, as
already noted, weakens the shadowing and narrows the uncertainty bands
but does not bring the central results at LO and NLO into better agreement.  
This conclusion is likely independent of the
specific final state as long as its production is dominated by low $x$ gluons.

There are, however, valid reasons for there to be a few percent difference
between the LO and NLO results seen with nDS and nDSg.  The LO CEM 
calculation is a $2 \rightarrow 1$ process with fixed values of $x_1$ and 
$x_2$ for a given $y$ since there is no $p_T$ scale in the calculation, only
in the factorization scale which can be adjusted to include an average value 
of $p_T$.  The NLO calculation, on the other hand, includes $2 \rightarrow 2$
and $2 \rightarrow 3$ processes (the LO+virtual and real contributions
respectively) to the NLO CEM result.  The NLO CEM calculation does not have
a fixed correspondence between $x_1$, $x_2$ and $y$ because of the $p_T$
scale.  This higher scale can also lead, on average, to a larger factorization
scale and a somewhat larger $x_2$ on average in the NLO calculation.  
All these differences can cumulatively lead to the $2-3$\%
effect between LO and NLO observed for the nDS sets.  Similar arguments can
also explain the differences between the LO CEM and LO CSM \cite{Lansberg}
results and are not directly attributable to the production mechanism per se.

\subsection{Comparison of Shadowing Parameterizations}
\label{SubSec:nPDFs_comp}

In this section, we compare the central EPS09 NLO results to those with
other shadowing parameterizations.

\subsubsection{$J/\psi$}

\begin{figure}[t]
\begin{center}
\includegraphics[width=0.45\textwidth]{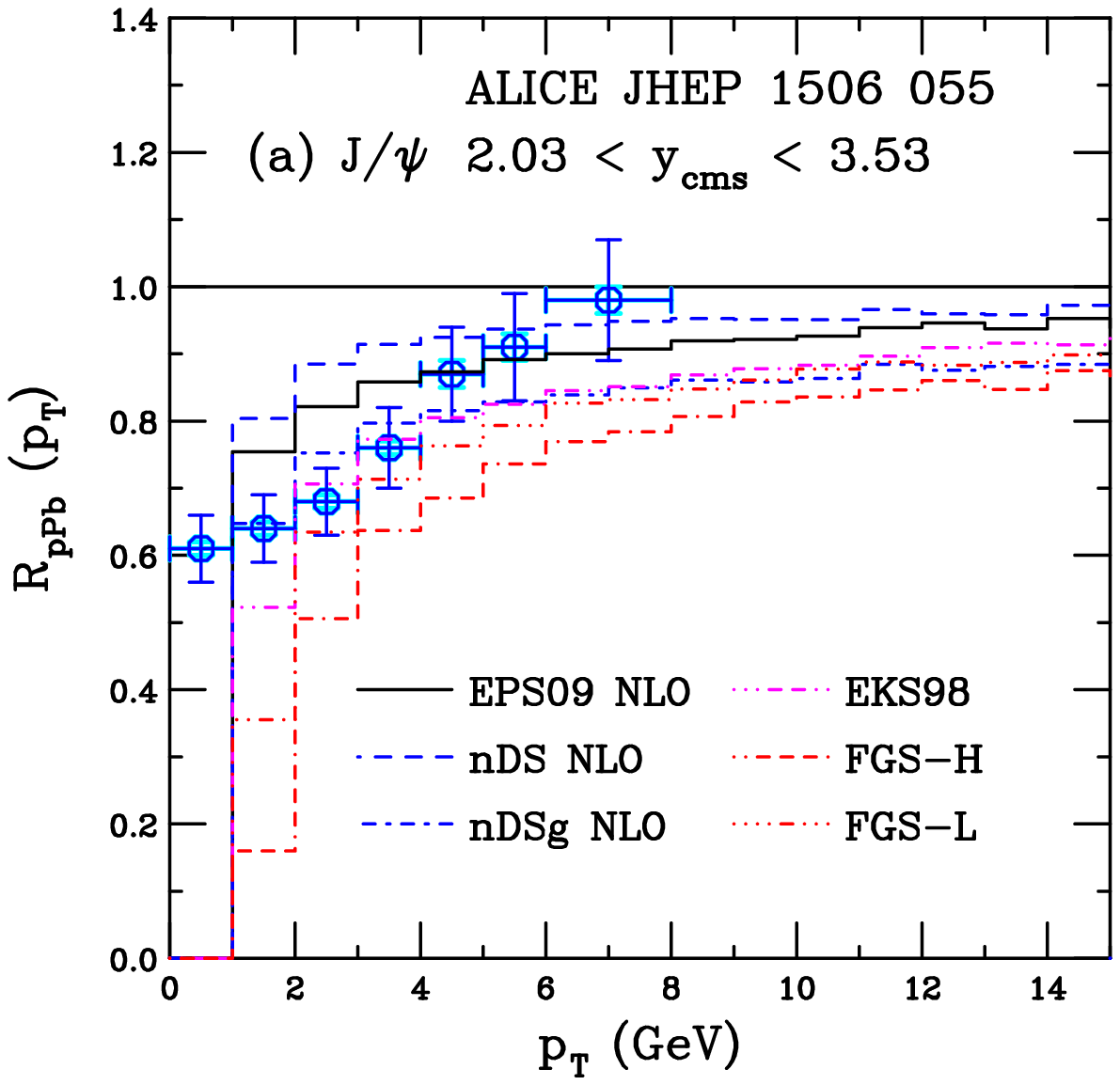}
\includegraphics[width=0.45\textwidth]{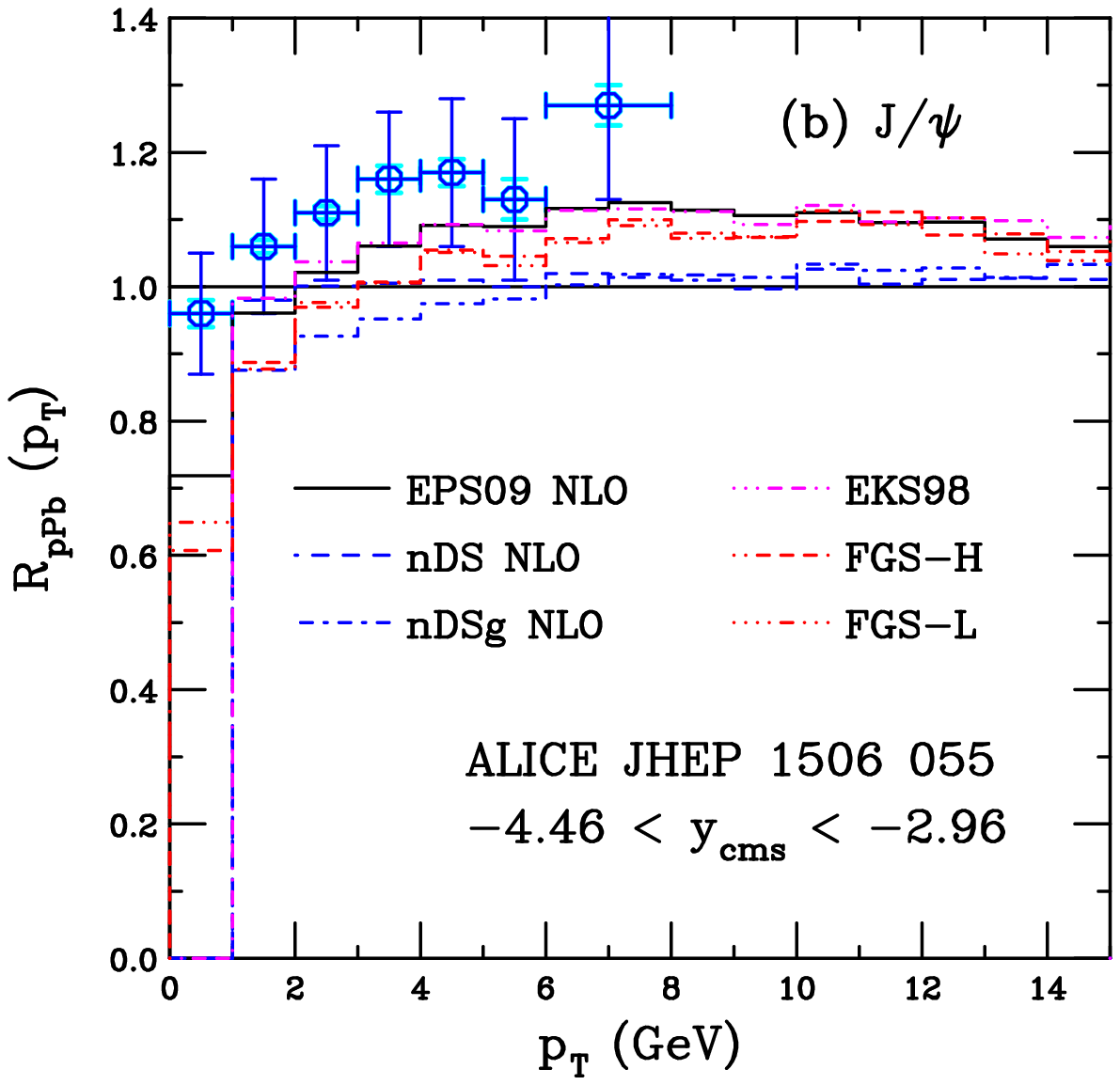} \\
\includegraphics[width=0.45\textwidth]{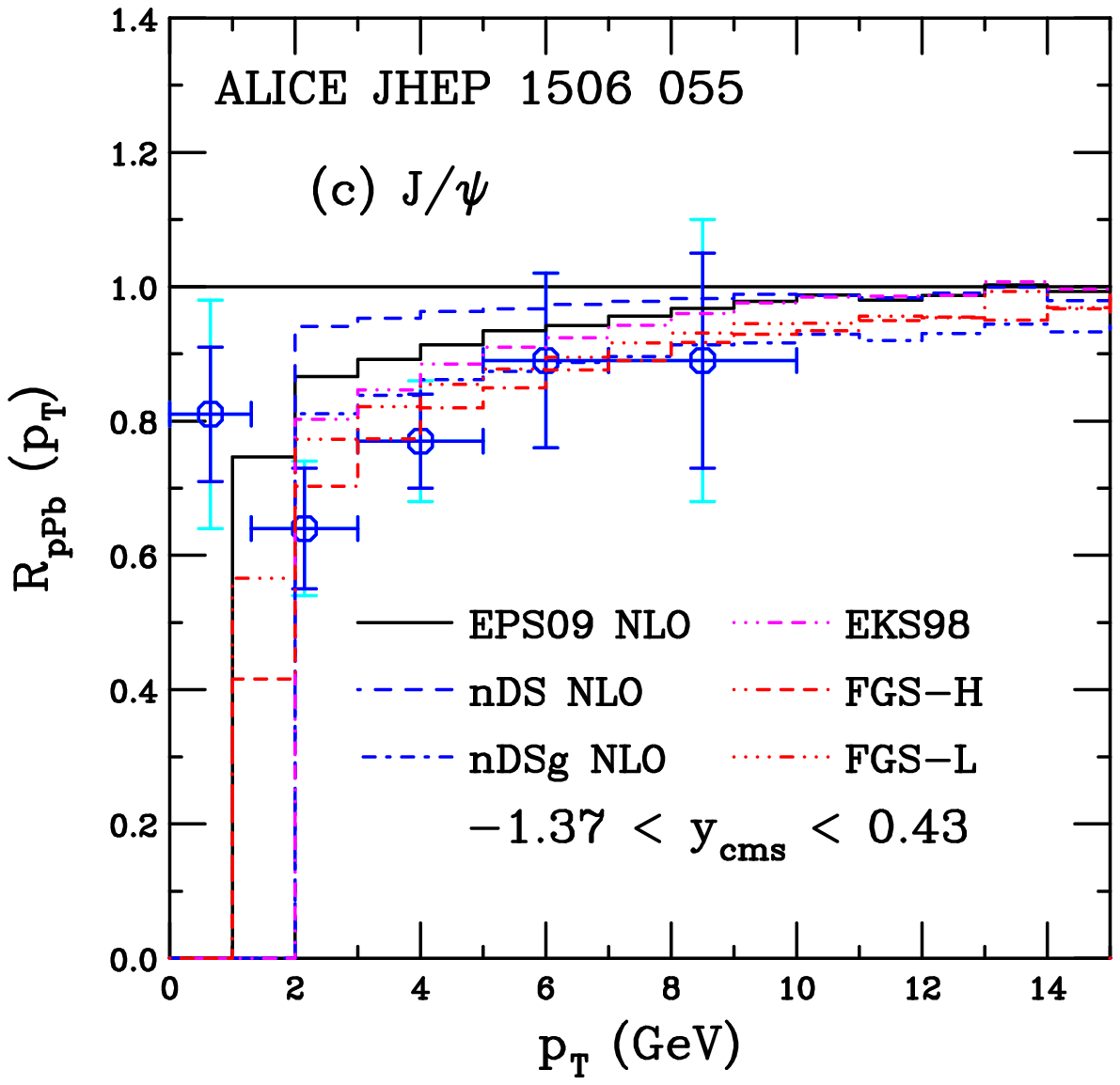}
\includegraphics[width=0.45\textwidth]{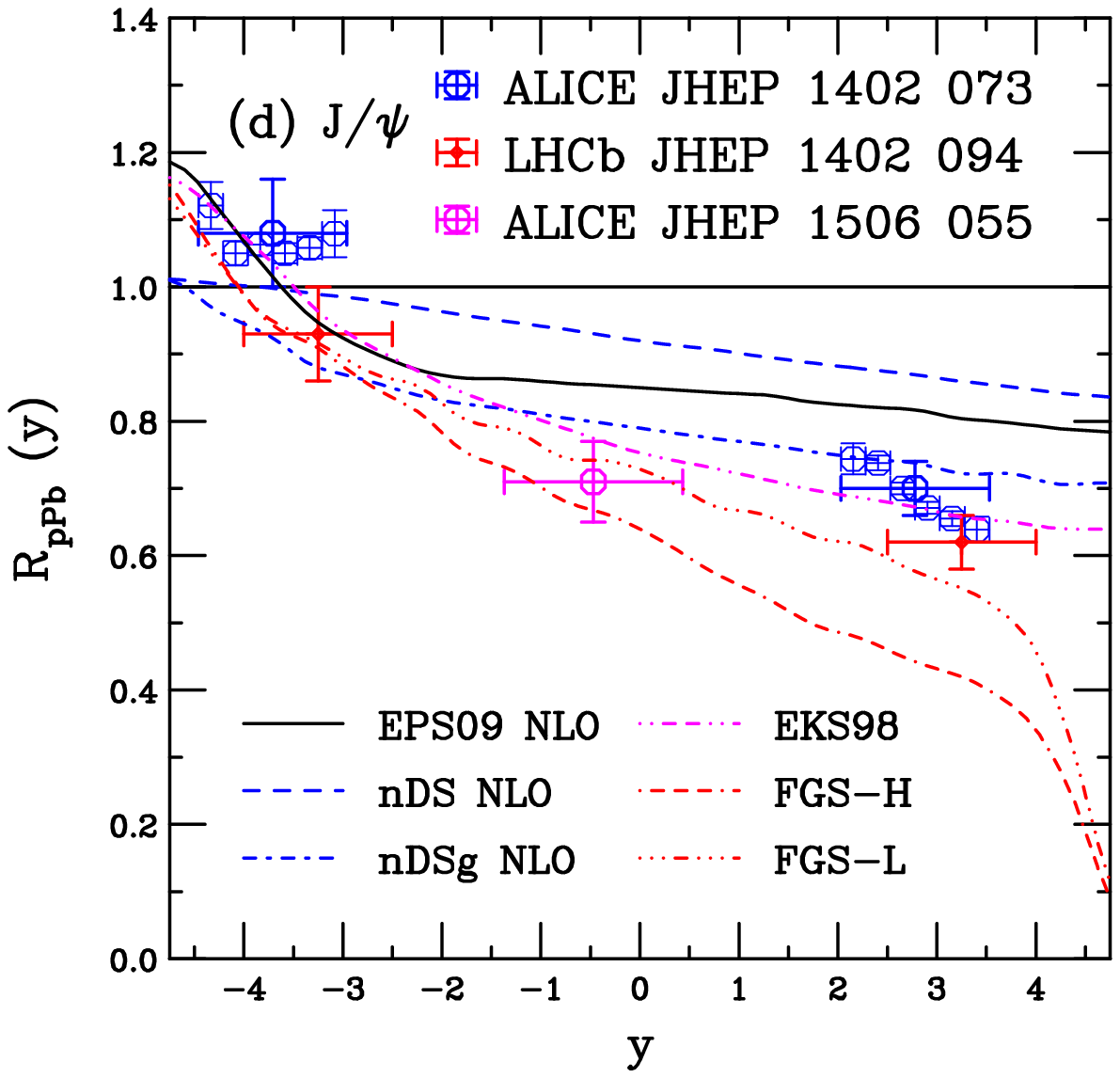}
\caption[]{(Color online)
The $J/\psi$ ratio $R_{p{\rm Pb}}(p_T)$ in the
ALICE acceptance at forward (a),
backward (b) and central (c) rapidity \protect\cite{ALICEpPbpsi_pT}.
The ratio $R_{p{\rm Pb}}(y)$ for ALICE \protect\cite{ALICEpPbpsi} and 
LHCb \protect\cite{LHCbpPbpsi} is shown in (d).
The results are shown for central 
EPS09 NLO (black),
nDS NLO (blue dashed), nDSg NLO (blue dot-dashed) and EKS98 LO (magenta
dot-dot-dash-dashed), FGS-H (red dot-dash-dash-dashed) and FGS-L (red
dot-dot-dot-dash).
}
\label{fig:nPDFsRpPb_Psi}
\end{center}
\end{figure}

\begin{figure}[t]
\begin{center}
%\vspace*{-0.05in}
\includegraphics[width=0.45\textwidth]{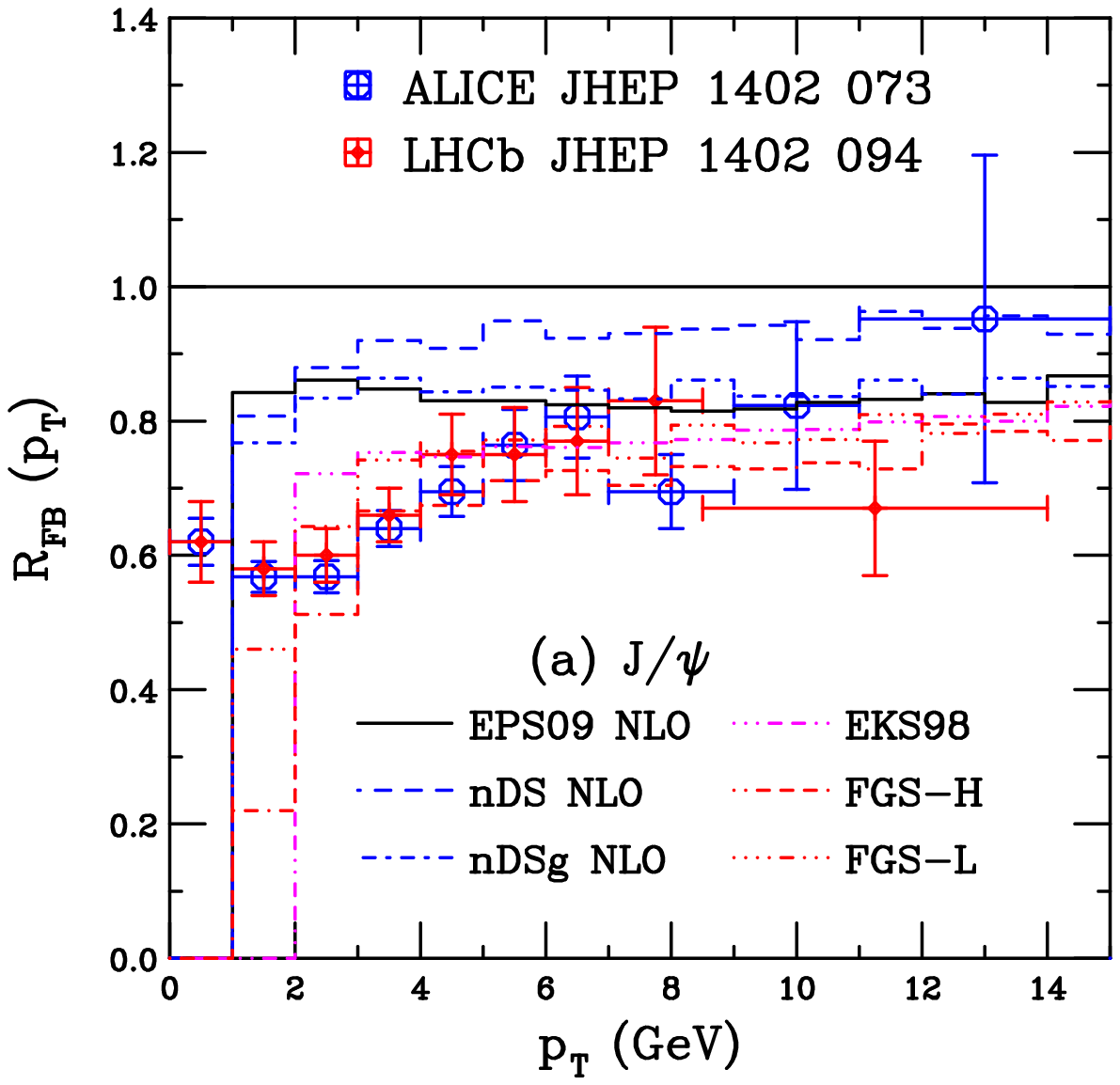}
\includegraphics[width=0.45\textwidth]{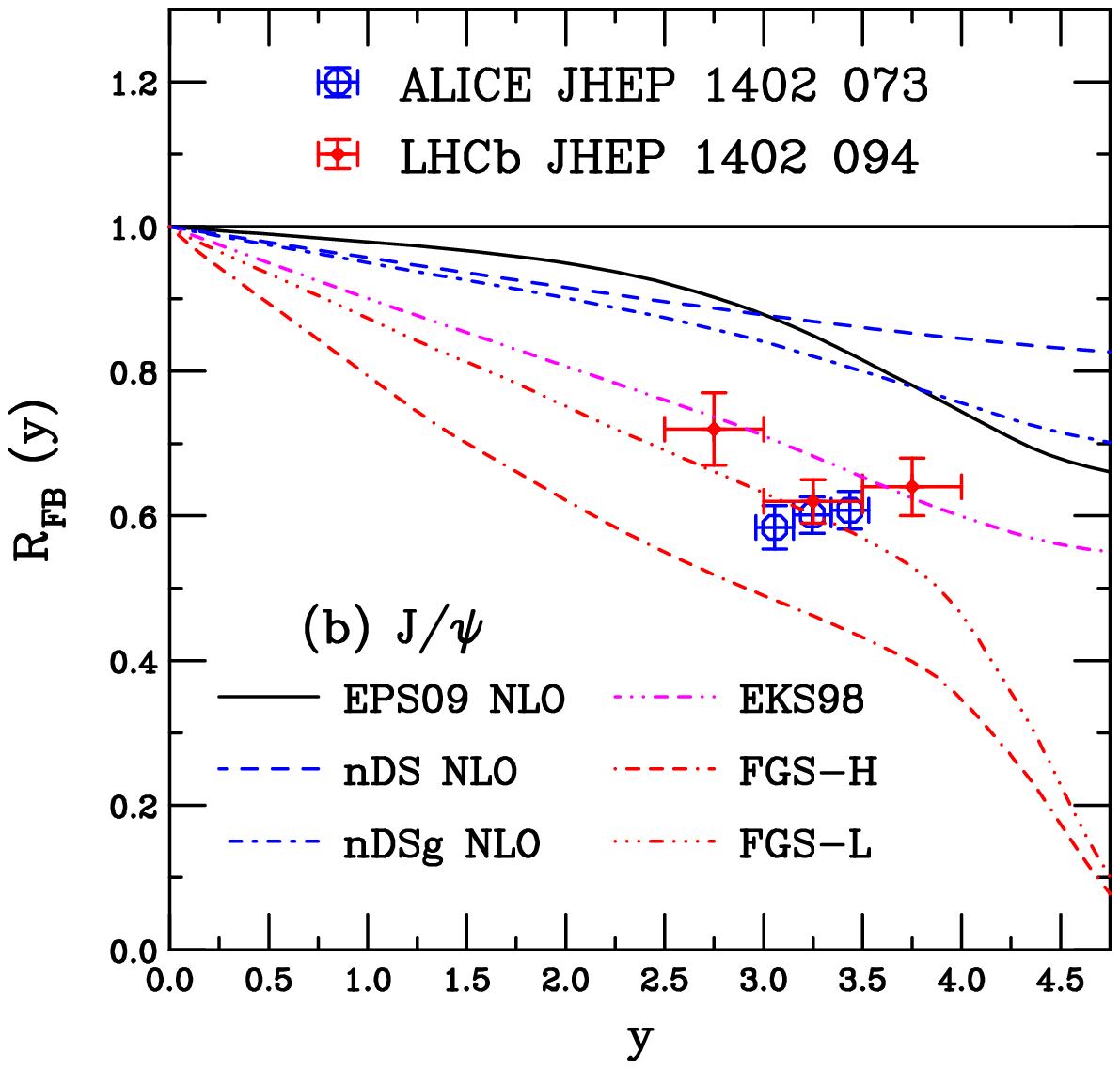}
\caption[]{
The $J/\psi$ forward-backward ratio $R_{FB}(p_T)$ in the ALICE overlap region 
(a) and $R_{FB}(y)$ (b) \protect\cite{ALICEpPbpsi}. The ratios are for central 
EPS09 NLO (black),
nDS NLO (blue dashed), nDSg NLO (blue dot-dashed) and EKS98 LO (magenta
dot-dot-dash-dashed), FGS-H (red dot-dash-dash-dashed) and FGS-L (red
dot-dot-dot-dash).
}
\label{fig:nPDFsRFB_Psi}
\end{center}
\end{figure}

Figures~\ref{fig:nPDFsRpPb_Psi} and \ref{fig:nPDFsRFB_Psi} compare the NLO
nDS, nDSg, FGS-H and FGS-L results to the central EPS09 NLO result (shown in 
black).  The results with the EKS98 LO set are also shown.  In the $p_T$
ratios, the EPS09 central set tends to underestimate the effect relative to the
data except at backward rapidity.  
Indeed, at forward and midrapidity, the nDS set is the only one that
underestimates the data more since it has the weakest gluon shadowing.  At
backward rapidity, the nDS result is close to unity over all $p_T$ while nDSg
exhibits some shadowing.  This is because these two sets have no gluon
antishadowing.  At forward rapidity, the FGS-H and FGS-L sets result in the
strongest effect, significantly stronger than the data in the case of FGS-H.
The set that comes closest to agreement with the data in all three rapidity
regions is EKS98 which only overestimates the shadowing effect for 
$p_T > 6$~GeV.

In all cases the $p_T$-dependent results for all the nPDF sets shown agree 
within 10\% for $p_T > 10$~GeV.  This can be expected because the evolution 
of the modifications is relatively strong for the gluons.  

As seen in 
Fig.~\ref{fig:nPDFsRpPb_Psi}(a), the spread between predictions is largest in
the forward region where the shadowing predictions differ most.  Indeed, in
the $p_T$ bin centered at 1.5~GeV, there is a factor of $\sim 8$ between
the values of $R_{p{\rm Pb}}$.  The gap is reduced to $\sim 1.3$
by $p_T \sim 3.5$~GeV.  The weakest $p_T$ dependence is given by the 
calculations with nDS and the central EPS09 NLO set.  The nDSg set also results
in a weak $p_T$ dependence but has a stronger overall shadowing effect.  
The strongest
shadowing comes from the FGS sets which overpredict the shadowing strength.
The best agreement with the ALICE data is obtained for the 
EKS98 LO set which has
no NLO counterpart.  We note that the measured $R_{p{\rm Pb}}$ approaches unity 
at $p_T \sim 7$~GeV while none of the shadowing parameterizations give a
result approaching unity over the entire $p_T$ range shown.

In the backward region, illustrated in Fig.~\ref{fig:nPDFsRpPb_Psi}(b), 
the results show some antishadowing, except for nDS and nDSg which specifically
exclude it.  All the calculations lie below the centroids of the data, partly
because the ALICE acceptance is centered on the side of the antishadowing peak
so that the calculated antishadowing is, on average, lower than that of the 
data.  Note also that the antishadowing peak is broad in $p_T$, with
some antishadowing remaining at $p_T \sim 15$ GeV.

The $p_T$-dependent results from the nPDF sets come closest 
together in the midrapidity bin, with a spread of at most 30\%, 
as seen in Fig.~\ref{fig:nPDFsRpPb_Psi}(c).
Although the data again suggest stronger shadowing at midrapidity than the
calculations, the uncertainties in the data are large since the number of
$J/\psi$ in this bin are lower by more than a factor of 100 than in the forward
and backward bins, see Table~\ref{ALICE-LHCb-yields}.  
In this rapidity range, the
shadowing effect is reduced to less than 10\% for $p_T > 10$ GeV with all sets.

The largest difference in the $R_{p{\rm Pb}}$ ratios is as a function of rapidity,
shown in Fig.~\ref{fig:nPDFsRpPb_Psi}(d), since the $p_T$-integrated ratios
access the lowest $x$ values.  The ratios are closest at backward rapidity,
not surprisingly, because the EPS09 NLO and EKS98 sets follow each other closely
for $x > 0.002$, see Fig.~\ref{fig:nPDFs}(a).  Since the FGS sets are 
somewhat narrower in the antishadowing $x$ region, this is manifested as an
apparent shift toward negative rapidity in the antishadowing peak for the 
$J/\psi$.  The EPS09 NLO and EKS98 are in closest agreement with the 
integrated ALICE rapidity bin which is larger than unity.  The ALICE data
were split into six smaller bins in both the forward and backward rapidity bins.
In the backward region, the smaller bins are almost independent of rapidity
which is not consistent with any of the shadowing results.  
Since the LHCb backward rapidity result is below unity, only the nDS result is
not in agreement with LHCb.  

The ratios separate 
further in the shadowing regions at mid and forward rapidity.
Since the ALICE $p_T$-integrated ratio at midrapidity is almost the same as
that at forward rapidity, albeit with somewhat larger uncertainty, the 
midrapidity point is only in agreement with the FGS-L and FGS-H sets.  This is
not surprising: because of their steep drop at low $x$, they show stronger
shadowing than EKS98 already at $x\sim 0.001$ and considerably overestimate 
the effect at forward rapidity where $x < 10^{-5}$.  
In the forward bin, the EKS98 and nDSg
results are in best agreement with the ALICE and LHCb results.  The ALICE
forward bin, when split into smaller bins, shows a linear dependence on 
rapidity but with a slope steeper than any of the shadowing calculations.
Overall the EKS98 parameterization, one of the oldest and LO only, agrees
best with the data.  Since it is similar in shape to the EPS09 LO central set,
shown in Fig.~\ref{fig:LOvsNLO}(a), this is not surprising.  The EPS09 LO
and EKS98 sets, based on CTEQ61L and CTEQ4L respectively, are the only sets
of those from EPS09 and nDS that are not based on valence-like gluon 
distributions in the proton and are thus subject to slower scale evolution.  

In Fig.~\ref{fig:nPDFsRFB_Psi}, the forward-backward ratios are shown as 
functions of $p_T$ and $y$.  The EPS09, nDS and nDSg ratios are nearly 
independent of $p_T$, thus above the data over most of the $p_T$ range.
The EKS98 and FGS sets have a somewhat stronger dependence on $p_T$ and thus
agree more closely with the forward-backward ratio at low $p_T$.  However, of
these, only the FGS-L and EKS98 sets are in agreement with the 
rapidity-dependent ratio.  All the sets, with the exception of the central
EPS09 NLO set, give forward-backward ratios that are linear in $y$.  As
already discussed this is due to the abrupt change in slope of EPS09 NLO at 
$x \sim 0.002$, see Fig.~\ref{fig:nPDFs}(a), at the transition from shadowing 
to antishadowing, absent from the other sets shown.  

Overall, the EKS98 LO set seems to agree best with the $J/\psi$ data, both 
$R_{p{\rm Pb}}$ and $R_{FB}$.  This
is somewhat less than satisfying since we are using a LO nPDF set with a NLO
cross section calculation.  In addition, it is one of the older sets employed
here.  However, it agrees rather well with EPS09 LO when calculated with LO
matrix elements,  even though the baseline
proton PDF and its behavior at low $x$ is very different.  We had previously
demonstrated that the ratio $R_{pA}(y)$, 
calculated at LO and NLO with the EKS98 
LO set, gave almost identical results at each order \cite{SQM04}.  This is not a
surprise since the $gg$ contribution dominates production both at LO and NLO
and the shift in $x$ and $\mu^2$ due to the different scales is small when 
integrated over $p_T$.  

\subsubsection{$\Upsilon$}

\begin{figure}[t]
\begin{center}
\includegraphics[width=0.45\textwidth]{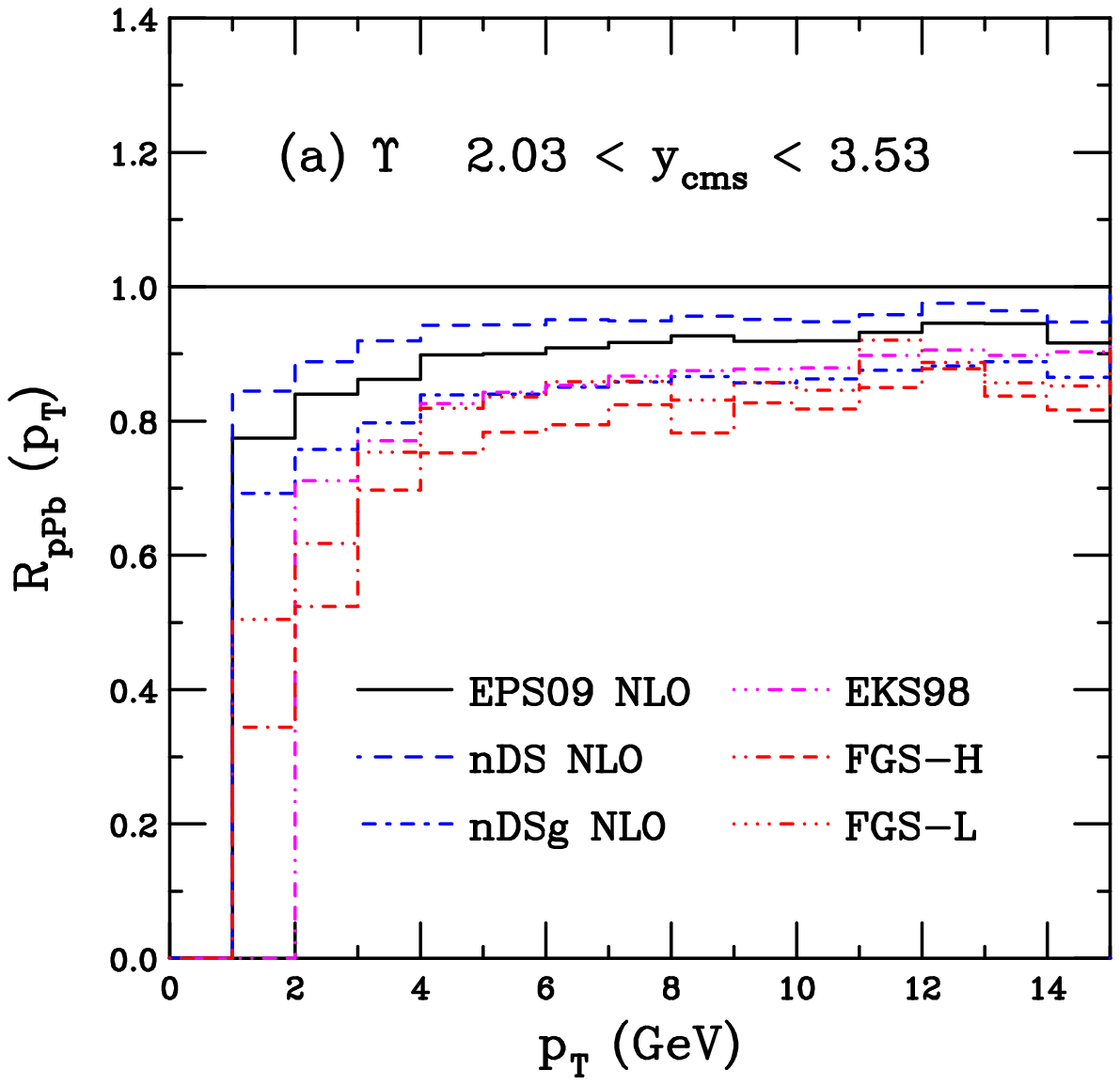}
\includegraphics[width=0.45\textwidth]{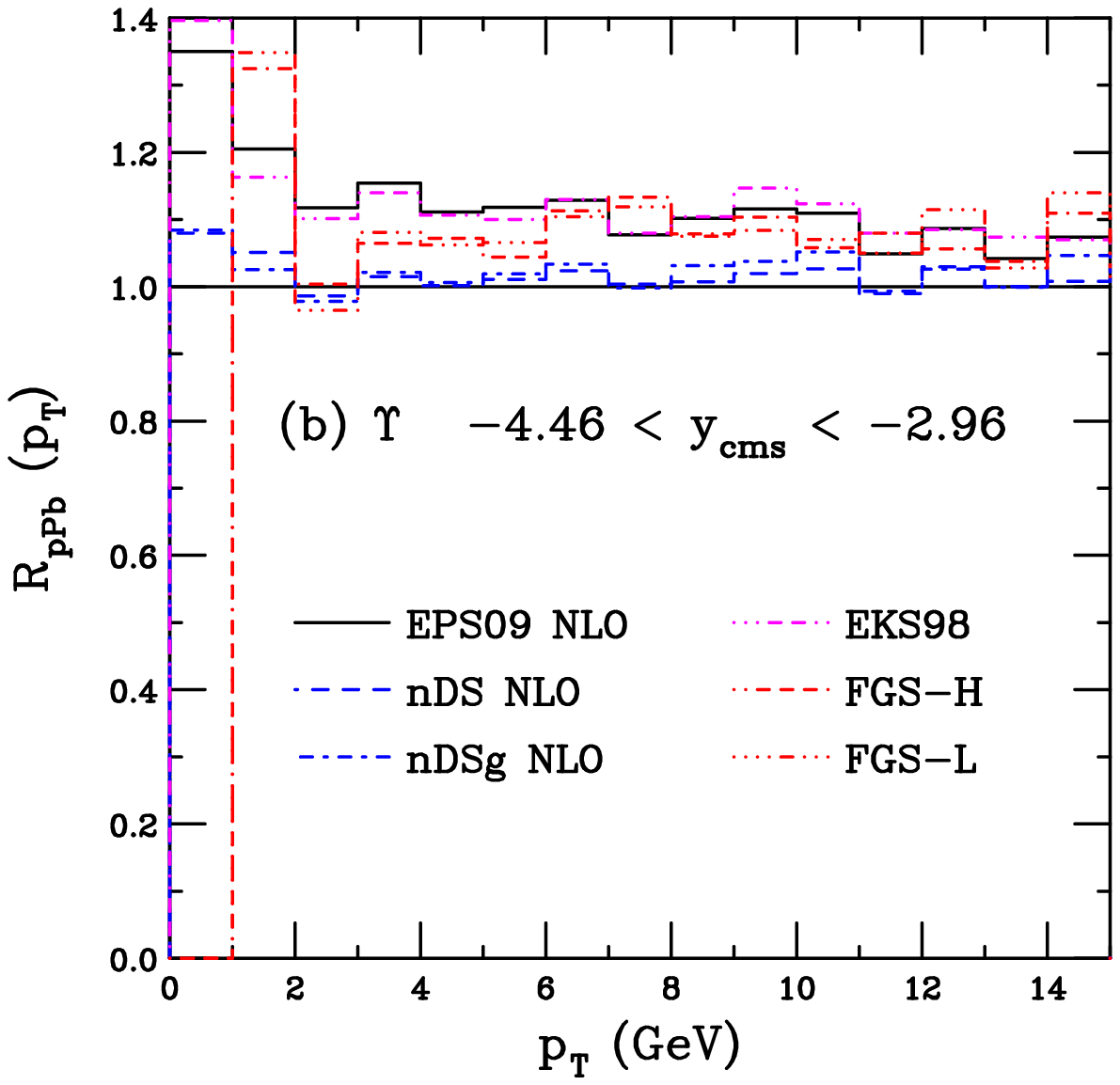} \\
\includegraphics[width=0.45\textwidth]{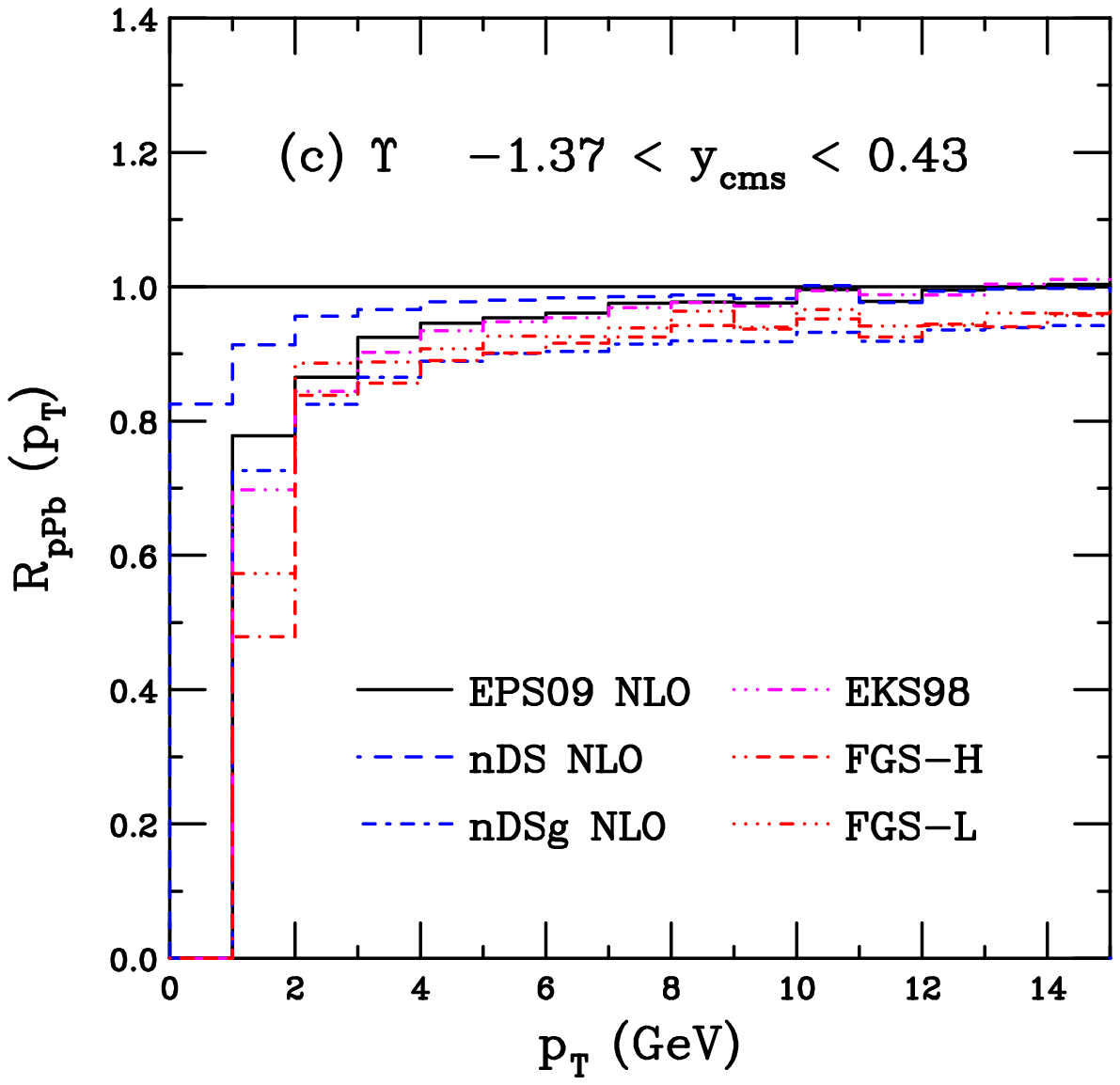}
\includegraphics[width=0.45\textwidth]{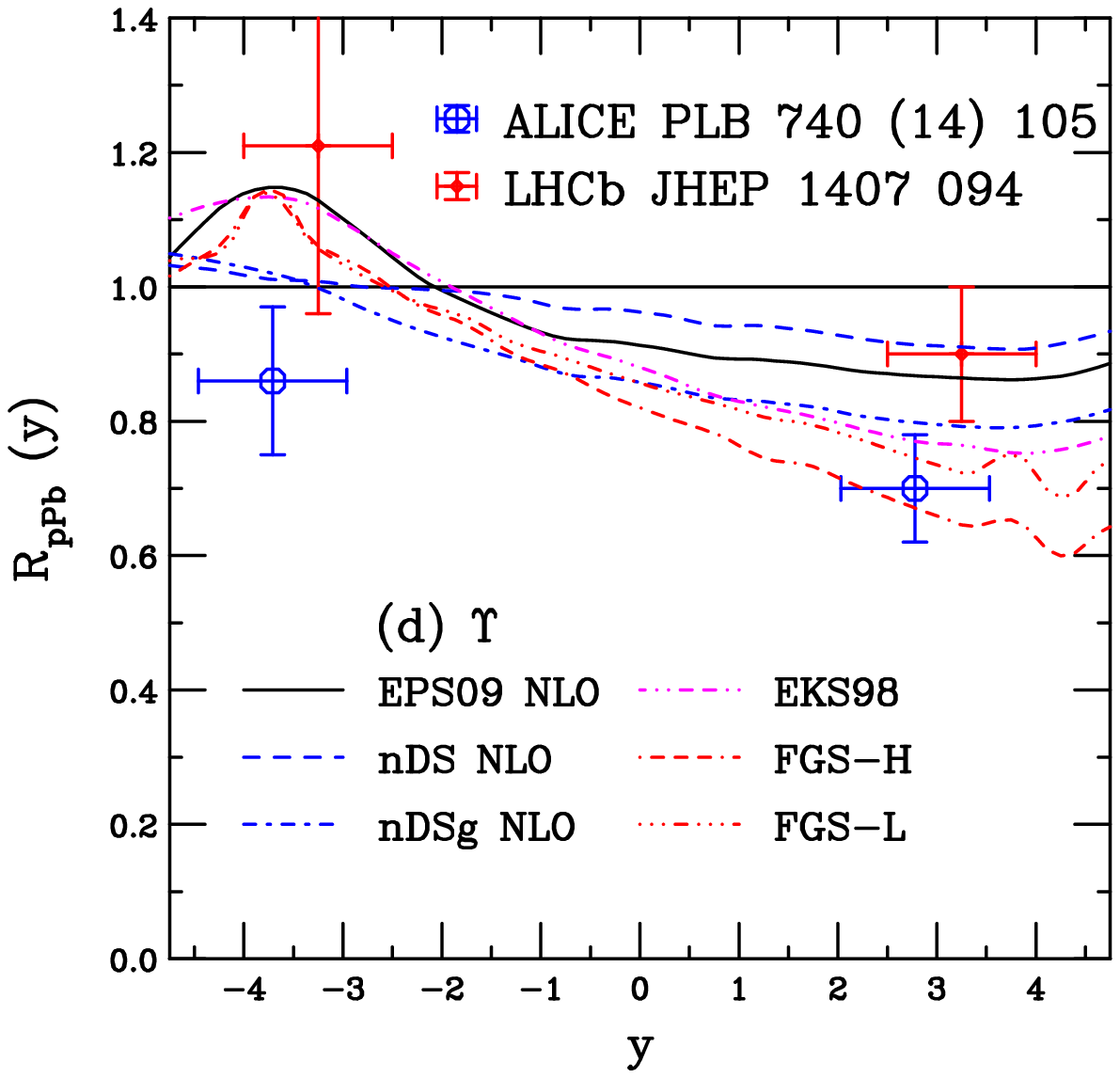}
\caption[]{(Color online)
The $\Upsilon$ ratio $R_{p{\rm Pb}}(p_T)$ in the
ALICE acceptance at forward (a),
backward (b) and central (c) rapidity.  The ratio 
$R_{p{\rm Pb}}(y)$ is compared to the ALICE \protect\cite{ALICEpPbUps} and
LHCb \protect\cite{LHCbpPbups} data in (d).  The ratios are for central 
EPS09 NLO (black),
nDS NLO (blue dashed), nDSg NLO (blue dot-dashed) and EKS98 LO (magenta
dot-dot-dash-dashed), FGS-H (red dot-dash-dash-dashed) and FGS-L (red
dot-dot-dot-dash).
}
\label{fig:nPDFsRpPb_Ups}
\end{center}
\end{figure}

\begin{figure}[t]
\begin{center}
%\vspace*{-0.05in}
\includegraphics[width=0.45\textwidth]{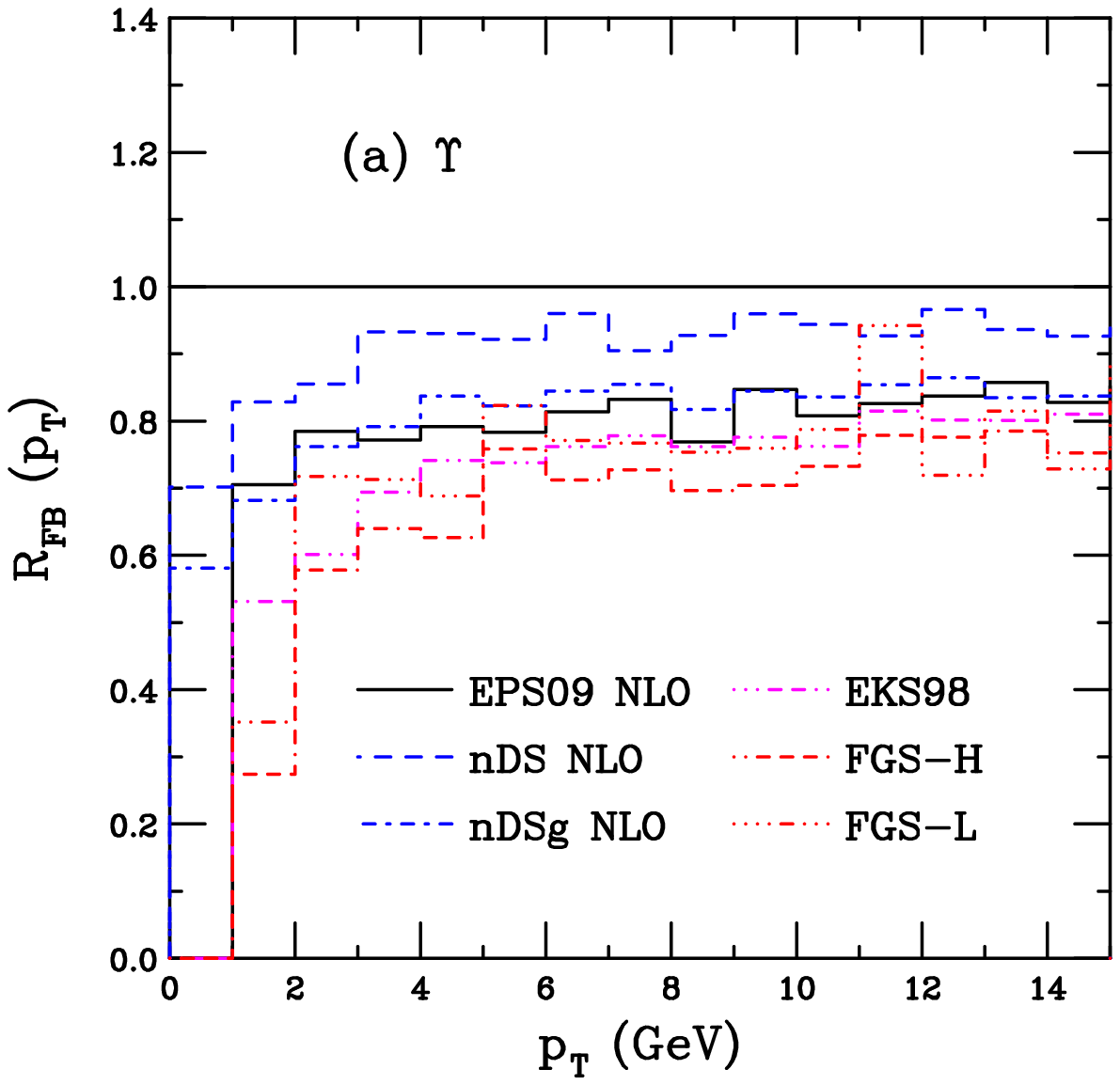}
\includegraphics[width=0.45\textwidth]{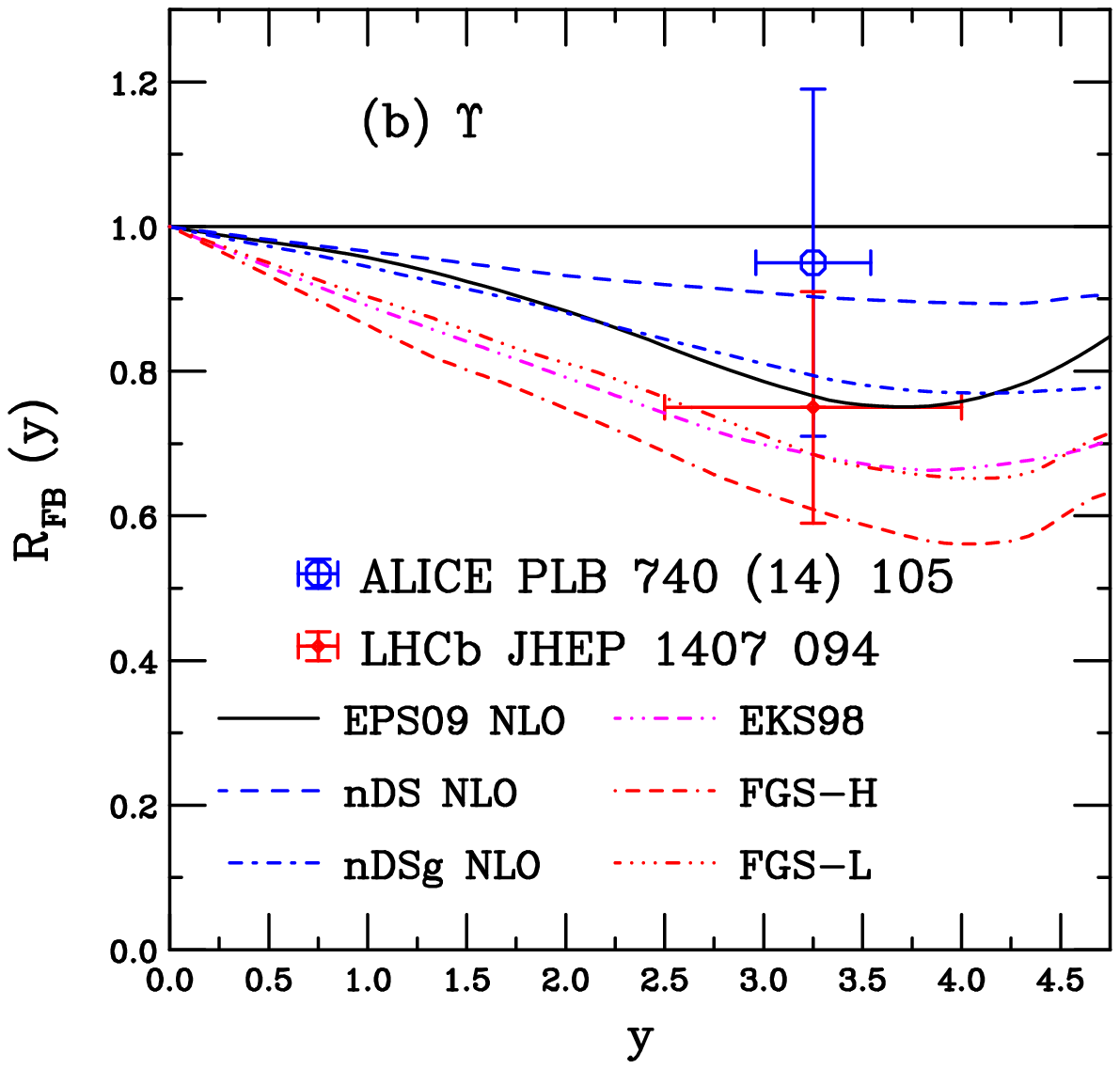}
\caption[]{
The $\Upsilon$ forward-backward ratio $R_{FB}(p_T)$ in the ALICE overlap region 
(a) and $R_{FB}(y)$ (b).  The ALICE \protect\cite{ALICEpPbUps} and
LHCb \protect\cite{LHCbpPbups} data are shown in (b).
The ratios are shown for central 
EPS09 NLO (black),
nDS NLO (blue dashed), nDSg NLO (blue dot-dashed) and EKS98 LO (magenta
dot-dot-dash-dashed), FGS-H (red dot-dash-dash-dashed) and FGS-L (red
dot-dot-dot-dash).
}
\label{fig:nPDFsRFB_Ups}
\end{center}
\end{figure}

Figures~\ref{fig:nPDFsRpPb_Ups} and \ref{fig:nPDFsRFB_Ups} present the 
results for $\Upsilon$ production in the same rapidity ranges.  
The $p_T$-dependent $R_{p{\rm Pb}}$ ratios shown in 
Fig.~\ref{fig:nPDFsRpPb_Ups}(a)-(c) exhibit somewhat
less spread than the counterpart
$J/\psi$ calculations.  The main difference is that now, at the higher scale
and correspondingly larger $x$ probed, the antishadowing peak for $\Upsilon$
production is fully within the acceptance of ALICE and LHCb in the backward
direction.  Thus $R_{p{\rm Pb}}(p_T) > 1$ over all $p_T$ in 
Fig.~\ref{fig:nPDFsRpPb_Ups}(b) with 
$R_{p{\rm Pb}}(p_T \sim 1.5\, {\rm GeV}) \sim 1.2$.

The rapidity-dependent ratio, $R_{p{\rm Pb}}(y)$, shown in 
Fig.~\ref{fig:nPDFsRpPb_Ups}(d) echoes the low $p_T$ results.  Even the nDS
and nDSg ratios are above unity in the backward rapidity region, albeit in
the backward edge of the experimental acceptance.  This is because, in the
high-$x$ region, these sets become large as $x \rightarrow 1$ with the turn on
of this effect beginning at $x \sim 0.5$ and the region of $x \sim 0.1$ is 
where the gluon distribution becomes larger than unity.  We note that even 
though the FGS gluon ratios are only somewhat narrower than the EPS09 NLO ratio 
for the same scale, see Fig.~\ref{fig:nPDFs}(b), for the actual $\Upsilon$
calculation shown here, the difference is enhanced.  At backward rapidity,
the sets are in agreement only with the higher LHCb point while they all
miss the ALICE point at $R_{p{\rm Pb}}(y) < 1$.  At forward rapidity, the
results spread more.  The strongest shadowing is again with the FGS results
but due to the larger $x$ probed for the $\Upsilon$, the $x < 10^{-5}$ region
is not entered in the rapidity range shown although it is approaching the
edge of this region, contributing to the fluctuations in the FGS curves in
Fig.~\ref{fig:nPDFsRpPb_Ups}(d).  As for the $J/\psi$, the EKS98 slope is
more linear for $y > -1$ 
than than the EPS09 NLO central value.  Thus the nDS and
nDSg results, with weak shadowing, as well as the EPS09 NLO results agree
best with the weaker shadowing reported by LHCb while the FGS results generally
agree better with the stronger shadowing reported by ALICE.  The EKS98
LO parameterization is between the two measurements and also includes 
antishadowing.  Therefore, it gives the best overall result for this
measurement given the quality of the currently available data.

The forward-backward ratios for $\Upsilon$ production are shown in 
Fig.~\ref{fig:nPDFsRFB_Ups}.  None of the results for the various nPDF sets 
exhibit a strong $p_T$ dependence.  The FGS and EKS98 results are generally
below those of EPS09 NLO over all $p_T$ while those of nDS and nDSg are higher.
These rather flat ratios can be attributed to the relative $p_T$-independence
for $p_T > 4$ GeV for $R_{p{\rm Pb}}(p_T)$ in Fig.~\ref{fig:nPDFsRpPb_Ups}(a) and
(b) while the stronger shadowing and antishadowing, respectively, in these
regions at $p_T < 4$ GeV essentially give the same forward-backward ratio as
at higher $p_T$.  

The rapidity dependence of the forward-backward ratio, 
Fig.~\ref{fig:nPDFsRFB_Ups}(b), is similar for all the nPDF sets even though
the magnitude of the effect varies.  Because all the ratios increase above
unity in the backward direction, the ratios all decrease almost linearly,
except for EPS09 NLO, with a minimum at $3.5 \leq y \leq 4.0$.  While most of
the ratios rise again at larger rapidity, the nDS and nDSg ratios only
flatten with rapidity above $y \sim 3.5$.

In summary, of all the nPDFs, the LO EKS98 set agrees best with the currently
available data.  However, this is a rather unsatisfying conclusion because
the LO set is used in a NLO calculation.  Even if the $x$ values do not
strongly differ between LO and NLO, the LO shadowing parameterization should
be a stronger function of $x$, with a larger overall shadowing magnitude, to
produce the same $R_{p{\rm Pb}}$ ratios order by order, as was previously
shown to be the case for the nDS and nDSg sets.  A re-evaluation of the
nPDF analysis, including appropriate LHC data, should be made 
in order to resolve the issue.

\subsection{Mass and Scale Dependence}
\label{SubSec:M_muvar}

In this section, we discuss the mass and scale dependence of the shadowing
ratios, using the EPS09 NLO set as an example.  Similar results are
expected for the other nPDFs but we choose EPS09 NLO here because we can compare
the shadowing uncertainty to that of the mass and scale.  Recall that for
charm-$J/\psi$ we employ
$(m,\mu_F/m, \mu_R/m) = (1.27 \pm 0.09 \, {\rm GeV}, 2.1^{+2.55}_{-0.85}, 1.6^{+0.11}_{-0.12})$ 
while for bottom-$\Upsilon$ production, we employ  
$(m,\mu_F/m, \mu_R/m) = (4.65 \pm 0.09 \, {\rm GeV}, 1.4^{+0.77}_{-0.49}, 1.1^{+0.22}_{-0.20})$.  

\begin{figure}[t]
\begin{center}
%\vspace*{-0.05in} 
\includegraphics[width=0.45\textwidth]{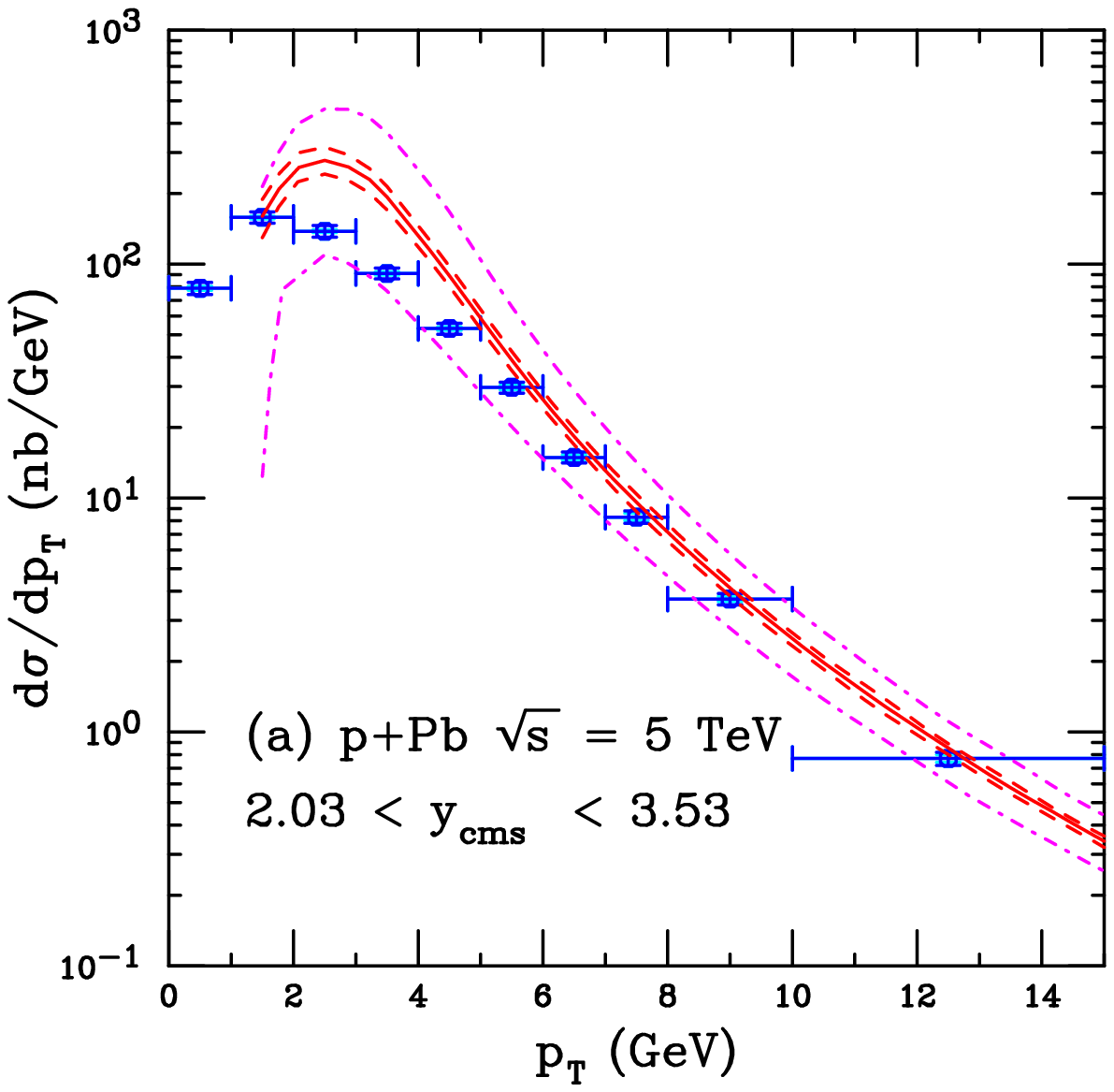}
\includegraphics[width=0.45\textwidth]{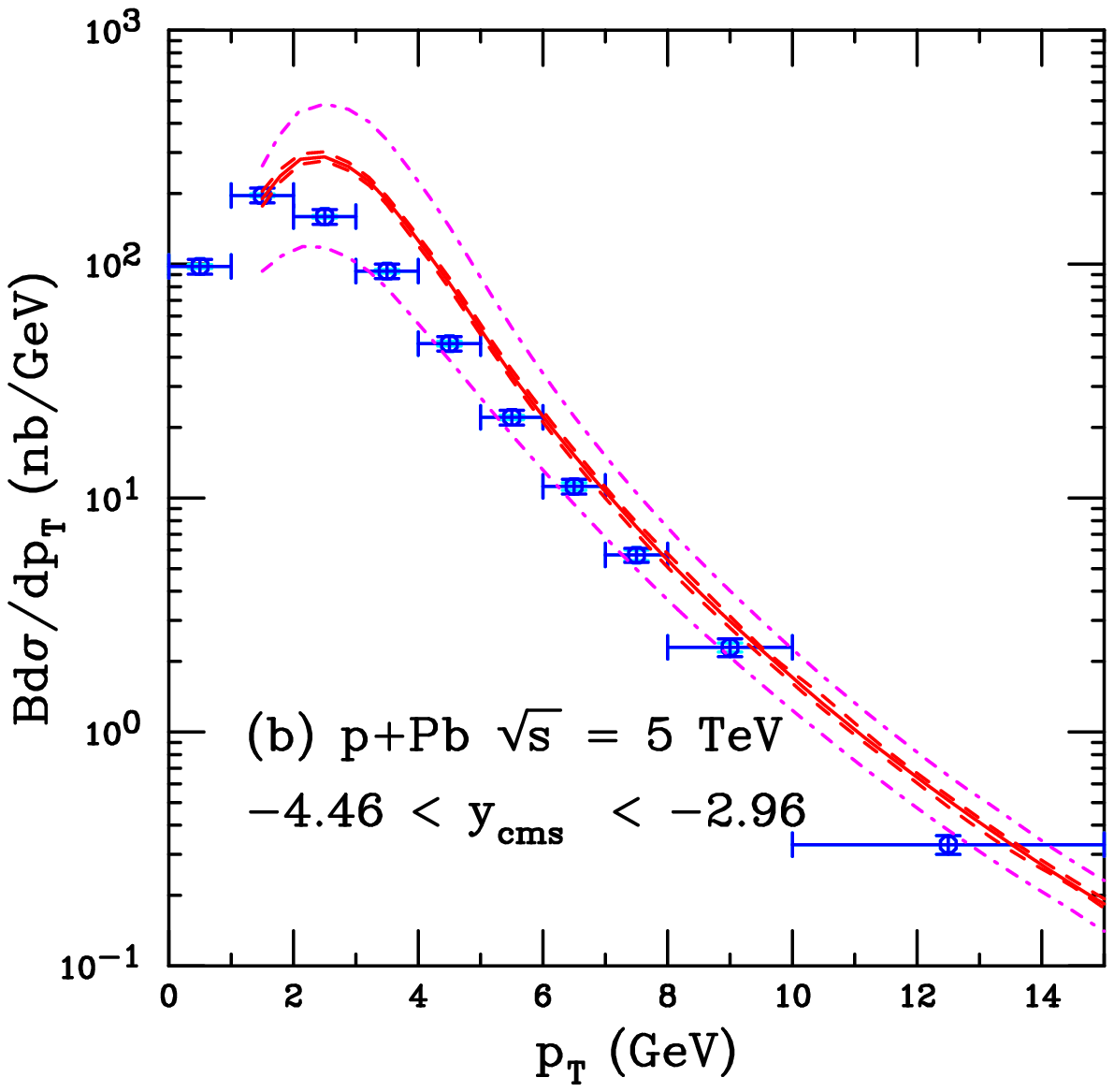} \\
\includegraphics[width=0.45\textwidth]{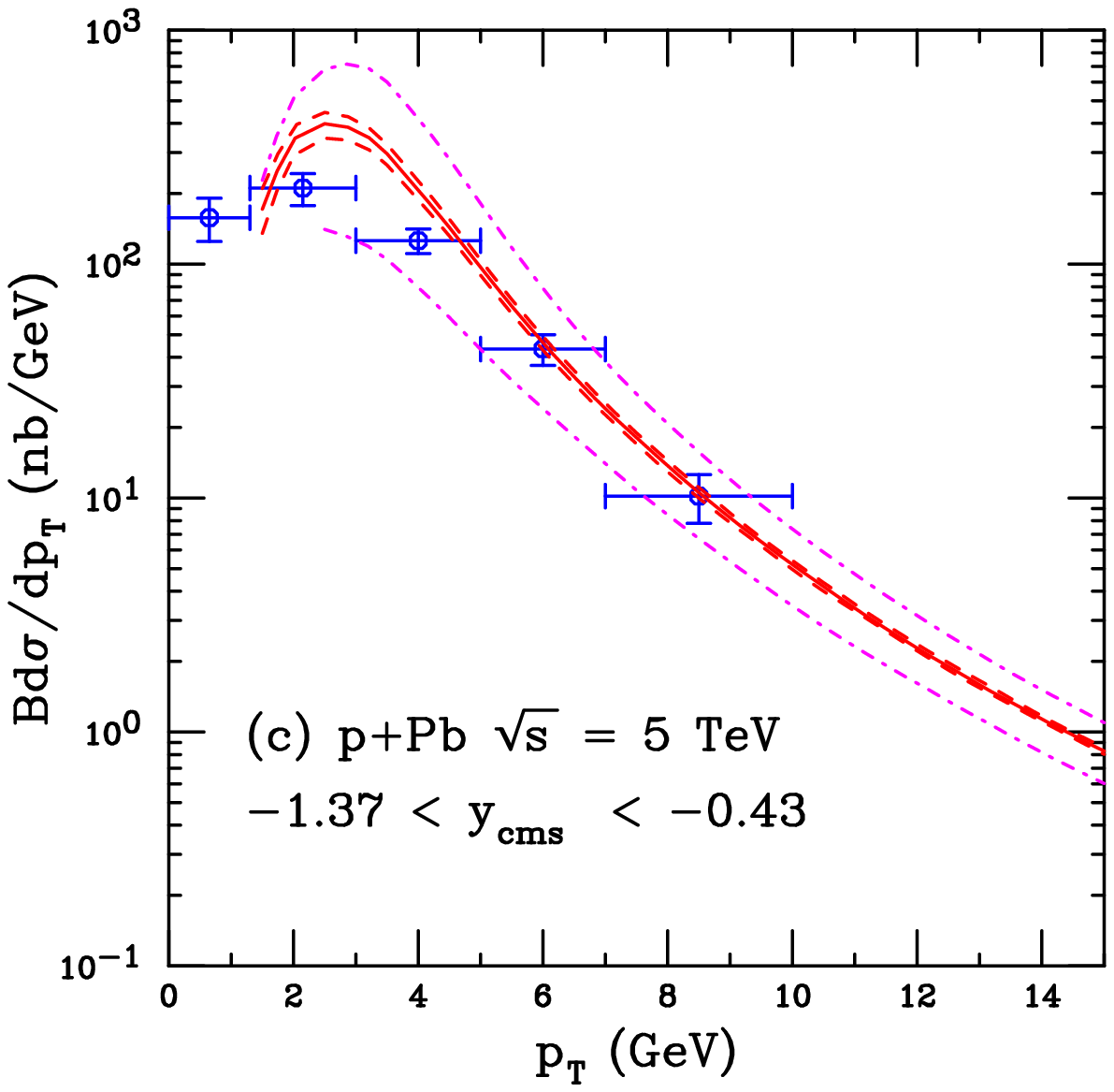}
%\vspace*{-0.4in}
\caption[]{(Color online)
The ALICE $J/\psi$ $p_T$ distributions \protect\cite{ALICEpPbpsi_pT} at 
forward (a), backward (b) and midrapidity (c) at NLO in the CEM 
\protect\cite{NVF}.  The solid red curve is the EPS09 NLO central value
while the dashed red curves are the EPS09 NLO uncertainties and the dot-dashed
magenta curves are the mass and scale uncertainties.
}
\label{fig:pPb_pTdist}
\end{center}
\end{figure}

The uncertainties of the EPS09 NLO nPDF sets compared to those of
the mass and scale uncertainties on $d\sigma/dp_T$ for the $J/\psi$
are illustrated in Fig.~\ref{fig:pPb_pTdist}.
Here the ALICE $p_T$ distributions in $p+$Pb collisions \cite{ALICEpPbpsi_pT}
are compared to our NLO CEM calculations.  The red solid and dashed curves are
the central EPS09 NLO results with the uncertainty band due to the EPS09
parameter variations.  This band is rather narrow on the scale of the plots
with the only clear separation between the curves at low $p_T$ where the 
shadowing effects are largest.  Clearly, a 20-40\% effect on $R_{p{\rm Pb}}(p_T)$
 on a linear scale in this region results in a narrow band on the 
$p_T$ distribution itself on a logarithmic scale.  We do not show the lowest
$p_T$ results because the underlying $p+p$ calculation does not accurately
numerically cancel the divergences in the negative weight Monte Carlo
\cite{MNRcode}.  The intrinsic $k_T$ kick used to obtain the shape of the $p_T$
distribution in the low-$p_T$ region was fixed at RHIC energies and
successfully compared to the $J/\psi$ $p_T$ distributions at 
$\sqrt{s} = 7$ TeV \cite{NVF}.

The magenta curves in Fig.~\ref{fig:pPb_pTdist} show the uncertainty due
to the mass and scale variations.  The mass and scale uncertainties are 
calculated 
based on results using the one standard deviation uncertainties on
the quark mass and scale parameters.  If the central, upper and lower limits
of $\mu_{R,F}/m$ are denoted as $C$, $H$, and $L$ respectively, then the seven
sets corresponding to the scale uncertainty are  $\{(\mu_F/m,\mu_F/m)\}$ =
$\{$$(C,C)$, $(H,H)$, $(L,L)$, $(C,L)$, $(L,C)$, $(C,H)$, $(H,C)$$\}$.    
The uncertainty band can be obtained for the best fit sets by
adding the uncertainties from the mass and scale variations in 
quadrature. The envelope containing the resulting curves,
\begin{eqnarray}
d\sigma_{\rm max}/dX & = & d\sigma_{\rm cent}/dX
+ \sqrt{(d\sigma_{\mu ,{\rm max}}/dX - d\sigma_{\rm cent}/dX)^2
+ (d\sigma_{m, {\rm max}}/dX - d\sigma_{\rm cent}/dX)^2} \, \, , \label{sigmax} \\
d\sigma_{\rm min}/dX & = & d\sigma_{\rm cent}/dX
- \sqrt{(d\sigma_{\mu ,{\rm min}}/dX - d\sigma_{\rm cent}/dX)^2
+ (d\sigma_{m, {\rm min}}/dX - d\sigma_{\rm cent}/dX)^2} \, \, , \label{sigmin}
\end{eqnarray}
defines the uncertainty on the cross section
where $X$ can be either $p_T$ or $y$.  The EPS09 uncertainty band is based
on the mass and scale set with the central mass value and 
$(\mu_F/m,\mu_F/m) = (C,C)$.

In Fig.~\ref{fig:pPb_pTdist}, the mass and scale uncertainties are clearly
significantly larger than those of the shadowing parameterizations alone,
still a factor of $\sim 2$ on the $J/\psi$ $p_T$ distributions over a wide
$p_T$ range.  Thus, even though the uncertainties on the $J/\psi$ cross section
in the CEM are reduced relative to those calculated previously, 
see Ref.~\cite{NVF}, they are still significantly larger than those of 
shadowing on the central mass and scale values.  One would naively expect the
uncertainties on $R_{p{\rm Pb}}$ and $R_{FB}$ to reflect the results shown
in Fig.~\ref{fig:pPb_pTdist} but, as we will show, it depends on the
method used to calculate the uncertainty.

We have tried several ways of estimating
the size of the mass and scale uncertainty on the ratios 
which we describe first before
comparing to the data.  We note that all of these calculations are based on
the EPS09 NLO central set rather than making a global nPDF + mass + scale
uncertainty by calculating the 30 error sets for EPS09 with the 8 additional
mass and scale combinations that define the mass and scale uncertainty on the
cross section for $p+p$ and $p+$Pb.

The most naive, labeled $m/\mu_F/\mu_R \, v1$ on 
Figs.~\ref{fig:MSuncRpPb_Psi}-\ref{fig:MSuncRFB_Ups}, 
is to take the ratios of $p+$Pb to $p+p$ for each
mass and scale combination and then locate the extrema of the ratios based
on the mass and scale
in each case, exactly as in Eqs.~(\ref{sigmax}) and (\ref{sigmin}) above. 
Thus the uncertainty band is based on the $R_{p{\rm Pb}}$ of each set.  Since
the ratios are all close to unity (within 20-40\% in most cases), and similar
to each other, this method would tend to underestimate the uncertainty.

The second, labeled $m/\mu_F/\mu_R \, v2$ on 
Figs.~\ref{fig:MSuncRpPb_Psi}-\ref{fig:MSuncRFB_Ups}, reflects the uncertainty
calculation of Eqs.~(\ref{sigmax}) and (\ref{sigmin}).  We calculate the
maxima and minima of $d\sigma/dX$, as in Fig.~\ref{fig:pPb_pTdist}, and then
form $R_{p{\rm Pb}}$ by dividing by the central value of the $p+p$ cross section
in the corresponding rapidity bin (for the $p_T$-dependent results) or
the $p_T$-integrated cross section (for the rapidity dependence).  Thus the
shadowing ratios for $v2$ are based on cross sections only.  We expect a larger
uncertainty band for this calculation, especially at low $p_T$.

We note, however, that 
there is actually no difference between the $v2$ and $v1$ calculations of
the uncertainty in the forward-backward ratios because we calculate
the forward-backward ratios for each mass and scale combination before
calculating the uncertainty to ensure that we make a one-to-one correspondence
in $R_{FB}$.  Thus, there is no $v2$ result in the plots of
$R_{FB}$.

The above methods of calculating the mass and scale uncertainty, relying
on Eqs.~(\ref{sigmax}) and (\ref{sigmin}), are not actually calculated the
same way as the EPS09 uncertainty band because they rely only on the extrema
of the mass and scale dependent cross sections from the central values rather
than adding the excursions from the central value in quadrature, as in the
EPS09 uncertainty calculation.  Therefore, finally, for $m/\mu_F/\mu_R \, v3$, 
we add the mass and scale uncertainties in quadrature, a la 
EPS09, and then form $R_{p {\rm Pb}}$ by dividing by the central $p+p$ cross
section.  Since this is a cumulative uncertainty rather than based on the
greatest excursion from the mean, we expect this to give the largest 
overall uncertainty.  
The $v3$ uncertainty band was calculated assuming that the appropriate 
$\mu_F/m$ and $\mu_R/m$ pairs
are $[(H,H),(L,L)]$, $[(H,C),(L,C)]$ and $[(C,H),(C,L)]$.    
Other choices could lead to different results.  (There is only
one possible pair for the mass uncertainty thus this part of the calculation
is the same for $v2$ and $v3$.)

Results using all three methods are shown for $J/\psi$ and $\Upsilon$
in the remainder of this section.

\subsubsection{$J/\psi$}

\begin{figure}[t]
\begin{center}
\includegraphics[width=0.45\textwidth]{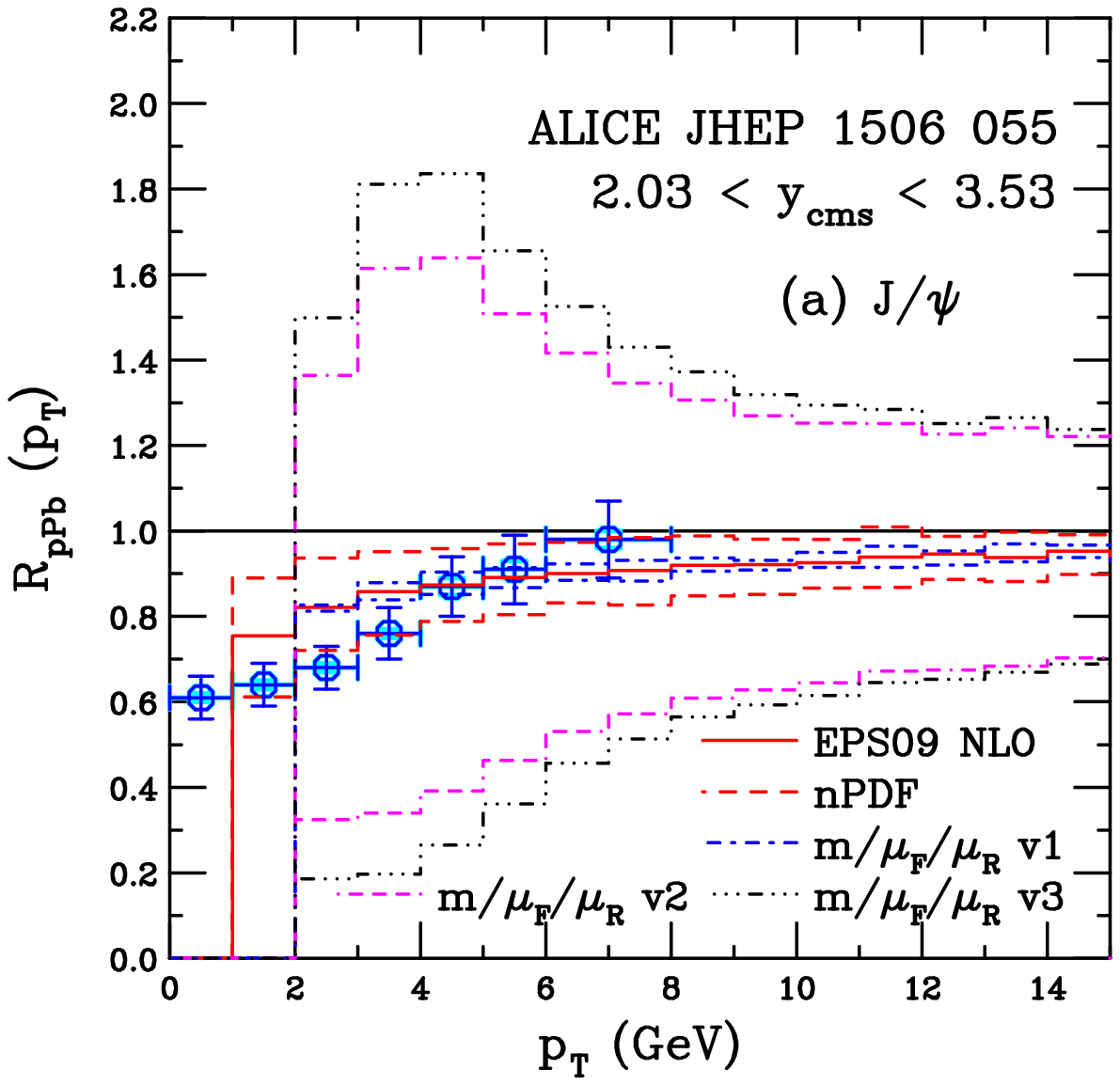}
\includegraphics[width=0.45\textwidth]{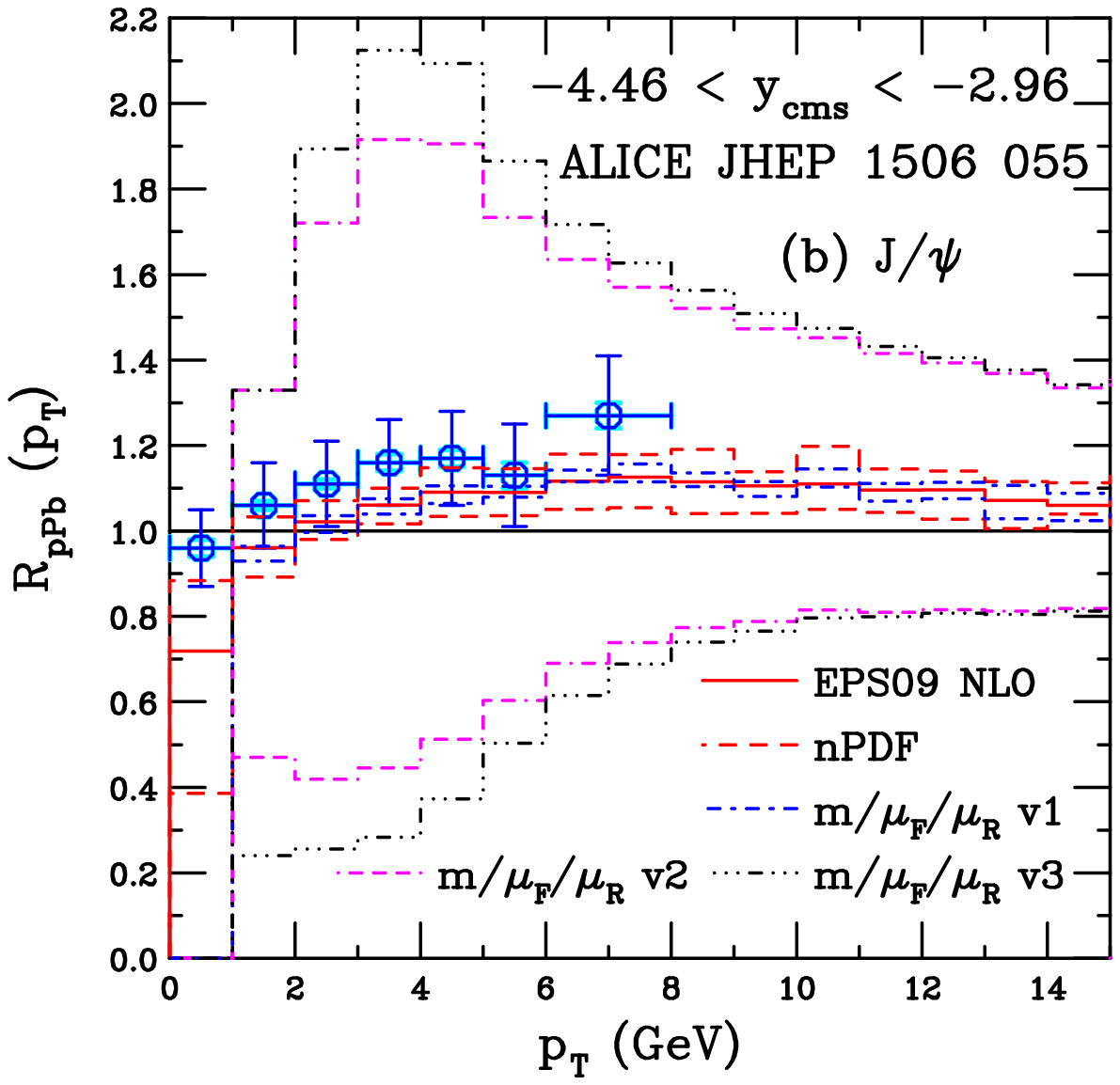} \\
\includegraphics[width=0.45\textwidth]{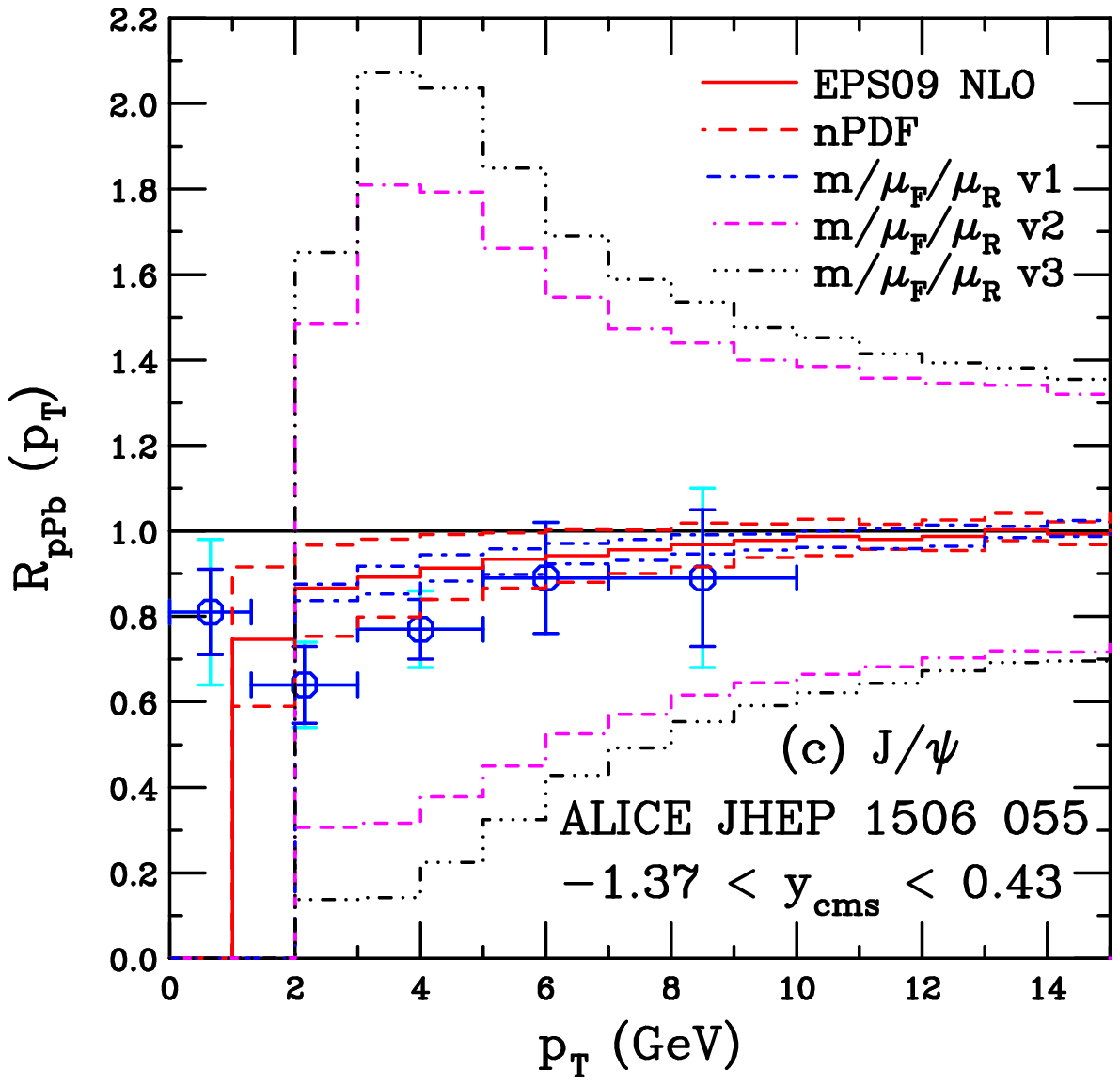}
\includegraphics[width=0.45\textwidth]{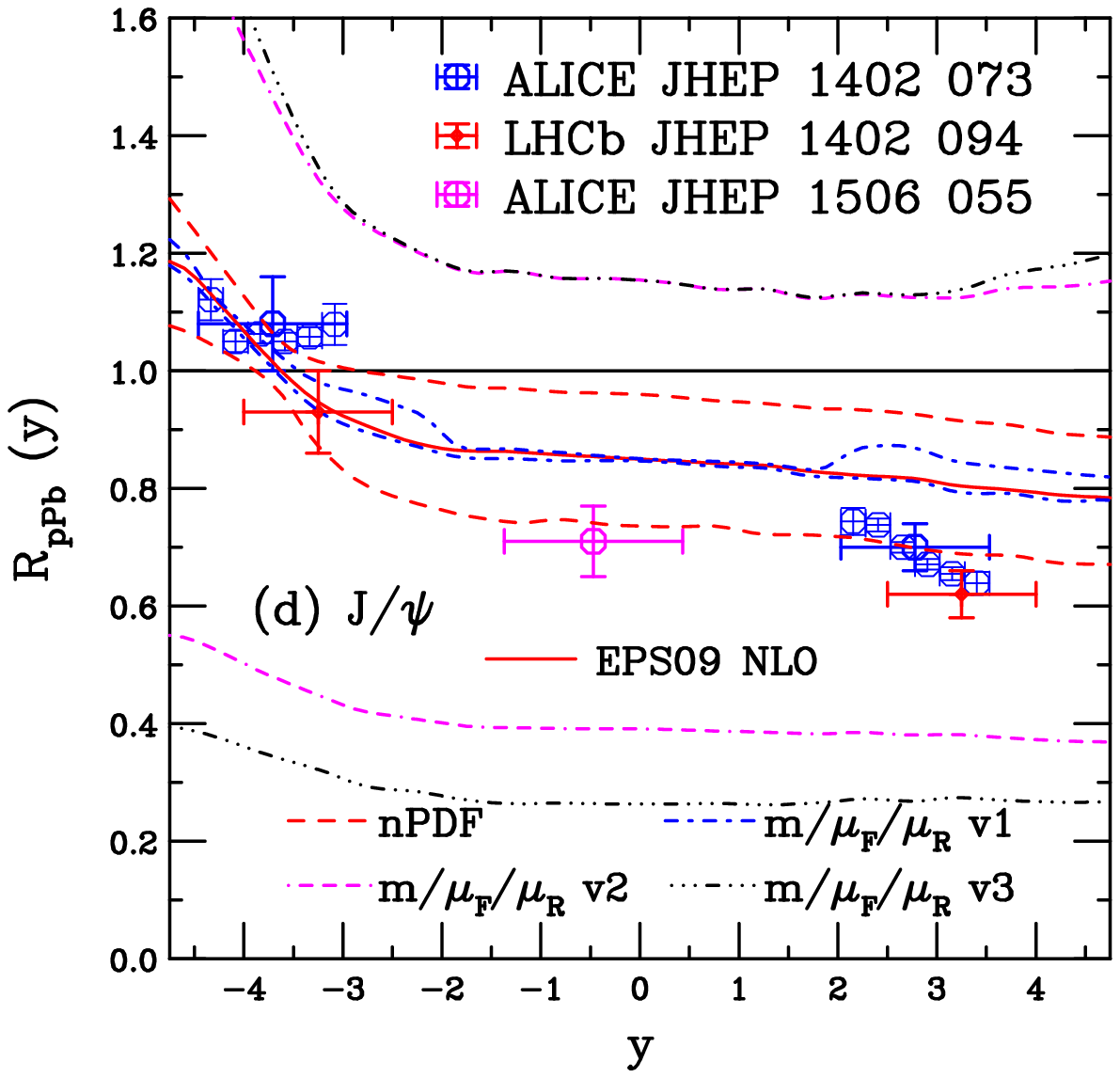}
\caption[]{(Color online)
The mass and scale uncertainties in the $J/\psi$ ratio $R_{p{\rm Pb}}(p_T)$ in the
ALICE acceptance at forward (a),
backward (b) and central (c) rapidity \protect\cite{ALICEpPbpsi_pT}.  The ratio 
$R_{p{\rm Pb}}(y)$ is compared to the  ALICE \protect\cite{ALICEpPbpsi} and
LHCb \protect\cite{LHCbpPbpsi} data in (d).  The EPS09 NLO uncertainty band is 
shown in red while the uncertainties calculated with method $v1$ in blue, 
$v2$ in magenta and $v3$ in black.
}
\label{fig:MSuncRpPb_Psi}
\end{center}
\end{figure}

Figure~\ref{fig:MSuncRpPb_Psi} shows the relative uncertainties for the EPS09
NLO band and the three ways of calculating the mass and scale uncertainty 
for $R_{p{\rm Pb}}$.  As expected, the $v1$ proceedure gives the smallest 
uncertainty since there is little variation in the individual values of 
$R_{p{\rm Pb}}$ for each mass and scale choice for the central EPS09 set.  
Therefore, the $v1$ band is narrower than that of
the EPS09 band itself, underestimating the uncertainty.  

On the other hand, the $v2$ and $v3$ bands are wider,
especially for $p_T < 5$ GeV, as expected.  Also, as mentioned previously, the
$v3$ band is broader than that of $v2$ over all $p_T$ although the two methods
merge at the highest $p_T$ values.  The $R_{p{\rm Pb}}(p_T)$ found for the $v2$
method is in agreement with what one might expect looking at the bands on
$d\sigma/dp_T$ in Fig.~\ref{fig:pPb_pTdist}.  The ratios as a function 
of $p_T$ are relatively regular even though, at least for $v1$ and $v2$, the
combination of factorization and renormalization scales that give the extreme 
values for that $p_T$ does depend on $p_T$.

\begin{figure}[t]
\begin{center}
\includegraphics[width=0.45\textwidth]{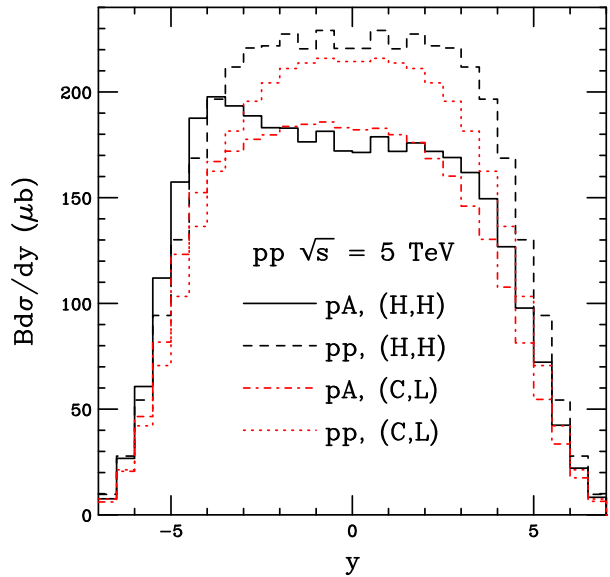}
\caption[]{(Color online) The $p+p$ and $p+$Pb $J/\psi$ rapidity distributions
for the $(H,H)$ $(C,L)$ sets showing the differences leading to the change
in the upper limit of the mass and scale uncertainties of method $v1$ around
midrapidity.
}
\label{fig:MSunc_ydist_Psi}
\end{center}
\end{figure}

The situation is somewhat different as a function of rapidity, in part because
the rapidity distributions fluctuate somewhat, especially at midrapidity.  The
size of the fluctuations depends on the scale choice and, depending on which
scale set determines the extreme value, these fluctuations can manifest 
themselves in $R_{p{\rm Pb}}(y)$.  This effect is obvious in
Fig.~\ref{fig:MSuncRpPb_Psi}(d) where the upper bound on $v1$ is visibly
above the central value of the ratio for $|y|> 2$ and almost on top of it for
$|y|< 2$.  The abrupt change of slope occurs because the maximum value of the
ratio is obtained with set $(H,H)$ away from midrapidity and with set $(C,L)$
at midrapidity.  This is illustrated in Fig.~\ref{fig:MSunc_ydist_Psi} where
the $p+p$ and $p+$Pb rapidity distributions are shown for each of these 
combinations.  The
change in the $p+$Pb rapidity distribution is a stronger function of $y$ for
the $(H,H)$ set than for the $(C,L)$ set.  There are also larger fluctuations
in the distribution at midrapidity for the $(H,H)$ set.  To emphasize this
difference, we present the rapidity distributions as histograms rather than
smoothed curves.  We note that these
fluctuations are not due to limited statistics in the calculation but are
more likely due to the uneven distribution of low $p_T$ negative-weight 
events in the MNR code at midrapidity.  The presence of these events may also
flatten the rapidity distribution in the LHC energy regime so that while the
$p+p$ rapidity distribution is within the mass and scale uncertainty band the
slope is somewhat flatter than the measured one, see Ref.~\cite{NVF}.

The $v2$ and $v3$ ratios are more separated and, because they are obtained by
taking the ratio to the $p+p$ distribution calculated with the central mass and
scale set, they are also smoother.  The `kink' in the central EPS09 NLO
ratio noted previously is sharpened for the upper limit of these bands and
reduced, but still present, in the lower limits.  The upper limits obtained
for $v2$ and $v3$ are very similar.  Indeed these upper limits are almost
indistinguishable for most of the illustrated rapidity range.  The numerical
values are generally different, with the $v3$ values larger than those of $v2$,
but the difference is not large enough to be visible on the scale
of the plot.  This similarity likely stems from the aforementioned
fluctuations in the rapidity distributions.  Since these fluctuations
tend to be smaller for the lower limits of $R_{p{\rm Pb}}(y)$, there is a large
separation between the lower bounds on $v2$ and $v3$.  

\begin{figure}[t]
\begin{center}
%\vspace*{-0.05in}
\includegraphics[width=0.45\textwidth]{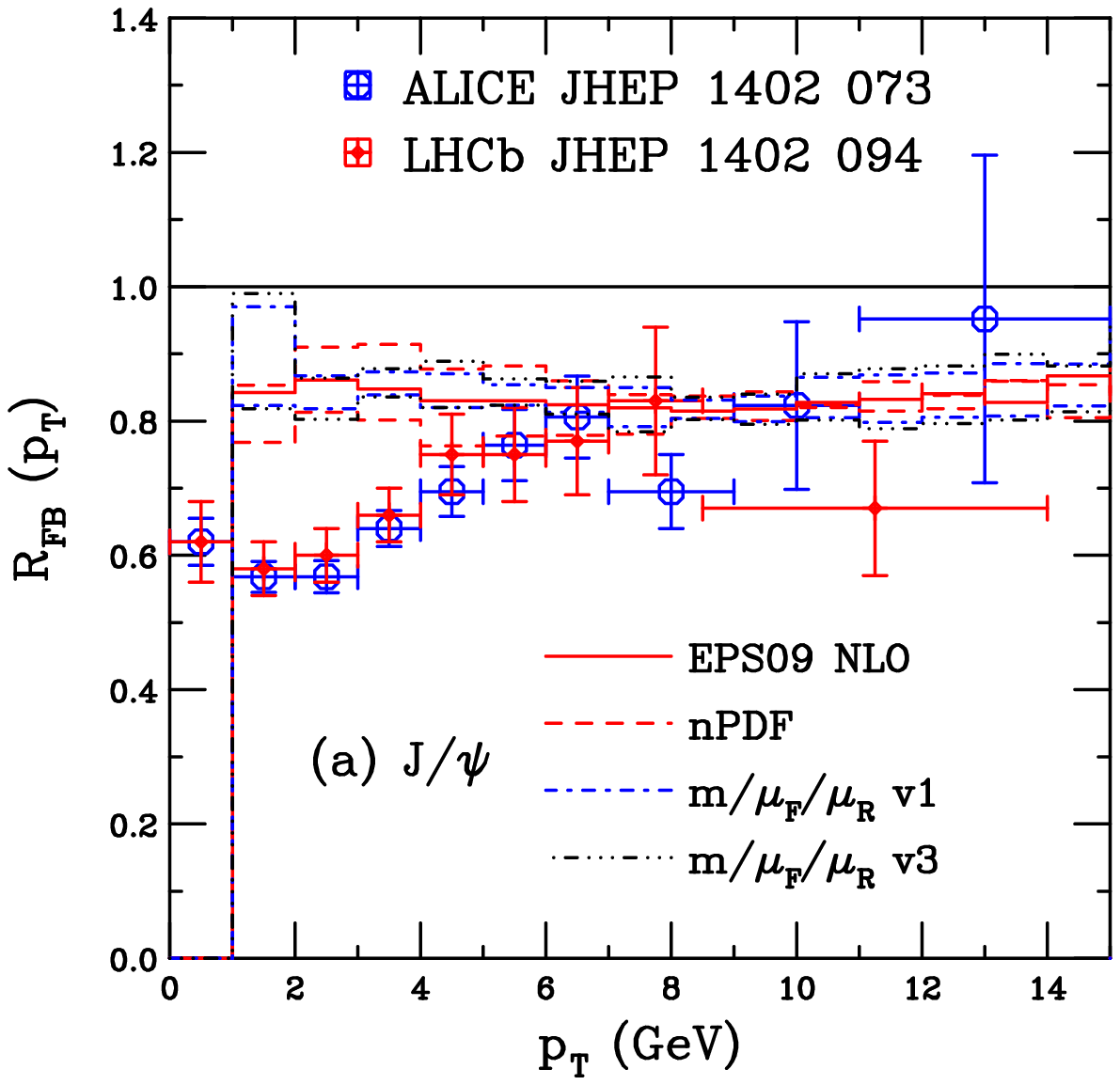}
\includegraphics[width=0.45\textwidth]{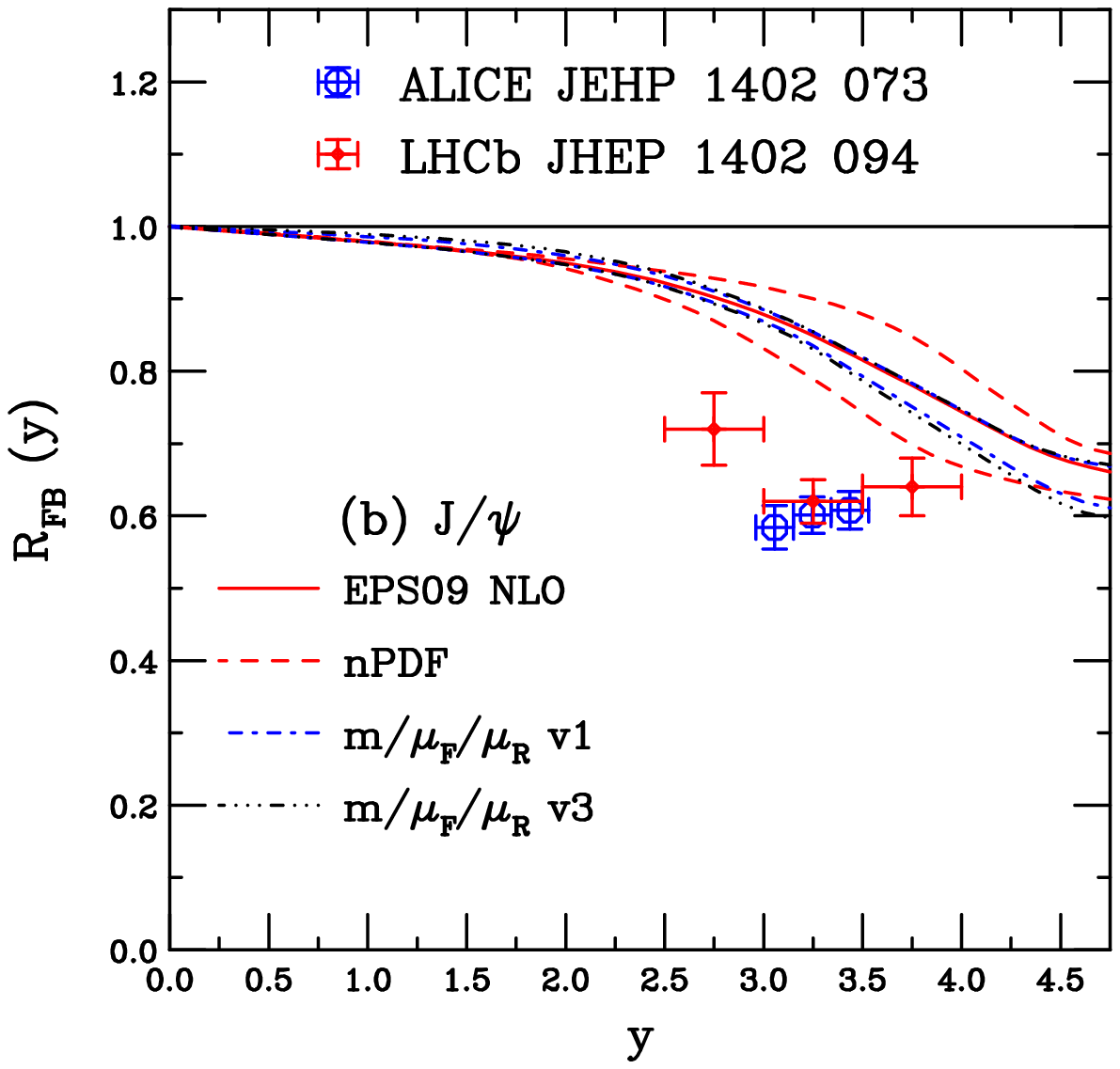}
\caption[]{(Color online)
The mass and scale uncertainties in the 
$J/\psi$ forward-backward ratio $R_{FB}(p_T)$ in the ALICE overlap region 
(a) and $R_{FB}(y)$ (b). 
The EPS09 NLO uncertainty band is 
shown in red while the uncertainties calculated with method $v1$ in blue
and $v3$ in black.  The  ALICE \protect\cite{ALICEpPbpsi} and
LHCb \protect\cite{LHCbpPbpsi} data are also shown.
}
\label{fig:MSuncRFB_Psi}
\end{center}
\end{figure}

In all the cases shown, except for the ratios calculated with $v1$, the
envelope of ratios described by the mass and scale uncertainties contains the
data.  This can be expected because this is the designed purpose of 
Eqs.~(\ref{sigmax}) and (\ref{sigmin}) for the $p+p$ distributions and, when
applied correctly, it does the same for the nuclear suppression factor
$R_{p{\rm Pb}}$ as well.

This is not necessarily the case for the forward-backward ratios, shown in
Fig.~\ref{fig:MSuncRFB_Psi}.  Indeed, in this case the ratios calculated with
methods $v1$ and $v3$ are nearly identical.  Recall that there is no
difference between $v1$ and $v2$ uncertainties for the forward-backward ratios.
The most interesting thing to note here is that the ratios of the mass and
scale uncertainty as a function of rapidity have different slopes than those of
the EPS09 NLO uncertainty.  The stronger rise in $R_{p{\rm Pb}}(y)$
at backward rapidity, combined with the weak dependence on rapidity in the
forward direction makes the upper limit on the mass and scale dependence
larger than the central EPS09 set at $y<2$.  At larger rapidities, these values
are almost coincident.  On the other hand, the weaker rise in $R_{p{\rm Pb}}(y)$
for the lower limit in the backward direction makes the lower limit of the
forward-backward ratio decrease more rapidly than the central set for
$y > 3$.

\subsubsection{$\Upsilon$}

\begin{figure}[t]
\begin{center}
\includegraphics[width=0.45\textwidth]{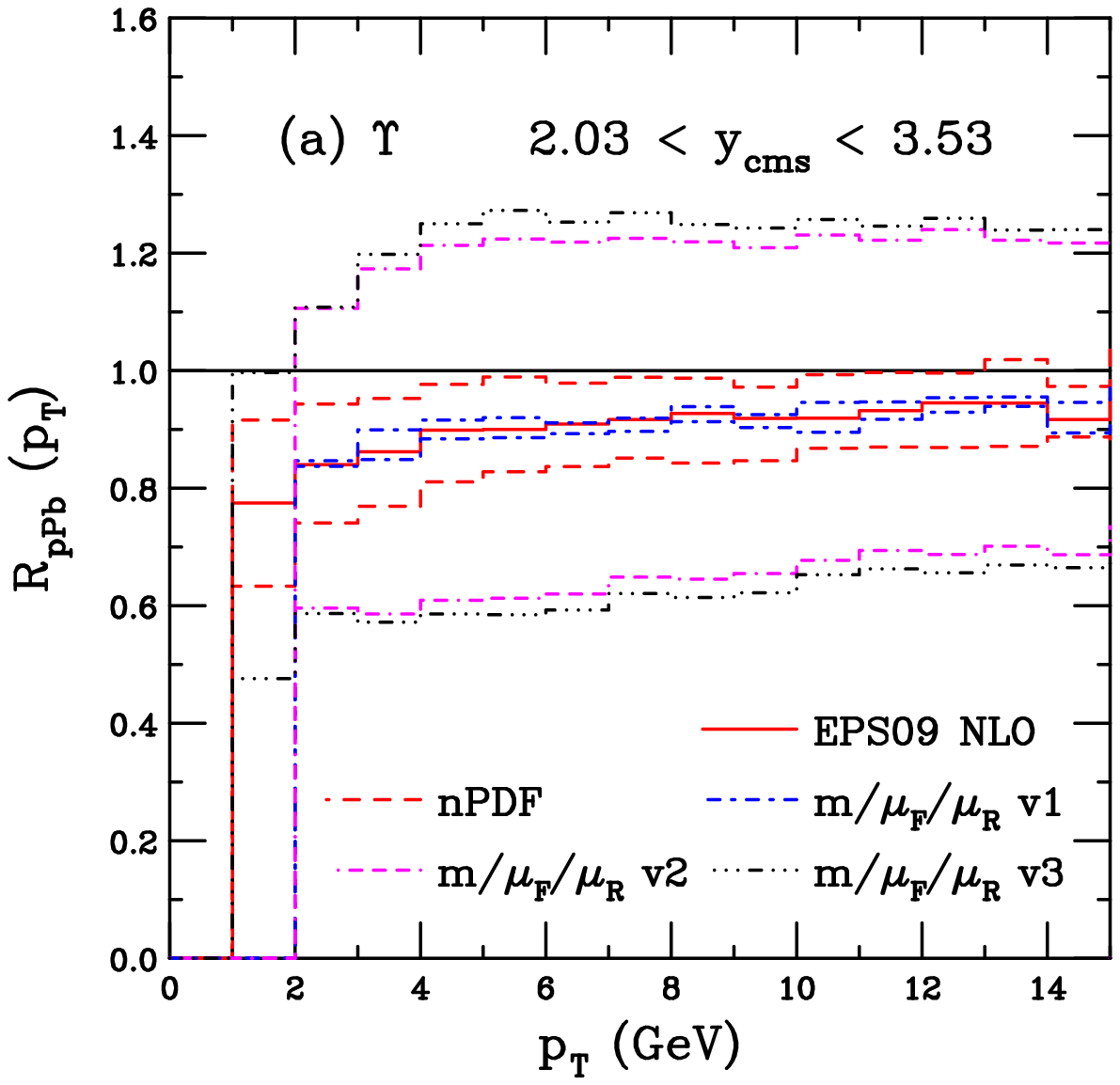}
\includegraphics[width=0.45\textwidth]{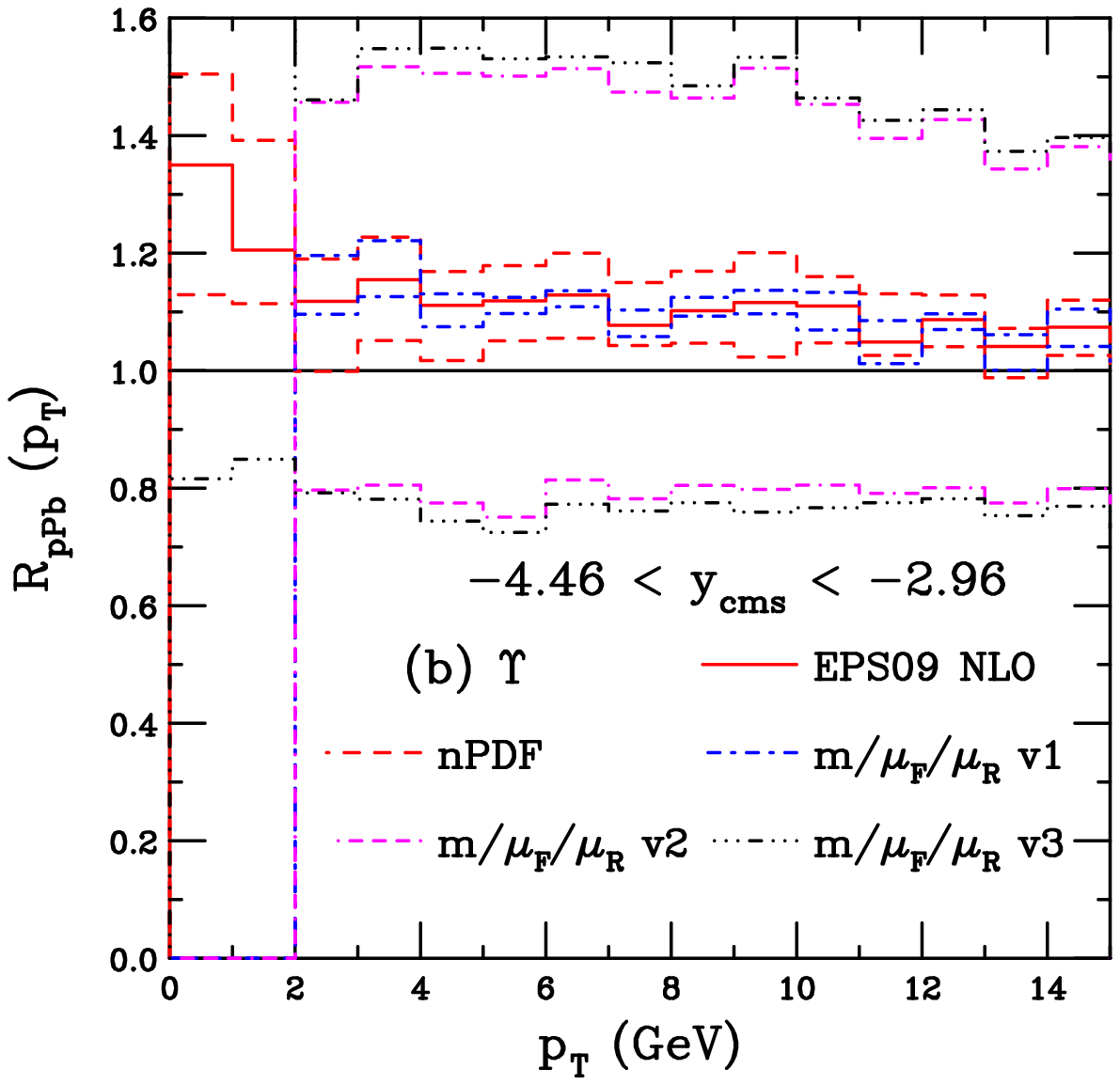} \\
\includegraphics[width=0.45\textwidth]{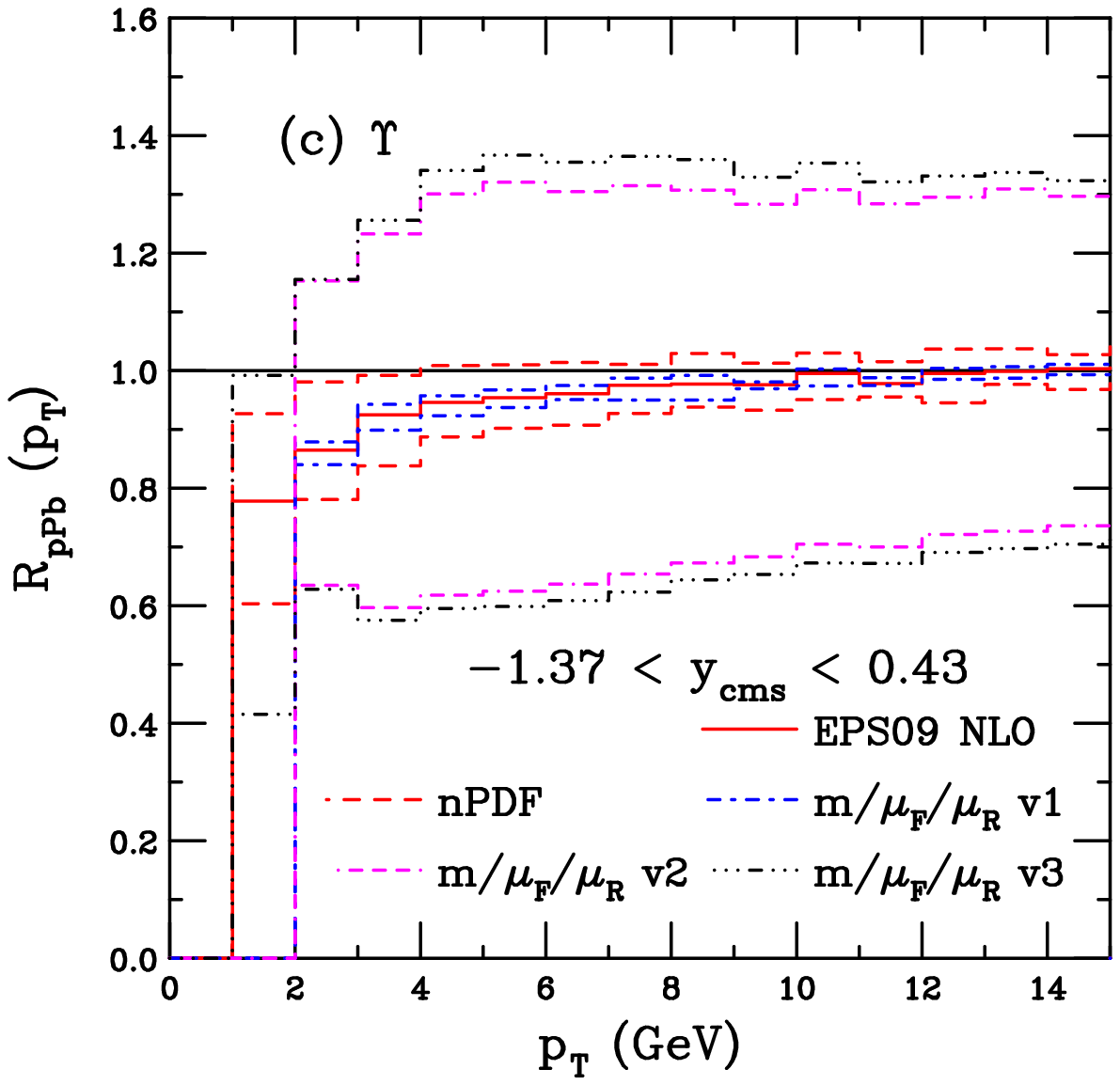}
\includegraphics[width=0.45\textwidth]{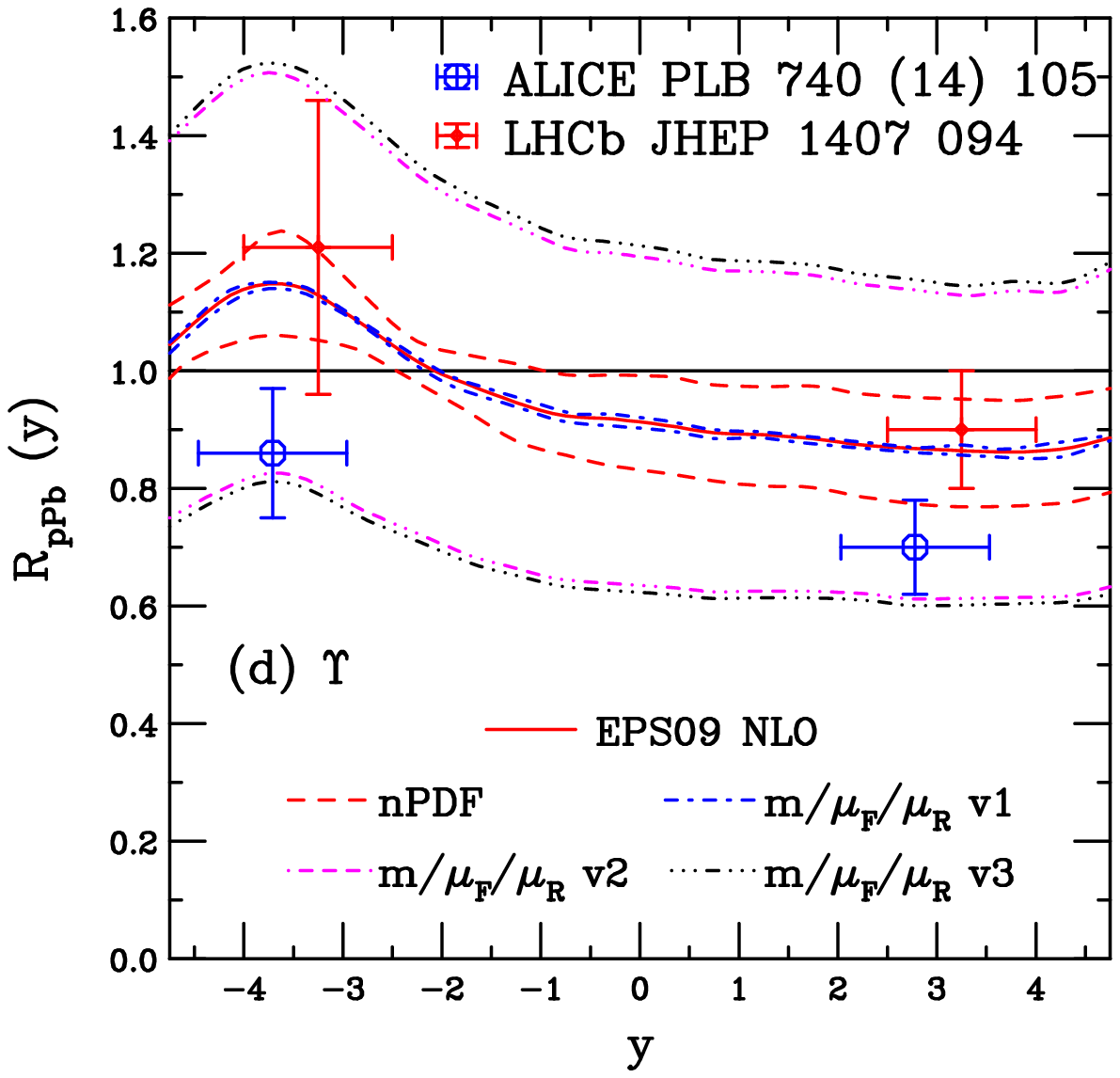}
\caption[]{(Color online)
The mass and scale uncertainties in 
the $\Upsilon$ ratio $R_{p{\rm Pb}}(p_T)$ in the
ALICE acceptance at forward (a),
backward (b) and central (c) rapidity.  The ratio 
$R_{p{\rm Pb}}(y)$ is compared to the  ALICE \protect\cite{ALICEpPbUps} and
LHCb \protect\cite{LHCbpPbups} data in (d).  The EPS09 NLO uncertainty band is 
shown in red while the uncertainties calculated with method $v1$ in blue, 
$v2$ in
magenta and $v3$ in black.
}
\label{fig:MSuncRpPb_Ups}
\end{center}
\end{figure}

\begin{figure}[t]
\begin{center}
%\vspace*{-0.05in}
\includegraphics[width=0.45\textwidth]{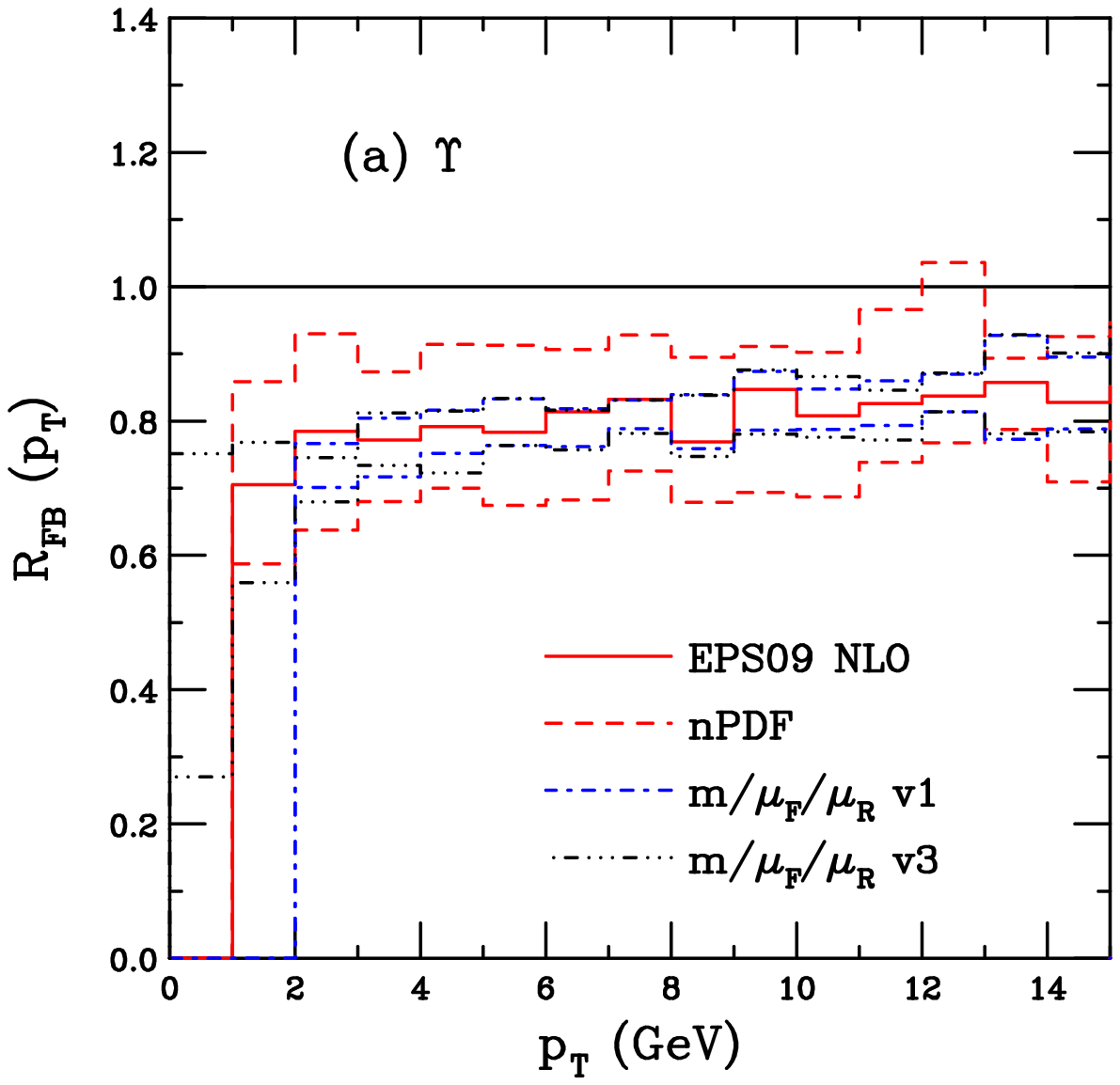}
\includegraphics[width=0.45\textwidth]{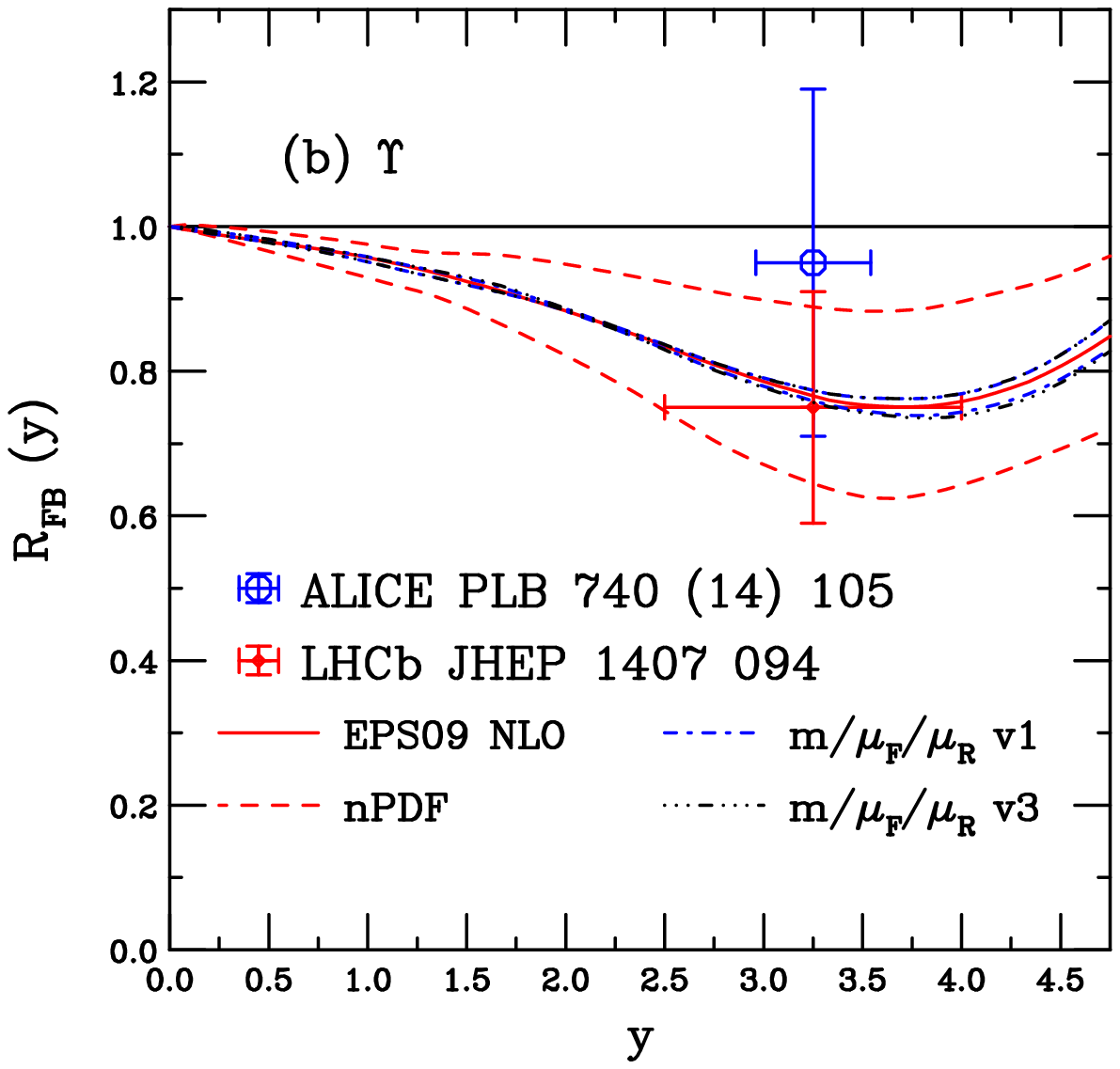}
\caption[]{(Color online)
The  mass and scale uncertainties in the $\Upsilon$ 
forward-backward ratio $R_{FB}(p_T)$ in the ALICE overlap region 
(a) and $R_{FB}(y)$ (b). 
The EPS09 NLO uncertainty band is 
shown in red while the uncertainties calculated with method $v1$ in blue and 
$v3$ in black.  The  ALICE \protect\cite{ALICEpPbUps} and
LHCb \protect\cite{LHCbpPbups} data are also shown in (b).
}
\label{fig:MSuncRFB_Ups}
\end{center}
\end{figure}

The results for the mass and scale uncertainties on the ratios for $\Upsilon$
production are presented in Figs.~\ref{fig:MSuncRpPb_Ups} and 
\ref{fig:MSuncRFB_Ups}.  Overall the trends are quite similar to those for
the $J/\psi$ even though the $p_T$-dependent results here are almost
independent of $p_T$.  
However, we note that the results for $R_{p{\rm Pb}}$ show much
smaller differences between $v2$ and $v3$ over the entire $p_T$ range, as well
as a function of rapidity.  In this case also, the shape of the limits on the
mass and scale uncertainty bands for $R_{p{\rm Pb}}(y)$ is the same as that of
the central EPS09 NLO value so the shape of that ratio has a weaker scale
dependence than the $J/\psi$.  The band on $R_{p{\rm Pb}}(y)$ is broad enough to
encompass the range of the low statistics $\Upsilon$ data.

The uncertainties on the forward-backward ratio in Fig.~\ref{fig:MSuncRFB_Ups}
are again small.  They are, in fact, considerably smaller than those due to 
EPS09 which, as we saw previously, is not because the uncertainties on the 
distributions in an individual rapidity bin, for $R_{FB}(p_T)$, are small.
Instead, these uncertainties are somewhat canceled in the ratio, even though
we still expect the largest excursion from the central value for $v3$.

\subsection{Factorization}
\label{SubSec:Fact}

The question of whether cold nuclear matter effects factorize has been much
discussed.  If so, then at the same energy, the product of the nuclear 
modification factors at forward and backward rapidity in $pA$ collisions
would give the predicted cold matter
result for $A+A$ collisions at the same energy.

In their latest paper on the $J/\psi$ $p+$Pb results 
\cite{ALICEpPbpsi_pT}, the ALICE Collaboration compared the product of their
$R_{p{\rm Pb}}(p_T)$ ratios in the forward and backward rapidity regions at
$\sqrt{s_{_{NN}}} = 5.02$ TeV, $2.03 < y_{\rm cms} < 3.53$ and
$-4.46 < y_{\rm cms} < -2.96$ respectively, to the $R_{\rm PbPb}(p_T)$ ratio 
in the region $2.5 < y < 4.0$ (symmetric around midrapidity) at 
$\sqrt{s_{_{NN}}} = 2.76$ TeV.  They did the
same for the square of the midrapidity ratio measured in 
$-1.37 < y_{\rm cms} < -0.43$ in $p+$Pb collisions at $\sqrt{s_{_{NN}}} = 5.02$ TeV
to $R_{\rm PbPb}$ in $|y|<0.8$ at $\sqrt{s_{_{NN}}} = 2.76$ TeV.  From these
comparisons they were able to obtain an estimate of the cold nuclear matter
effects on Pb+Pb collisions, assuming that the ratios in the forward and 
backward regions factorize.  They then used their $p+$Pb measurements at the
higher energy to form the ratio $S_{J/\psi} = R_{\rm PbPb}/(R_{p{\rm Pb}}(+y) \times
R_{p{\rm Pb}}(-y))$.  From this comparison, they were able to deduce that, at 
least for their dimuon measurement, the Pb+Pb data are significantly more
surpressed than expected for cold matter effects alone \cite{ALICEpPbpsi_pT}.

However, the comparison in Ref.~\cite{ALICEpPbpsi_pT} is made for different 
rapidity regions due to the asymmetric beams in $p+$Pb collisions.  In addition,
the difference in the nucleon-nucleon center of mass energy is almost a factor
of two.  These two effects will cause the $x$ values probed in $p+$Pb and
Pb+Pb collisions to be shifted so that the correspondence is not exact.

In this section, we compare the $R_{p{\rm Pb}}$ and $R_{\rm PbPb}$ ratios at
LO (rapidity only) and NLO (rapidity and $p_T$ in symmetric forward and backward
regions) at the same energy for $p+$Pb and Pb+Pb so that the correspondence 
between the $x$ values should be exact.  In addition, since we employ the same
$\sqrt{s_{_{NN}}}$, there no rapidity shift in $p+$Pb relative to Pb+Pb.  
We can thus
check this factorization does indeed explicitly hold for shadowing effects
in the CEM.  In our calculations here, we employ the central EPS09 sets.  
The results should be similar for other nPDF sets.

\begin{figure}[t]
\begin{center}
%\vspace*{-0.05in}
\includegraphics[width=0.45\textwidth]{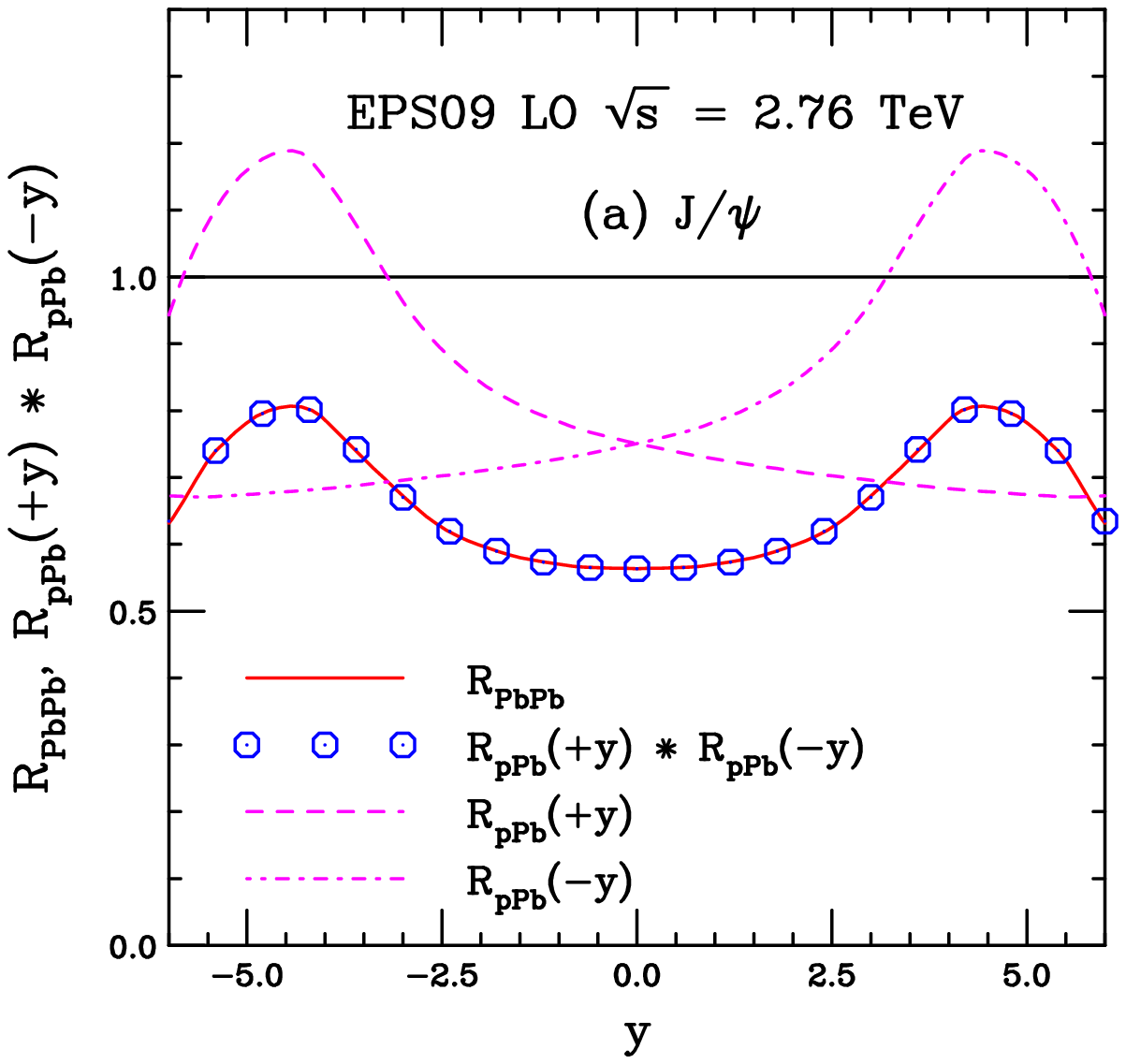}
\includegraphics[width=0.45\textwidth]{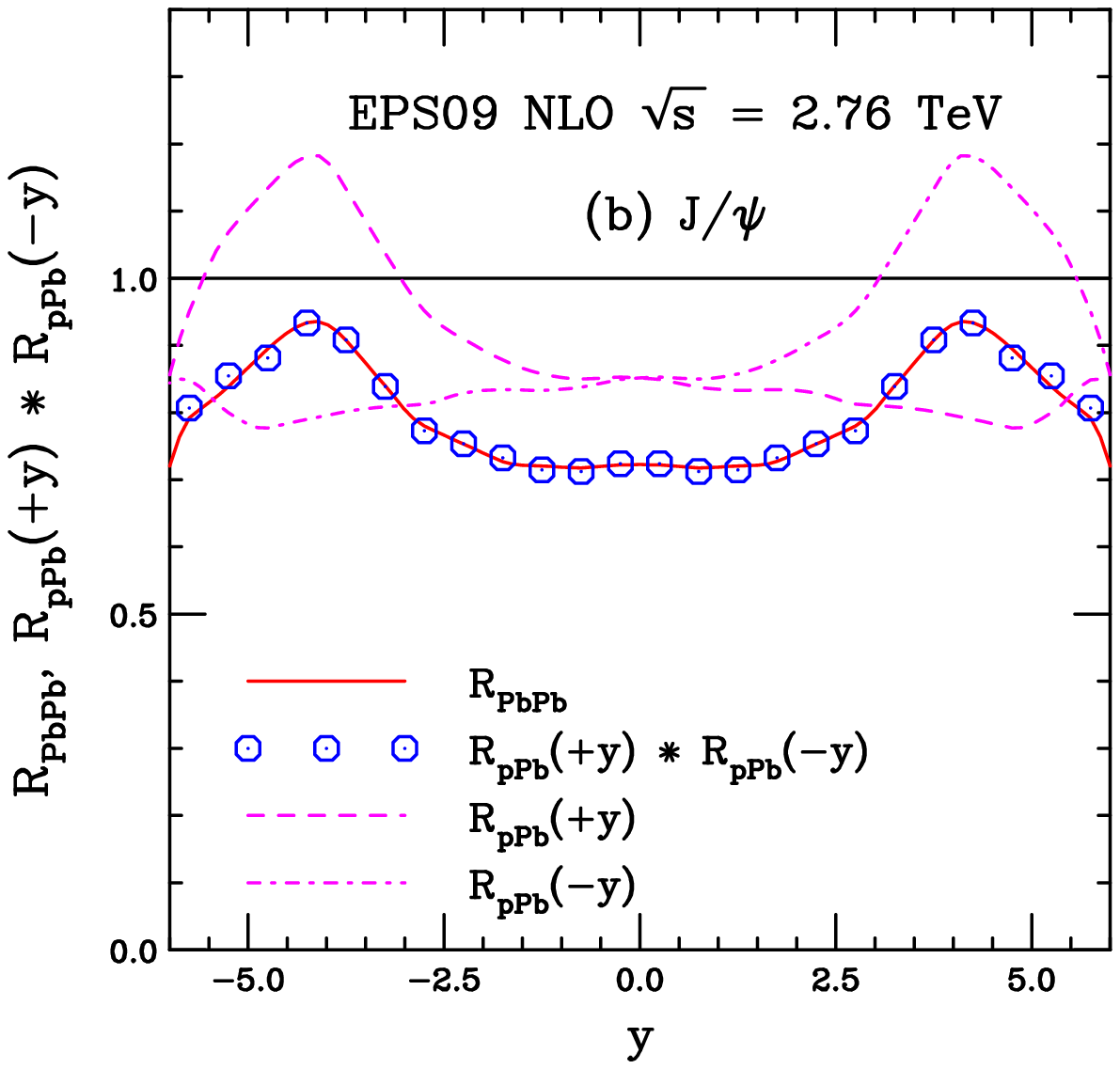} \\
\includegraphics[width=0.45\textwidth]{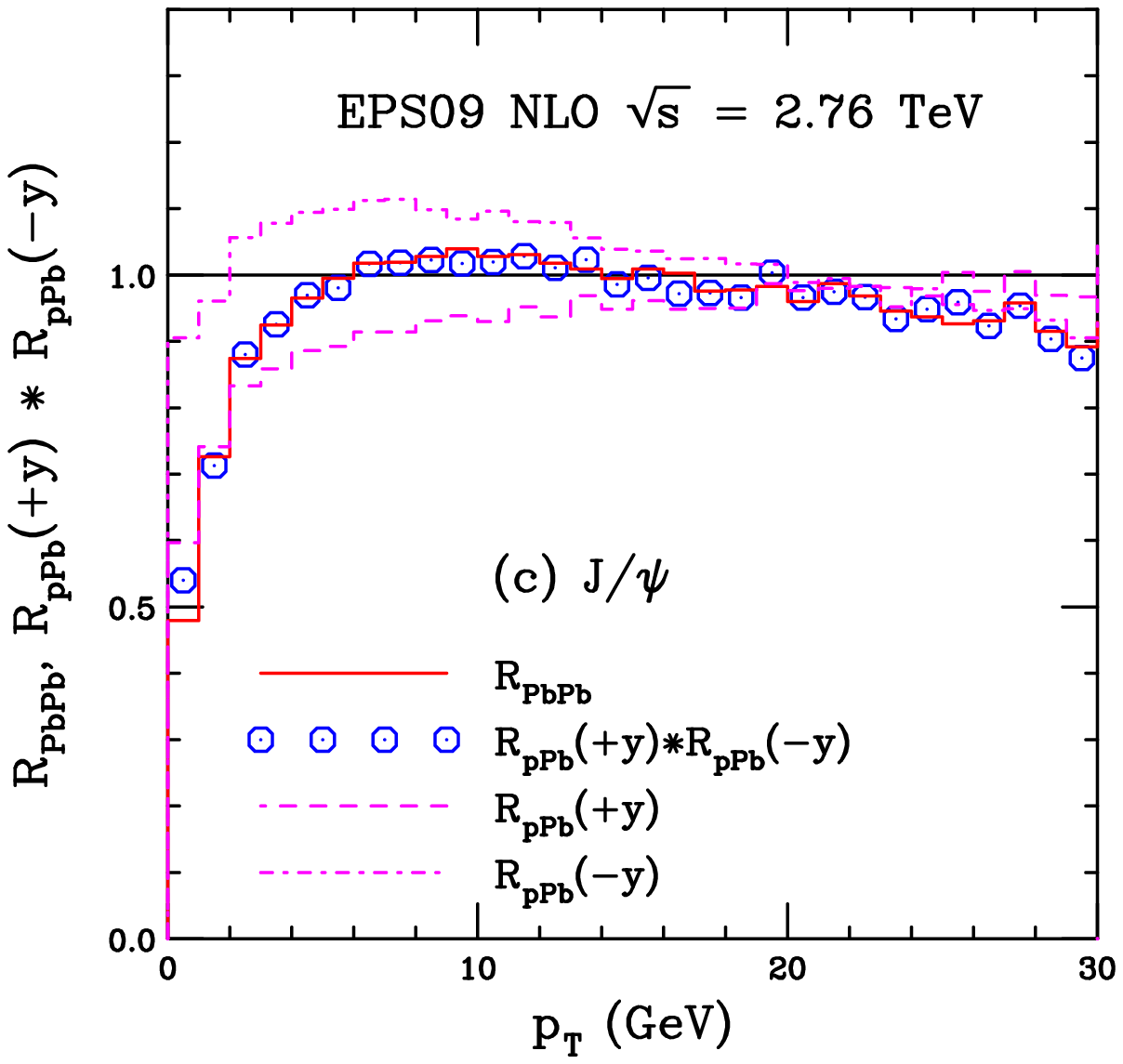}
\caption[]{(Color online)
The $J/\psi$ 
$R_{AA}$ (red) ratio is compared to the product $R_{pA}(+y) \times R_{pA}(-y)$
(points) along with the individual $pA$ ratios at forward (dashed) and backward
(dot-dashed) rapidity.  Results are compared for the rapidity distributions
at LO (a) and NLO (b) as well as for the $p_T$ dependence at NLO
(c).
}
\label{fig:fact_Psi}
\end{center}
\end{figure}

\begin{figure}[t]
\begin{center}
%\vspace*{-0.05in}
\includegraphics[width=0.45\textwidth]{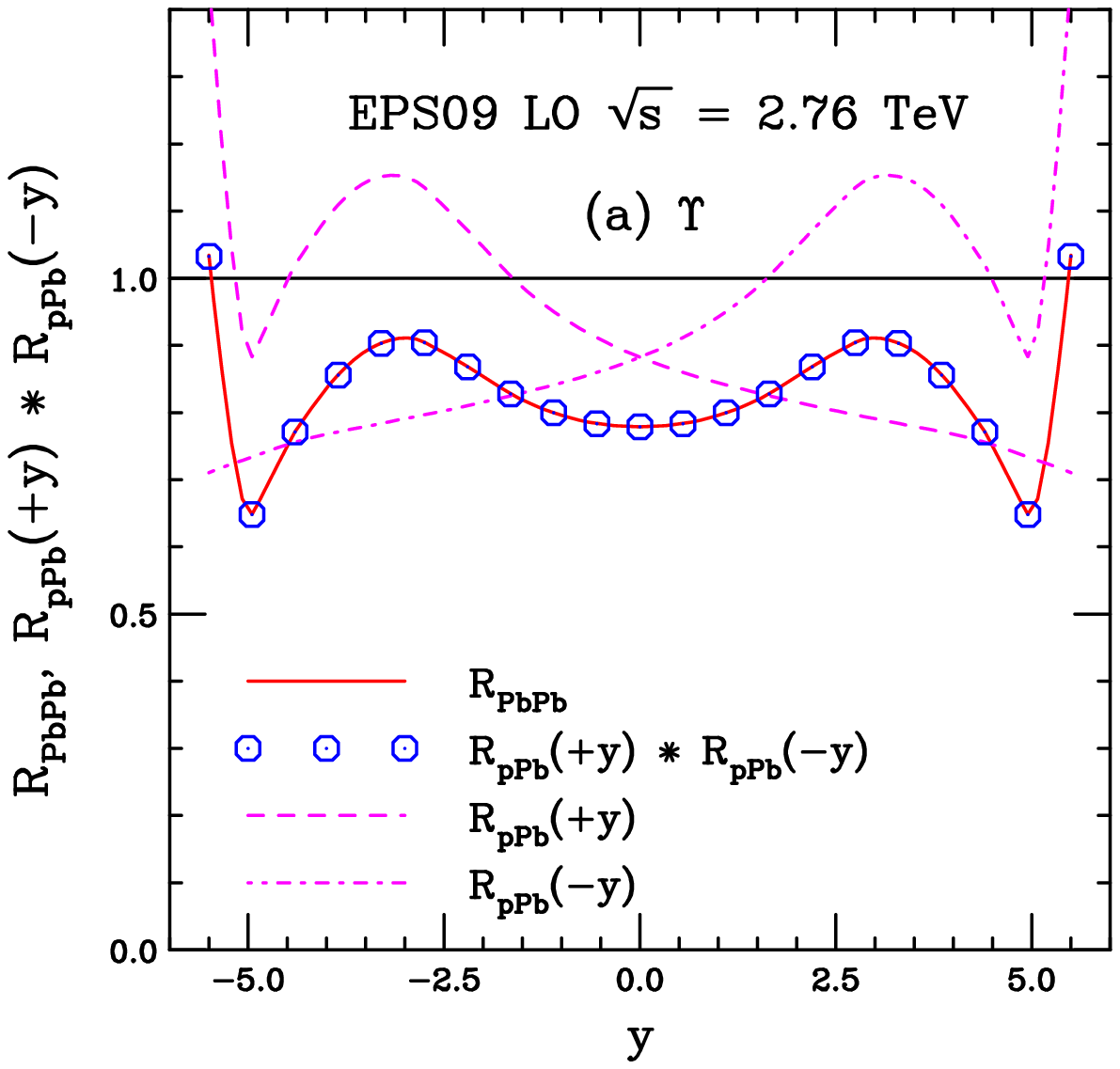}
\includegraphics[width=0.45\textwidth]{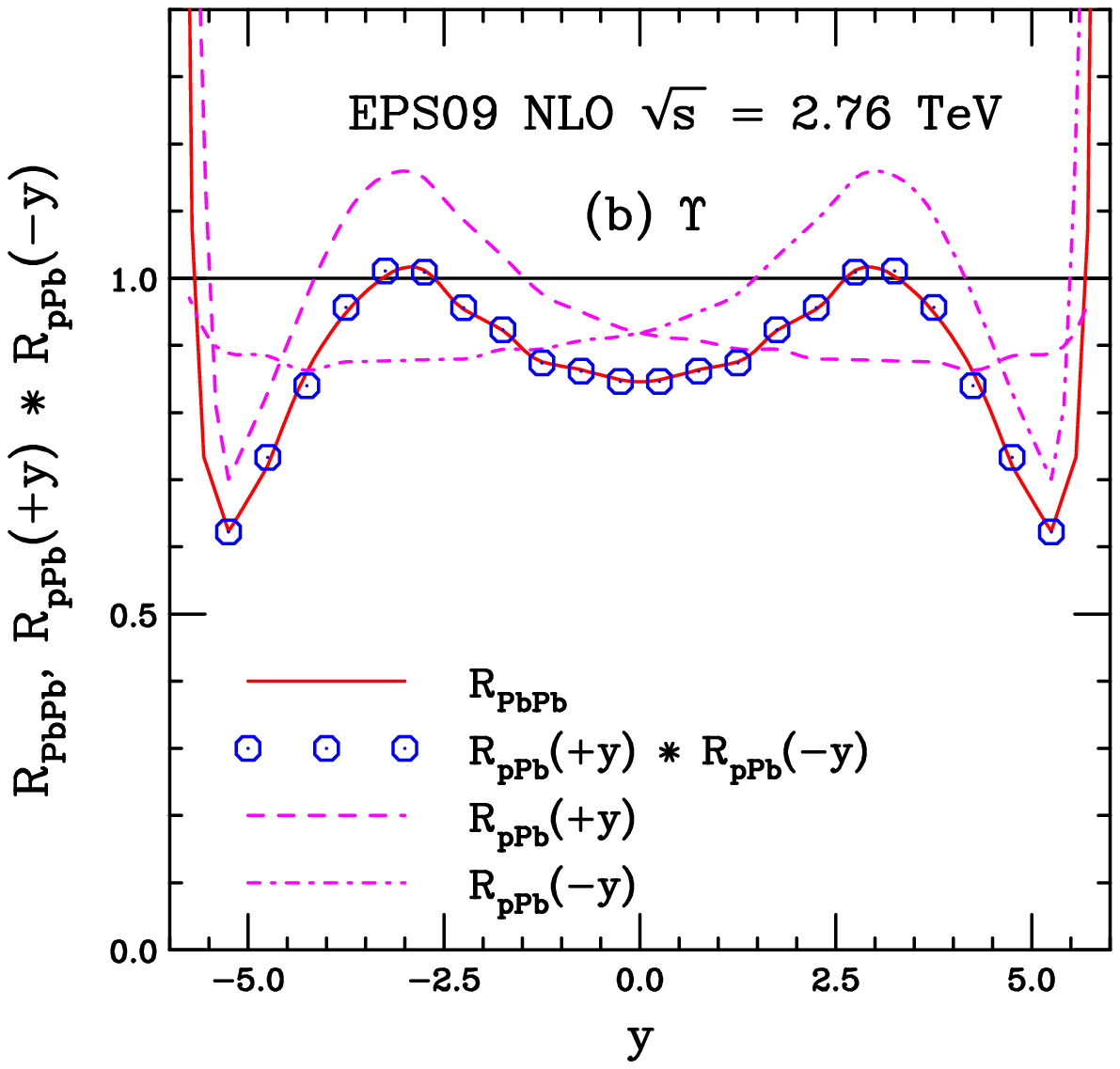} \\
\includegraphics[width=0.45\textwidth]{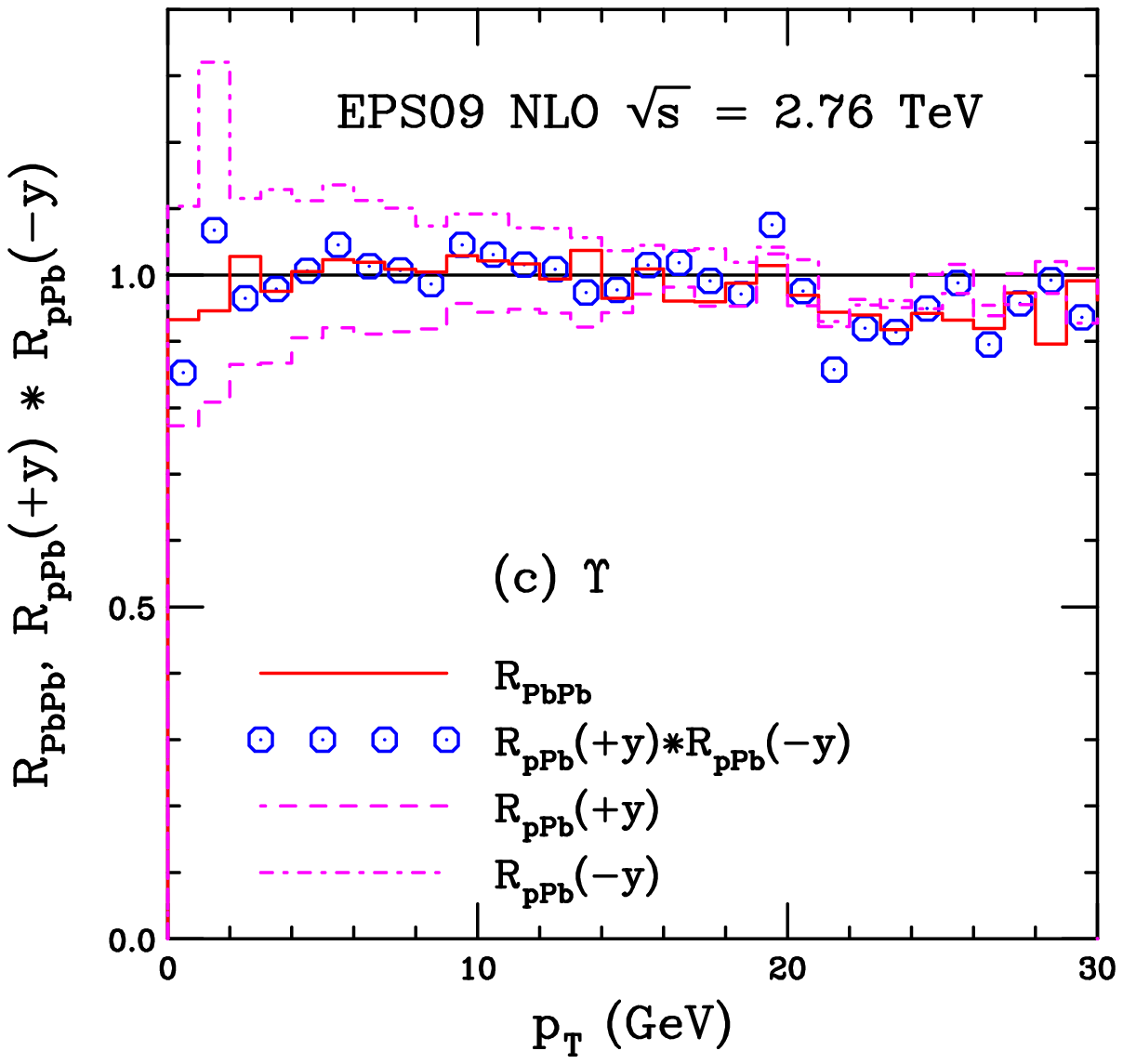}
\caption[]{(Color online)
The $\Upsilon$ 
$R_{AA}$ (red) ratio is compared to the product $R_{pA}(+y) \times R_{pA}(-y)$
(points) along with the individual $pA$ ratios at forward (dashed) and backward
(dot-dashed) rapidity.  Results are compared for the rapidity distributions
at LO (a) and NLO (b) as well as for the $p_T$ dependence at NLO
(c).
}
\label{fig:fact_Ups}
\end{center}
\end{figure}

It is straightforward to make this comparison at leading order in the
color evaporation model (CEM) 
where the $p_T$ of the $Q \overline Q$ pair is zero
and the $x_1$ and $x_2$ values are related to the quarkonium rapidity by
$x_1, x_2 = (M/\sqrt{s})\exp(\pm y)$.  As long as factorization is assumed
for the parton densities, the individual shadowing ratios applied to each also
factorize.  The result is compared for the EPS09 LO central shadowing set
in Figs.~\ref{fig:fact_Psi}(a) and \ref{fig:fact_Ups}(a)
for $\sqrt{s_{NN}} = 2.76$ TeV.  
In these $2 \rightarrow 1$ kinematics, the relation 
$R_{AA}(y) = R_{pA}(y) \times R_{pA}(-y)$ is exact.  

Here the dashed and dot-dashed curves show the ratios $R_{p{\rm Pb}}$ at positive
and negative rapidities respectively.  (In this case, the $-y$ refers to
the ion beam moving toward positive rapidity while $+y$ is the standard 
assumption that the proton beam moves toward positive rapidity, the convention
adopted throughout this paper.)  The blue dots represent the product
of $R_{p{\rm Pb}}(+y)$ and $R_{p{\rm Pb}}(-y)$.  At LO, they lie exactly on top
of the red solid curve calculated for Pb+Pb.

A consequence of the lower energy is that the antishadowing peak moves close
to midrapidity.  This effect is most clearly seen for the $\Upsilon$ in
Fig.~\ref{fig:fact_Ups}(a) where the drop into the EMC region and the subsequent
rise into the Fermi motion region at high $x$ is clearly visible.

It is also obvious, both for the $J/\psi$ and $\Upsilon$, that the
Pb+Pb ratio gives stronger shadowing at midrapidity than at forward (and
backward) rapidity since, at $y \sim 0$, the ratios $R_{p{\rm Pb}}$ at $\pm y$ are
both less than unity while, at $2.5 < |y| < 4.0$, these ratios are in the
antishadowing region.  Therefore, the combination of the two in Pb+Pb
interactions is always less than unity with stronger shadowing at midrapidity.

This was also the case for RHIC cold matter $A+A$ calculations at 
$\sqrt{s_{_{NN}}} =200$ GeV \cite{rhicii}.
More suppression is predicted at
$y=0$ than at forward and backward rapidities with all the shadowing
parameterizations for both $J/\psi$ and $\Upsilon$ production.
At RHIC, the $A+A$ data are more suppressed at forward rapidity than at
central rapidity, both in the minimum bias data as a function of rapidity
and as a function of collision centrality, as quantified by the number of
participant nucleons.  Standard models of shadowing alone or shadowing with
absorption by nucleons in cold nuclear matter or shadowing combined with
dissociation in a quark-gluon plasma leads to strong suppression at central
rapidities.  However, $J/\psi$ regeneration by dynamical coalescence of $c$ and 
$\overline c$ quarks in the medium \cite{QQ_yellow,bob}
is biased toward central rapidities and could
lead to more suppression at forward rapidity relative to central rapidity
since the rapidity distribution of $J/\psi$ production by coalescence is
expected to be narrower than the initial $J/\psi$ rapidity distribution 
\cite{bob}.  Thus, with coalescence, there should be more suppression at forward
$y$ than at midrapidity.  Statistical coalescence \cite{Andronic}, which is
based only on the energy density of the medium, which is higher at $y \sim 0$,
would also support this picture.

The same trend has been observed at the LHC, see Ref.~\cite{StrongDoc} and 
references therein.  
Coalescence production of the $J/\psi$ should be 
more important than at RHIC since more $c \overline c$ pairs are created in a
central Pb+Pb collision at $\sqrt{s_{_{NN}}} = 2.76$ TeV.  We can also expect
$\Upsilon$ production by coalescence at the LHC to be similar to that expected
for the 
$J/\psi$ at RHIC since the $b \overline b $ production cross section at the LHC
will be similar to the $c \overline c$ production cross section at RHIC
\cite{rhicii}.

At next-to-leading order, the assumption of factorization of shadowing is less 
straightforward because quarkonium production in the CEM includes a large
contribution from $2 \rightarrow 3$ diagrams so that the correlation between
the initial momentum fractions $x_1$ and $x_2$ with the rapidity of the 
quarkonium state is weaker, particularly for high $p_T$.  
However, factorization is seen to still hold
as a function of rapidity at NLO, as shown in
Figs.~\ref{fig:fact_Psi}(b) and \ref{fig:fact_Ups}(b), 
calculated with the EPS09 NLO central set, 
also at $\sqrt{s_{NN}} = 2.76$ TeV.  There are some small fluctuations in
the points relative to the curve but these are within the size of the points and
are therefore negligible.

Note, however, that there is a significant 
difference in the shape of the shadowing ratios as a function of rapidity
between the LO and NLO results of Figs.~\ref{fig:fact_Psi}(a) and (b) as well
as between Figs.~\ref{fig:fact_Ups}(a) and (b).    
As discussed in Sec.~\ref{SubSec:LOvsNLO},
this is due to the differences in the EPS09
LO and NLO sets themselves and gives some indication of the uncertainty 
inherent in the extraction of the nuclear gluon density.  

Finally, we turn to the $p_T$ dependence at NLO, unavailable in the CEM at LO.
We might expect to see the largest deviations in this comparison because of
the different kinematics in $2 \rightarrow 2$ and $2 \rightarrow 3$ interactions
in the CEM at NLO.  At high $p_T$, the $2 \rightarrow 3$ kinematics is dominant.
However, the agreement between the factorized product and the direct Pb+Pb
calculation is very good for both $J/\psi$ and $\Upsilon$ up to quite high
$p_T$, see Figs.~\ref{fig:fact_Psi}(c) and \ref{fig:fact_Ups}(c).  
(The result is shown up to $p_T = 30$ GeV.)  There are more fluctuations in the
$\Upsilon$ calculations.  Despite this, the agreement is still very good.

The comparisons shown here, at the same energy for both $p+$Pb and Pb+Pb
collisions, demonstrate that the cold matter effects due to shadowing in 
heavy-ion collisions can be effectively deduced from proton-nucleus collisions,
preferably at the same energy.

\subsection{Comparison to RHIC results}
\label{SubSec:RHIC}

\begin{figure}[t]
\begin{center}
\includegraphics[width=0.45\textwidth]{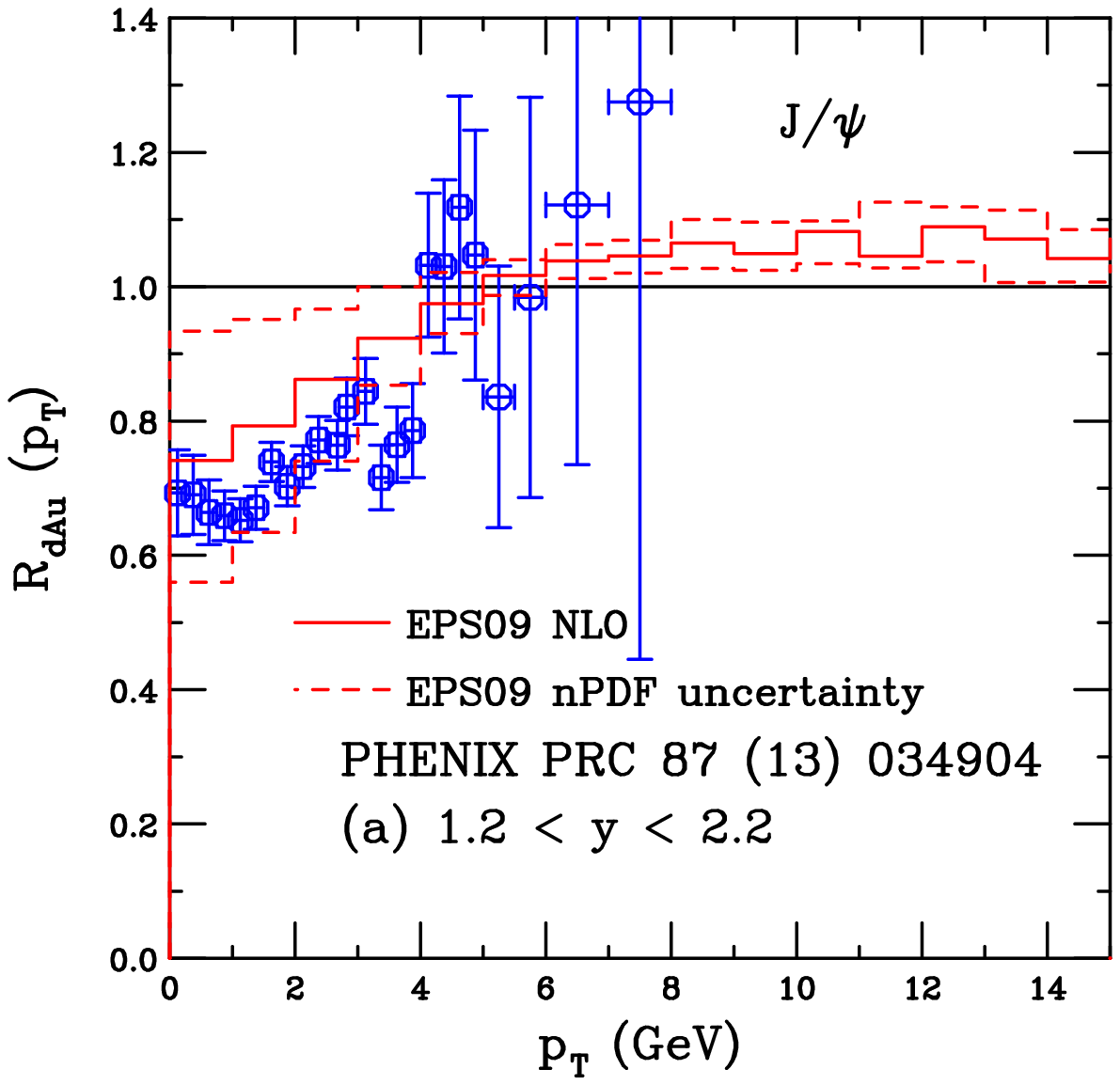}
\includegraphics[width=0.45\textwidth]{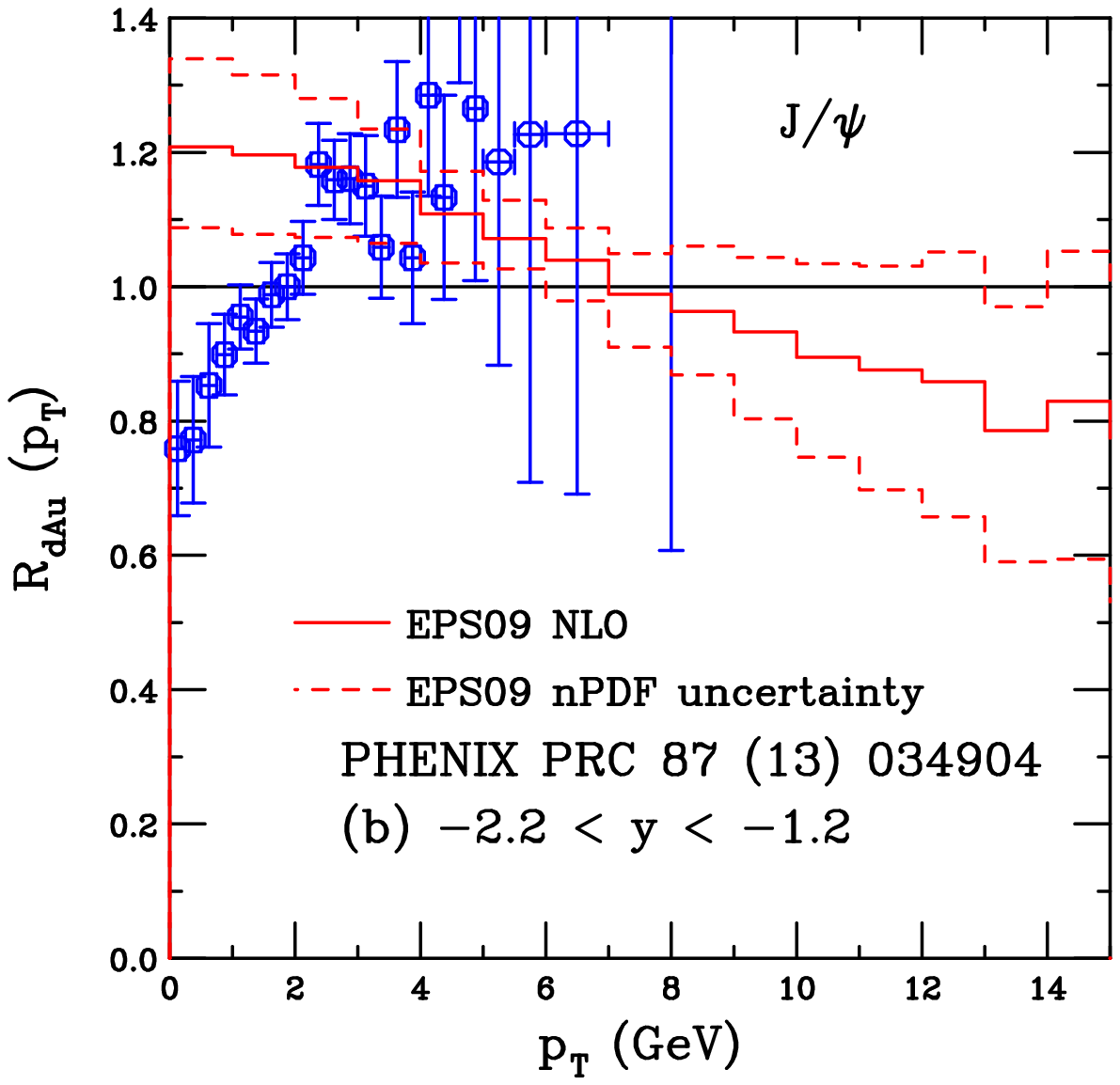} \\
\includegraphics[width=0.45\textwidth]{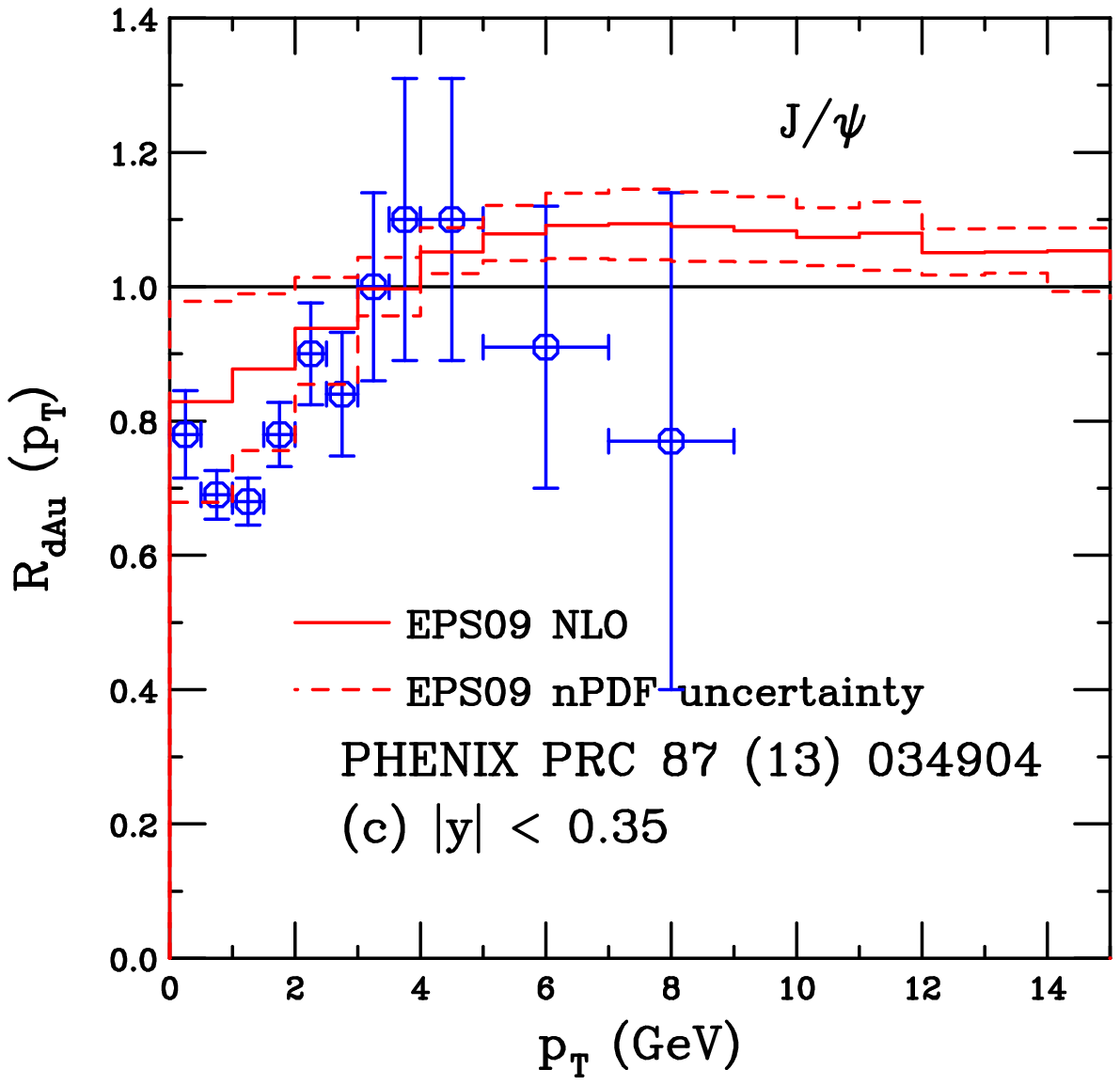}
\includegraphics[width=0.45\textwidth]{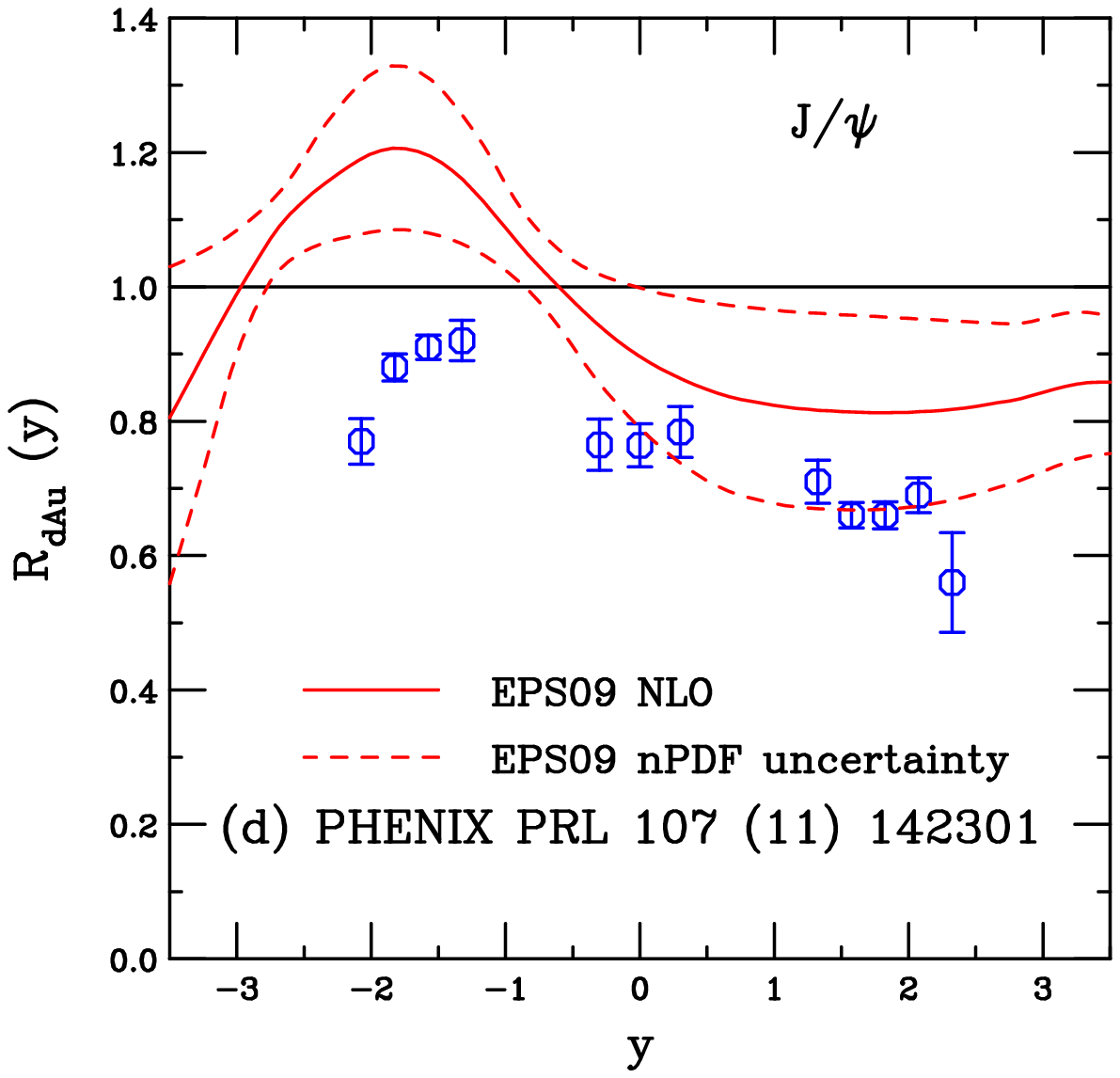}
\caption[]{(Color online)
The $J/\psi$ ratio $R_{{\rm dAu}}(p_T)$ from PHENIX at forward (a),
backward (b) and central (c) rapidity \protect\cite{PHENIX}.  The ratio 
$R_{{\rm dAu}}(y)$ \protect\cite{PHENIX_ydep} is shown (d).
The EPS09 NLO uncertainty band  
is compared to the data.
}
\label{fig:RdAu_Psi}
\end{center}
\end{figure}

\begin{figure}[t]
\begin{center}
\includegraphics[width=0.45\textwidth]{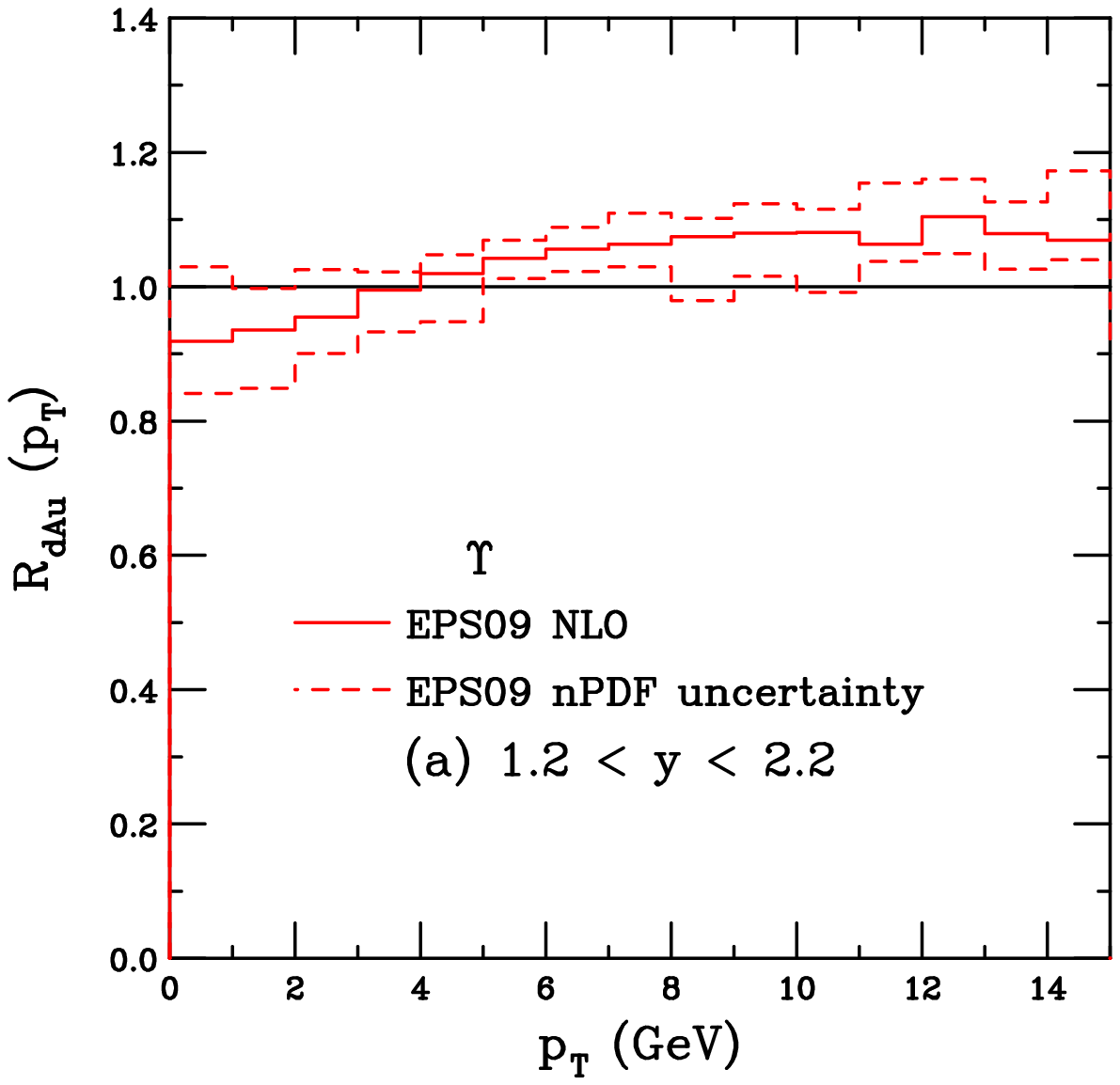}
\includegraphics[width=0.45\textwidth]{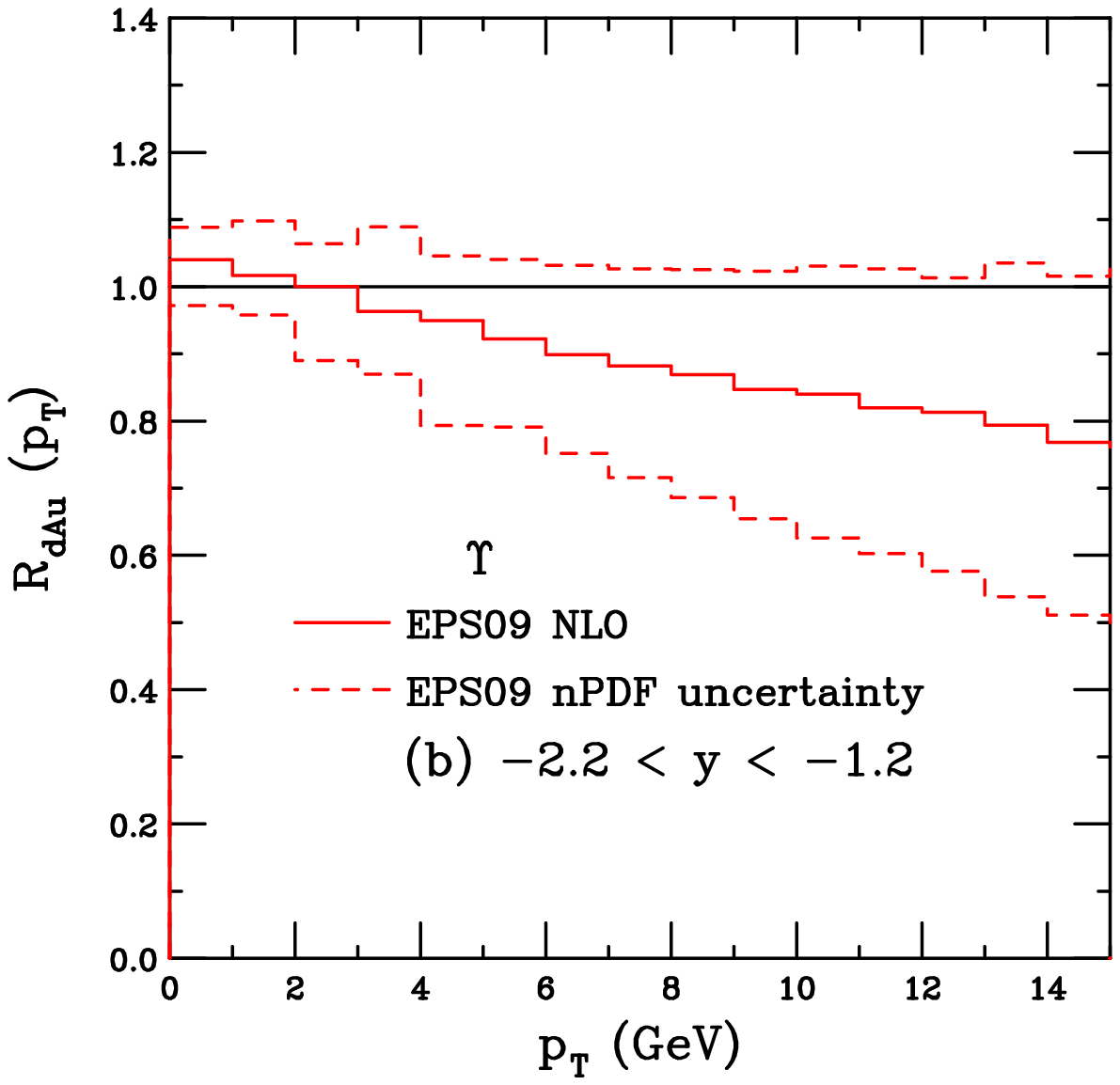} \\
\includegraphics[width=0.45\textwidth]{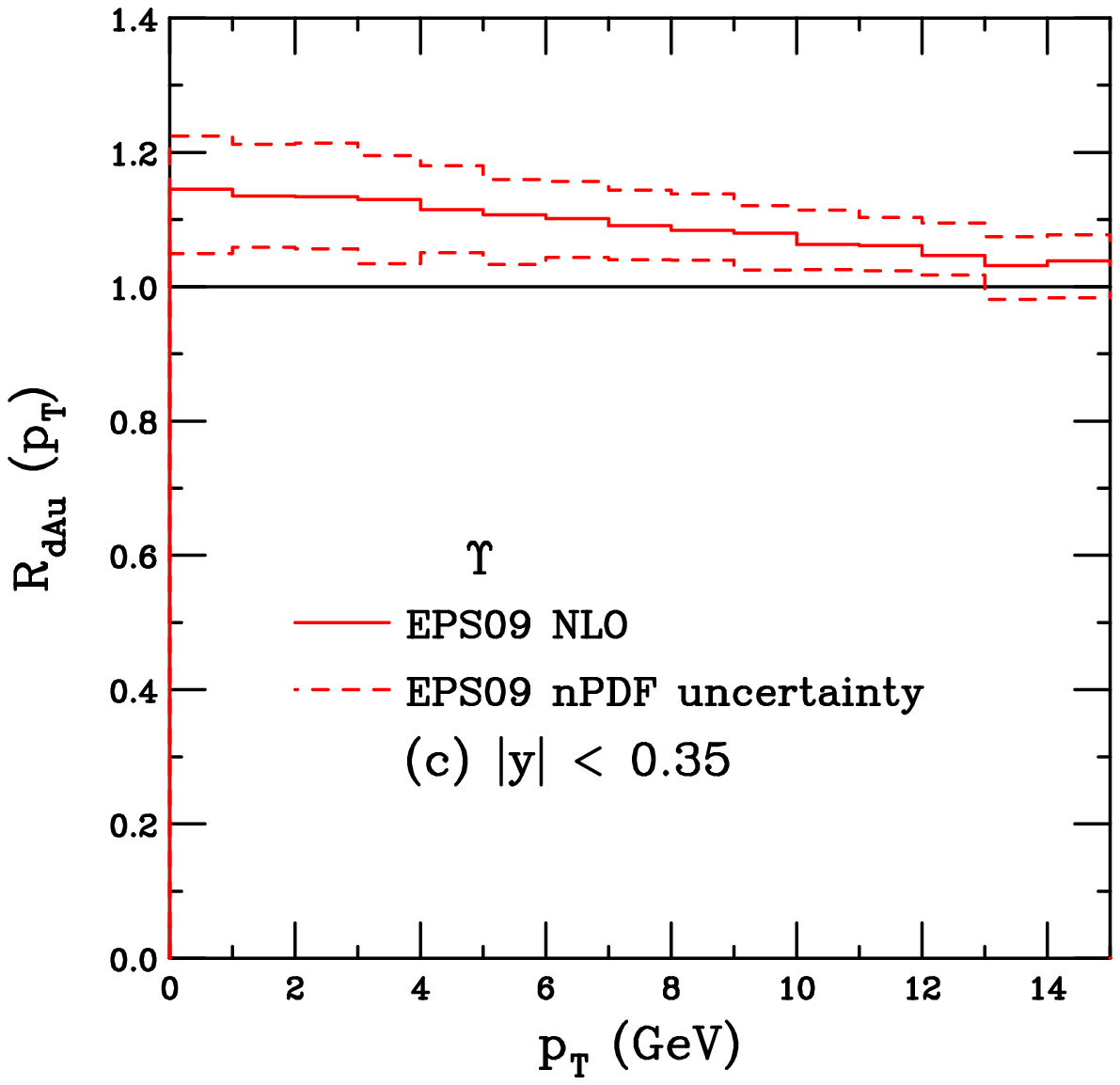}
\includegraphics[width=0.45\textwidth]{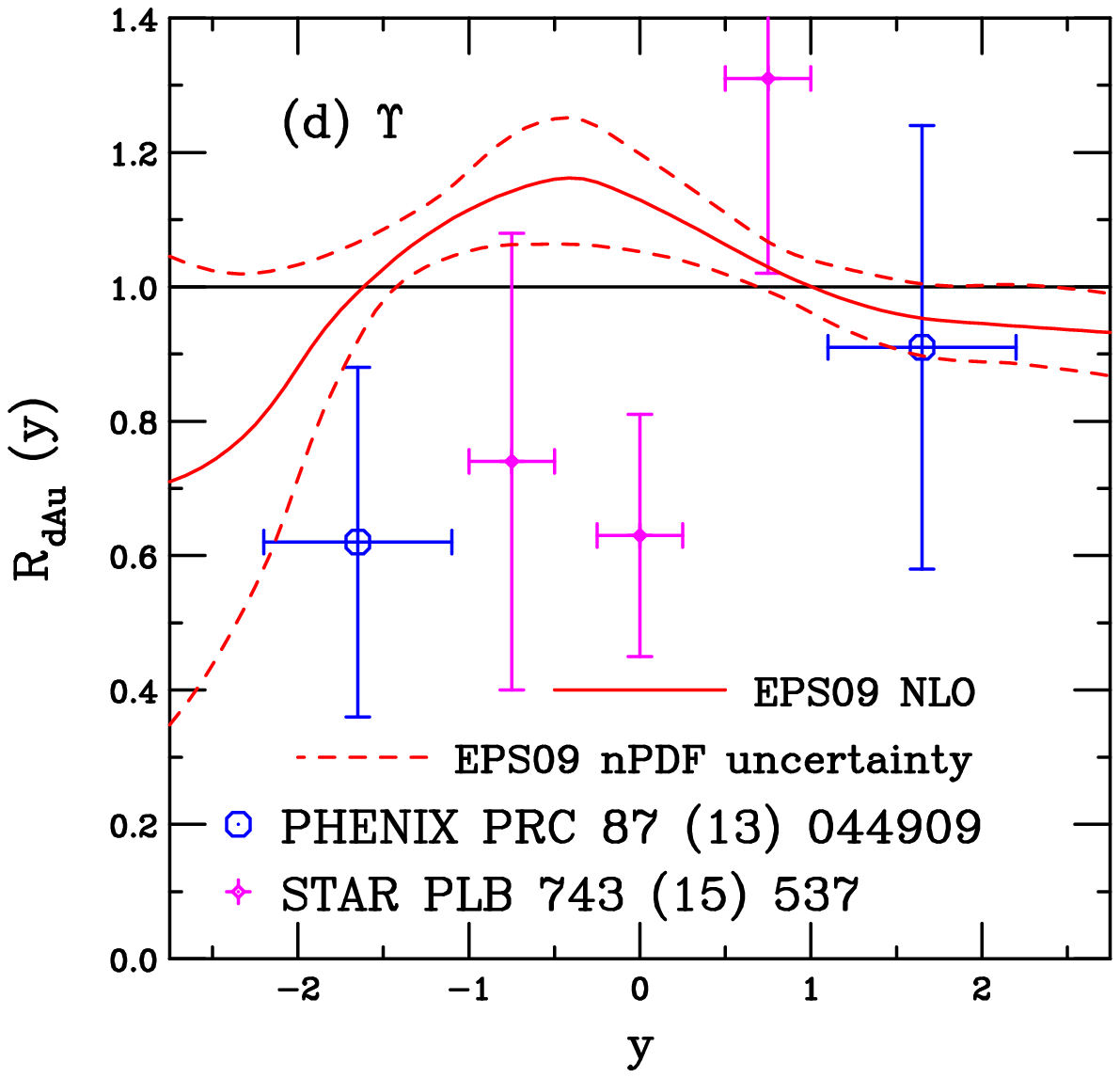}
\caption[]{(Color online)
The $\Upsilon$ ratio $R_{{\rm dAu}}(p_T)$ for 
RHIC at forward (a),
backward (b) and central (c) rapidity.  The ratio 
$R_{{\rm dAu}}(y)$ is also shown (d).
The EPS09 NLO uncertainty band 
is compared to the PHENIX \protect\cite{PHENIX_Ups} and 
STAR \protect\cite{STAR_Ups} data on $R_{\rm dAu}(y)$ data in (d).
}
\label{fig:RdAu_Ups}
\end{center}
\end{figure}

Finally, we compare the EPS09 NLO calculations to the RHIC $J/\psi$
and $\Upsilon$ data in Figs.~\ref{fig:RdAu_Psi} and \ref{fig:RdAu_Ups}.
No absorption is included here, even though the absorption cross section may
be non-negiglble at RHIC energies.  The EPS09 LO calculations were employed
by the PHENIX Collaboration \cite{PHENIX} to
extract the putative absorption cross section required to make a shadowing 
and absorption scenario agree with the data.  The NLO calculations are
presented as a function of $p_T$ and rapidity together for the first time.

The $J/\psi$ $p_T$-dependent data, shown in Fig.~\ref{fig:RdAu_Psi}(a)-(c),
follows approximately the same trend at forward, backward and midrapidity.
At $p_T \sim 0$, $R_{\rm dAu} \sim 0.7-0.8$.  The ratio then increases with $p_T$
until becoming compatible with unity, albeit it with rather poor statistics,
at $p_T \geq 4$ GeV.  At forward and midrapidity the calculations agree
with the trends of the data rather well.  At low $p_T$, the $x$ range is in
the shadowing region but moves toward the antishadowing region as $p_T$
increases, see the shape of $R_{\rm dAu}(y)$ in Fig.~\ref{fig:RdAu_Psi}(d).
However, at backward rapidity at RHIC, the calculations suggest that, at low
$p_T$, the antishadowing region is probed while, at higher $p_T$, the EMC
region is reached.  Thus, for $p_T < 2$ GeV, the trend of the data and the
calculations are opposite.  For $p_T > 2$ GeV, the uncertainties in the data
and the calculations are large enough for the results to be compatible.

A recent calculation studied the centrality dependence of gluon shadowing.  
The impact parameter dependence, together with a rapidity-dependent
absorption cross section, was extracted from the centrality dependence of
$R_{\rm dAu}(y)$ \cite{FMV}.  The large absorption cross section required at
backward rapidity was explained in the context of an expanding color octet
that reaches its final-state size at backward rapidity but has a negligible
effect at forward rapidity because the primordial $J/\psi$ passes through the
target before reaching its full size \cite{Arleo}.  Presumably, this would
also have the desired effect on the $p_T$ dependence in the backward region
since low $p_T$ $J/\psi$'s will be strongly absorbed in this region while those
at higher $p_T$ will again pass through the transverse direction of the target
before fully forming, maintaining the agreement of the calculation with the 
data at higher $p_T$.

Finally, we discuss the $\Upsilon$ results shown in Fig.~\ref{fig:RdAu_Ups}.
There is only limited data for the $\Upsilon$, including the low statistics
PHENIX \cite{PHENIX_Ups} and STAR \cite{STAR_Ups}
data presented in Fig.~\ref{fig:RdAu_Ups}(d).  The uncertainties on the
measurements, along with the large uncertainties in the calculation at backward
rapidity, allow the calculation and the data to agree within the errors,
except for the STAR point at $y \sim 0$.  Although it has the lowest statistical
uncertainty, it exhibits strong suppression in a rapidity region where
antishadowing is predicted.
The $p_T$ dependence has not yet been measured.  

The calculated $p_T$ dependence in the forward region, 
Fig.~\ref{fig:RdAu_Ups}(a), is already entering the 
antishadowing region for $p_T > 6$ GeV.  At midrapidity, the entire $p_T$
range is in this region so that $R_{\rm dAu}(p_T) > 1$ for all $p_T$.  On the
other hand, at backward rapidity, most of the $p_T$ region is, in fact, in the
EMC region where the uncertainies in the calculated $R_{\rm dAu}(p_T)$ are large,
see Fig.~\ref{fig:RdAu_Ups}(b).  

It is unlikely that the situation for $\Upsilon$ can be
improved with the current RHIC detectors.  However, the future sPHENIX
detector \cite{sPHENIX}
is expected to be able to separate the three $\Upsilon$(S) states.
This improved mass resolution, together with the higher efficiency expected
for sPHENIX and higher luminosity of the sPHENIX runs, means that the
$p_T$ dependence could be measured, if a d+Au run is made during the sPHENIX
lifetime.

\section{Summary}

The results with EPS09 NLO shadowing agree with the measured $R_{p{\rm Pb}}$ 
within the uncertainties of both the calculation and the data.  However,
the forward-backward ratios, independent of the interpolated $p+p$ baseline,
are not as well described.  The older EKS98 LO parameterization, although
not applied consistently in a NLO calculation, does the best job of
describing all the data.  

The shadowing parameterizations used in our study exhibit a wide
range of behavior for the nuclear gluon density at low $x$, an $x$ region
outside the current range of the fits from fixed-target nDIS
data at higher $x$ and low $\mu^2$.  If nuclear data
were available from high energy
$e+A$ collisions, the nuclear gluon densities could be
more precisely pinned down by global analyses of
the scale dependence of the nuclear structure functions.  In hadroproduction,
direct photon or open charm production, dominated by gluon-induced processes
but without the  additional complexities of nuclear absorption, could be
utilized to study the nuclear gluon density.  The LHC data, with the lower $x$
reach, should be
incorporated into a new fit of the nuclear parton densities.
%Any new $e+A$ data before an electron ring is available at the LHC will be at lower energies than previously available at HERA, reducing the potential overlap of the low $x$ range between an electron-ion collider and the LHC. 

We note that, since we have assumed
absorption is negligible at the LHC and include no other cold nuclear matter
effect, the uncertainties on the ratios can be obtained from the EPS09 NLO
bands shown in the figures.  However, if other effects are incorporated, a more
extensive error analysis, including the uncertainties on these other effects,
is necessary.

\section*{Acknowledgements}

The numerical values of the ratios shown in this paper are available from
the author.  We thank R. Arnaldi, W. Brooks, K. J. Eskola, and E. Scomparin
for discussions.

This work was performed under the auspices of the U.S.\
Department of Energy by Lawrence Livermore National Laboratory under
Contract DE-AC52-07NA27344.  The author would also like to acknowledge the
Institute for Nuclear Theory at the University of Washington in Seattle for
the hospitality at the beginning of this work.

\end{document}